\documentclass[12pt]{article}

\global\arraycolsep=1pt
\usepackage{amsmath,amssymb,amsthm,latexsym,epic,color}

\usepackage[dvipdfmx]{graphicx,xcolor}

\usepackage{ytableau}
\ytableausetup{aligntableaux=top,mathmode,boxsize=1.2em}

\def\young(#1){\ytableaushort{#1}}
\def\yng(#1){\tiny {\ydiagram{#1}}}
\newcommand{\beq}{\begin{equation}}
\newcommand{\eeq}{\end{equation}}
\newcommand{\beqa}{\begin{eqnarray}}
\newcommand{\eeqa}{\end{eqnarray}}
\newcommand{\CR}{\nonumber \\}

\newcommand{\barsigma}{\overline{\sigma}}
\numberwithin{equation}{section}

 \newtheorem{dfn}{Definition}[section]
 
 \newtheorem{prp}[dfn]{Proposition}
 \newtheorem{lem}[dfn]{Lemma}

 \newtheorem*{ack}{Acknowledgements}

\begin{document}
\begin{titlepage}
%%
%\begin{flushright}
%18 September, 2025
%\end{flushright}

\begin{center}
\Large{\bf Super Macdonald polynomials and BPS state counting on the blow-up}
\end{center} 
\bigskip
\bigskip

\begin{center}
\large 
Hiroaki Kanno$^{a,b,}$\footnote{kanno@math.nagoya-u.ac.jp}, 
Ryo Ohkawa$^{c,}$\footnote{ohkawa@kurims.kyoto-u.ac.jp}, and
Jun'ichi Shiraishi$^{d,}$\footnote{shiraish@ms.u-tokyo.ac.jp} \\

\bigskip

$^a${\small {\it Graduate School of Mathematics, Nagoya University,
Nagoya 464-8602, Japan}}\\
$^b${\small {\it Kobayashi-Maskawa Institute (KMI), Nagoya University,
Nagoya 464-8602, Japan}} \\
$^c${\small {\it Research Institute for Mathematical Sciences, Kyoto University, Kyoto
606-8502, Japan}} \\
$^d${\small {\it Graduate School of Mathematical Sciences, University of Tokyo, Komaba, Tokyo 153-8914, Japan}}
\end{center}
\bigskip

\begin{abstract}
We explore the relation of the super Macdonald polynomials
and the BPS state counting on the blow-up of $\mathbb{P}^2$,
which is mathematically described by framed stable perverse coherent sheaves.
Fixed points of the torus action on the moduli space of BPS states
are labeled by super partitions. From the equivariant character of 
the tangent space at the fixed points we can define the Nekrasov factor 
for a pair of super partitions, which is used for the localization computation 
of the partition function. The Nekrasov factor also allows us to compute matrix elements
of the action of the quantum toroidal algebra of type $\mathfrak{gl}_{1|1}$
on the $K$ group of the moduli space. We confirm that these matrix elements are
consistent with the Pieri rule of the super Macdonald polynomials.
\end{abstract}
\end{titlepage}

%%%%%%%%%%%%%%%%%%%%%%%%%%%%%%%%%%%%%%%%%%%%%%%%%%%%%%%%%%%%%%%%%%%%%

\setcounter{tocdepth}{1}
\tableofcontents
\setcounter{footnote}{0}
\section{Introduction}

The Macdonald polynomials play a significant role in the instanton counting 
of the five dimensional supersymmetric gauge theory and topological strings. 
The building block of the partition function, called refined topological vertex, is expressed 
in terms of appropriate specialization of the Macdonald polynomials \cite{Awata:2005fa,Iqbal:2007ii,Awata:2008ed}. 
The underlying algebraic structure behind such a construction is the $\mathfrak{gl}_1$ 
quantum toroidal algebra, which is also called Ding-Iohara-Miki (DIM) algebra. 
In fact the Macdonald polynomials provide a basis of a level zero (vertical) representation of the DIM algebra
\cite{FT,FFJMM,Awata:2011ce}. 
Recall that the Macdonald polynomials $P_{\lambda}(x;q,t)$ give
an orthogonal basis of the ring of symmetric polynomials in $x_1, x_2, \ldots$,
over the field $\mathbb{F}=\mathbb{Q}(q,t)$ of rational functions in $q$ and $t$ \cite{MacD}. 
They are labeled by the set of partitions $\mathcal{P} \ni \lambda$.
The generating function of the number of partitions of a fixed integer $N=|\lambda|$ agrees 
with the character of the Fock space of a free boson;
\begin{equation}
\sum_{\lambda  \in \mathcal{P}} \mathfrak{q}^{|\lambda|} =  \prod_{k=1}^\infty \frac{1}{1-\mathfrak{q}^{k}}.
\end{equation}
Hence the level zero representation of DIM algebra is naturally identified with the Fock module of a free boson.

The super Macdonald polynomial is a generalization of the Macdonald polynomial.
Namely, the super Macdonald polynomials $\mathcal{M}_\Lambda(x,\theta;q,t)$ are polynomials 
in $x_1, x_2, \ldots$ and the Grassmann variables $\theta_1, \theta_2 , \ldots$, which are
invariant under the exchange $(x_i, \theta_i) \leftrightarrow (x_j, \theta_j)$.\footnote{$x$ and $\theta$ are
permutated simultaneously.} 
They are labeled by the set of super partitions $\Lambda$, which are non-increasing sequences of 
elements in $\mathbb{Z}_{\geq 0}/2$;
\cite{Blondeau-Fournier:2011sft,Blondeau-Fournier:2012exj,Alarie-Vezina:2019ohz,Galakhov:2024cry}.
The generating function of the number of super partitions of a fixed level $\ell \in \mathbb{Z}_{\geq 0}/2$ 
agrees the character of the tensor product of the Fock spaces of a free boson and a free NS fermion.
Recently it is pointed out that the super Macdonald polynomials give a basis of a level zero representation
of the quantum toroidal algebra of type $\mathfrak{gl}_{1|1}$ \cite{Galakhov:2024zqn,Galakhov:2025phf}.
For simplicity we will call the quantum toroidal algebra of type $\mathfrak{gl}_{1|1}$, super DIM algebra in this paper.
The super DIM algebra has generators $E_{i,k}, F_{i,k}, K_{i, \pm r}^{\pm}$
and a central element $C$, where $i \in \mathbb{Z}_2, k \in \mathbb{Z}$ and $r \in \mathbb{Z}_{\geq 0}$.\footnote{
Some of the monomials in $K_{i,0}^\pm$ are also central.}
When $C=1$, we say the representation has level zero. 
In this case the Cartan generators $K_{i, \pm r}^{\pm}$ are mutually commuting and
the super Macdonald polynomials are characterized as simultaneous eigenstates of $K_{i, \pm r}^{\pm}$. 
Furthermore, the Pieri rule of  the super Macdonald polynomials is derived from the action of 
the zero modes $E_{i,0}$ \cite{Galakhov:2024zqn}. 

%%%%%%%%%%%%%%%%%%%%%%%%%%%%%%%%%%%%%%%%%%%%%%%%%%%%%%%%%%%%%%%%%%%%%%%%%%%%%%%%%%%%%%%%%%%%%%%%%%%%%%%%%%%%%%%%%%%%%

In this paper we explore the relation of the super Macdonald polynomials 
to the counting problem of the framed sheaves on the blow-up of the complex projective space $\mathbb{P}^2$. 
It should be mentioned that in the rank one case the relation of the Hilbert scheme on the blow-up
to the super affine Yangian of type $\mathfrak{gl}_{1|1}$ and the Fock space of a free fermion was also discussed 
in \cite{Zhao:2024myn}. 
Recall that the fixed points of the torus action on the moduli space of instantons (framed torsion free sheaves) 
on $\mathbb{P}^2$ are in one to one correspondence with the partitions \cite{Nakajima:2003pg}. 
Our work is inspired by the fact that the fixed points of the torus action on the moduli space of 
framed stable perverse coherent sheaves on the blow-up of $\mathbb{P}^2$ are similarly 
labeled by the super partitions. 
More precisely, let us consider the blow-up $p : \widehat{\mathbb{P}}^2 \longrightarrow \mathbb{P}^2$ 
at a point $0$ with the exceptional curve $C$. 
For a stability parameter $m \in \mathbb{Z}$ and homological data $c = (r, c_1, \mathrm{ch}_2)
\in H^{0}(\widehat{\mathbb{P}}^2) \oplus H^{2}(\widehat{\mathbb{P}}^2) \oplus H^{4}(\widehat{\mathbb{P}}^2)$, 
let $\widehat{M}^m(c)$ be the moduli space of coherent sheaves $E$ with $\operatorname{ch}(E) =c$ 
such that  $E(-mC)$ is stable perverse coherent (See \cite{Nakajima:2008ss} for a precise definition). 
We have
\begin{lem}[\cite{Nakajima:2008ss} Lemma 5.1]
The torus fixed points in $\widehat{M}^0(c)$ are in bijection to
$r$-tuples of pairs $(Y_\alpha, S_\alpha)$ of a Young diagram $Y_\alpha$ 
and a subset $S_\alpha$ of removable boxes such that\footnote{The second Chern
character is $\operatorname{ch}_2 = \frac{1}{2}( c_1^2 -2 c_2)$. 
Recall that $C$ is a $(-1)$-curve; $([C], [C]) =-1$. }
$$
\sum_\alpha |S_\alpha| = (c_1, [C]), \qquad \sum_\alpha |Y_\alpha| 
= - \int_{\widehat{\mathbb{P}}^{2}}\operatorname{ch}_2 + \frac{1}{2} (c_1, [C]).
$$
\end{lem}

As we will explain in section \ref{sec:superY}, $\Lambda_\alpha = (Y_\alpha, S_\alpha)$ can be identified with 
an $r$-tuple of super partitions. A box in $S_\alpha \subset Y_\alpha$ is called marked. 
In \cite{Nakajima:2008ss} the equivariant character of the tangent space at a fixed point $\Lambda_\alpha$
is computed. From the formula of the equivariant character we can define the Nekarsov factor\footnote{
In this paper we always consider $K$-theoretic Nekrasov factor for the up-lift to five dimensional gauge theories.}
for a pair of super partitions as a generalization of the standard Nekrasov factor 
which is a basic block for the instanton partition function and topological string amplitudes on
toric Calabi-Yau three-folds. 
\begin{dfn}
For a pair of super partitions $\Lambda_\alpha=(Y_\alpha, S_\alpha)$ and $\Lambda_\beta =(Y_\beta, S_\beta)$,
we define the Nekrasov factor by 
$$
 \mathsf{N}_{\Lambda_\alpha, \Lambda_\beta} (u \vert q, t) 
 = {\prod_{\mathsf{s} \in Y_\alpha \setminus S_\alpha}}^{\hspace{-2mm}\prime}
  (1- u t^{-\ell_{Y_\beta}(\mathsf{s})} q^{-a_{Y_\alpha\setminus S_\alpha}(\mathsf{s})-1})
 {\prod_{\mathsf{t} \in Y_\beta}}^{\prime} (1- u t^{\ell_{Y_\alpha\setminus S_\alpha}(\mathsf{t})+1} q^{a_{Y_\beta}(\mathsf{t})}),
$$
where ${\prod}^\prime$ means restricting the product to the set of the relevant boxes
(see sections \ref{sec:superY} and \ref{sec:framed-sheaves} for the definition)
and $a_Y(\mathsf{s})$ and $\ell_Y(\mathsf{s})$ are the arm length and the leg length of $\mathsf{s}$ with respect to $Y$. 
We have introduced the spectral parameter $u$ for later convenience.
\end{dfn}

When the difference of the size of a pair of partitions is $\pm 1$,
the Nekrasov factor gives us matrix elements of the generators $E_k$ and $F_k$ 
of the DIM algebra, whose geometric action on the $K$ group of the moduli space is
defined by the correspondence \cite{Nakajima-AMS}. 
They are called matrix elements in the fixed point basis.
On the other hand the level zero representation of the DIM algebra is constructed
by the semi-infinite tensor product of the vector representations and Macdonald 
polynomials give a basis of the the level zero representation.
It is known that the fixed point basis from the geometry of the moduli space 
corresponds to the integral form of the Macdonald polynomials.
In this paper we show that an exactly parallel claim holds for the super Macdonald 
polynomials. 
For a super partition $\Lambda=(Y,S)$, we define 
\begin{equation}
\widetilde{c}_\Lambda(q,t) := \prod_{\mathsf{s} \in \mathcal{B}(\Lambda)}
 (1- t^{\ell_{Y/S}(\mathsf{s})+1} q^{a_{Y}(\mathsf{s})}),
\end{equation}
where $\mathcal{B}(\Lambda)$ denotes the set of relevant boxes in $\Lambda=(Y,S)$.
Then the main result in this paper is
\begin{prp}
\label{main-prp}
Let $ \Lambda + \boxtimes_k$ denote
the super partition obtained by adding a marked box in the  (bosonic) $k$-th row
and $\Lambda(\boxtimes_k \to \square_k)$ denote the super partition obtained by 
replacing the marked box in the  (fermionic) $k$-th row by an unmarked box. 
Then
\begin{align}\label{main-equality1}
&\operatorname{Res}_{u=1}
\left(\frac{\mathsf{N}_{\Lambda, \Lambda}(u \vert q,t)}{\mathsf{N}_{\Lambda,  \Lambda + \boxtimes_k}(u \vert q,t)} \right)
\frac{\widetilde{c}_{ \Lambda + \boxtimes_k}(q,t) }{\widetilde{c}_\Lambda(q,t) } \CR
& \qquad = (-q)^{F(k)}(t-1) t^{k-1}\cdot \mathsf{f}^{(k)}(q,t)^{-1} \prod_{i=1}^{k-1} \frac{1 - t^{k-i-1}q^{\lambda_i -\lambda_k -\barsigma_i}}
{1- t^{k-i}q^{\lambda_i -\lambda_k - \barsigma_i}}, \\
\label{main-equality2}
&\operatorname{Res}_{u=1}
\left(\frac{\mathsf{N}_{\Lambda, \Lambda}(u \vert q,t)}{\mathsf{N}_{\Lambda, \Lambda(\boxtimes_k \to \square_k)}(u \vert q,t)} \right)
\frac{\widetilde{c}_{\Lambda(\boxtimes_k \to \square_k)}(q,t) }{\widetilde{c}_\Lambda(q,t) } \CR
& \qquad = (-1)^{F(k)} (t-1) \cdot \mathsf{f}^{(k)}(q,t) \prod_{i=1}^{k-1}\frac{1 - t^{k-i+1}q^{\lambda_i -\lambda_k}}
{1- t^{k-i}q^{\lambda_i - \lambda_k}},
\end{align}
where $F(k)$ is the number of fermionic rows above the $k$-th row
and $\mathsf{f}^{(k)}(q,t)$ is the monomial factor defined by \eqref{f-framing}. 
\end{prp}
After substituting $(q,t) \to (q^2, t^2)$, up to monomial factors \eqref{main-equality1} and  \eqref{main-equality2} agree 
with the coefficients of the Pieri formulas \eqref{matrix-elem-super1} and \eqref{matrix-elem-super2} 
for the super Macdonald polynomials, 
which are derived from the action of the zero modes $E_{i,0}$ of the super DIM algebra. 
%$F(k)$ is the number of fermionic rows above the $k$-th row and
The sign factor $(-1)^{F(k)}$ is a characteristic feature of the super Macdonald case, 
due to the fact that $E_{i,0}$ are fermionic generators. It is amusing that we can obtain such
a sign factor from the geometric construction of the representation by the correspondence.

We note that both $\mathsf{N}_{\Lambda,  \Lambda + \boxtimes_k}(u \vert q,t)$ and 
$\mathsf{N}_{\Lambda, \Lambda(\boxtimes_k \to \square_k)}(u \vert q,t)$ have a zero at $u=1$. 
This is the reason why we need to take the residue at $u=1$ to reproduce the coefficients of the Pieri formula. 
From the computation of the equivariant character reviewed in Appendix \ref{App:correspondence}, we expect 
that the residue of the ratio of the Nekrasov factors in Proposition \ref{main-prp} gives the matrix elements of 
the zero modes $E_{i,0}$ in the fixed point basis up to the overall normalization of the currents $E_i(z)$. 
Proposition \ref{main-prp} shows that this is indeed the case. Namely the base change from the fixed point basis 
to the super Macdonald basis is achieved by multiplying the scaling factor $\widetilde{c}_\Lambda(q,t)$. 
More precisely speaking, in order to eliminate the monomial factor $t^{k-1}$ in \eqref{main-equality1}
we have to make a further scaling by $t^{n(\Lambda^{\circledast})}$. 
See \eqref{t-framing} for the definition of $n(\lambda)$ for a Young diagram $\lambda$.

The problem of instanton counting on the blow-up $\widehat{\mathbb{C}}^{2}
=\lvert \mathcal{O}_{\mathbb{P}^{1}}(-1) \rvert$
is closely related to the curve countings
on the resolved conifold $\lvert \mathcal{O}_{\mathbb{P}^{1}}(-1) 
\oplus 
\mathcal{O}_{\mathbb{P}^{1}}(-1) \rvert$, 
where 
$\mathcal{O}_{\mathbb{P}^{1}}(-1)$
denotes the tautological bundle
on $\mathbb{P}^{1}$, and 
$\lvert \mathcal{O}_{\mathbb{P}^{1}}(-1) \rvert$
and 
$\lvert \mathcal{O}_{\mathbb{P}^{1}}(-1) 
\oplus 
\mathcal{O}_{\mathbb{P}^{1}}(-1) \rvert$
are the total spaces of vector bundles
$\mathcal{O}_{\mathbb{P}^{1}}(-1)$
and $\mathcal{O}_{\mathbb{P}^{1}}(-1) 
\oplus 
\mathcal{O}_{\mathbb{P}^{1}}(-1)$.
In fact, the quiver for the moduli space of the framed sheaves on the blow-up (see Figure \ref{blowupADHM}) 
is obtained from the dimensional reduction of the conifold quiver in Figure \ref{conifold-quiver} 
with the framing sector as in Appendix \ref{App:quiver-reduction}.
The framing sector gives a trivialization
called a {\it framing}
over the infinity line $\ell_{\infty}=\widehat{\mathbb{P}}^{2}
\setminus \widehat{\mathbb{C}}^{2}$
in the study \cite{Nakajima:2003pg} of the instanton moduli
over $\widehat{\mathbb{P}}^2$, 
where $\widehat{\mathbb{P}}^2$ is the blow-up of $\mathbb{P}^2$.
Deleting an arrow from the conifold
quiver in the dimensional reduction corresponds to choosing 
one of two fibers over 
$\mathbb{P}^1$ in the resolved conifold $\lvert \mathcal{O}_{\mathbb{P}^{1}}(-1) 
\oplus 
\mathcal{O}_{\mathbb{P}^{1}}(-1) \rvert$. 
We see the chamber-and-wall structures for both framed quivers coincide 
(see \cite{Nagao-Nakajima} and \cite{Nakajima:2008eq}).
From the view point of the representations of the quantum toroidal algebra, 
the BPS states on the resolved conifold are described by 
the pyramid partition \cite{Szendroi:2007nu,Jafferis:2008uf,Chuang:2009crq}
in the melting crystal model \cite{Ooguri:2009ijd}. 
As pointed out in \cite{Noshita:2021dgj} the level zero representation employed in this paper 
is identified as a subcrystal representation of the pyramid partition. 
This is parallel to the relation of the MacMahon representation and the Fock representation
in the case of the quantum toroidal algebra of type $\mathfrak{gl}_1$.

%%%%%%%%%%%%%%%%%%%%%%%%%%%%%%%%%%%%%%%%%%%%%%%%%%%%%%%%%%%%%%%%%%%%%%%%%%%%%%%%%%%%%

The present paper is organized as follows;
In section 2 we introduce super partitions and super Young diagrams.
We explain that a super partition is described by a pair $(Y,S)$
of a Young diagram and a subset $S$ of the removable boxes in $Y$.
We also define the (ir)relevant boxes of a super partition $(Y, S)$, 
which plays an important role in the following sections.
In section 3, we define the quantum toroidal algebra of type $\mathfrak{gl}_{1|1}$
as a quiver quantum toroidal algebra. We review a construction of 
the level zero representations. By using the Drinfeld coproduct of the algebra,
we can construct the super Fock representation by taking a semi-infinite tensor 
product of the vector representations. One of the significant features is
that the super Fock representation becomes a shifted representation,
where the mode expansion of the Cartan currents $K_i^{\pm}(z)$ is changed 
by a regularization of the infinite product. 
The super Macdonald polynomials give a basis of the the super Fock representation 
and we derive the Pieri rule from the action of the generators $E_{i,0}$.
In section 4, after reviewing a quiver description of the moduli space of framed locally free sheaves 
on the blow-up, we introduce the Nakrasov factor for a pair of super partitions.
Finally section 5 is devoted to a proof of Proposition \ref{main-prp}.
We show that the matrix elements derived from the Nekrasov factor agree
with the Pieri rule of the super Macdonald polynomials by the diagonal 
base change given by the factor $\widetilde{c}_\Lambda(q,t)$ that is conjectured to give integral
forms of the super Macdonald polynomials. 
Some of technical details and examples are presented in Appendices.

%%%%%%%%%%%%%%%%%%%%%%%%%%%%%%%%%%%%%%%%%%%%%%%%%%%%%%%%%%%%%%%%%%%%%%%%%%%%%%%%%%%%%%%

\section{Super partitions and super Young diagrams}
\label{sec:superY}

In this section we review the definitions of super partition and super Young diagram following \cite{Galakhov:2024cry} 
and \cite{Galakhov:2024zqn} (see also \cite{Blondeau-Fournier:2011sft,Blondeau-Fournier:2012exj}).
A super partition is a generalization of a partition, where the set of integers $\mathbb{Z}$ is replaced by
the set of half-integers $\mathbb{Z}/2$. Namely it is a non-increasing sequence of non-negative elements in $\mathbb{Z}/2$;
\begin{equation}
\Lambda_1 \geq \Lambda_2 \geq \ldots \geq \Lambda_i \geq  \ldots \geq \Lambda_{\ell(\Lambda)}, \qquad \Lambda_i \in \mathbb{Z}_{> 0}/2,
\end{equation}
where $\ell(\Lambda)$ is the number of non-zero components in $\Lambda$.\footnote{In this paper we use the notation $\Lambda$
for a super partition to distinguish it from the (ordinary) partition, which will be denoted by $\lambda$.}
We introduce a natural $\mathbb{Z}_2$ grading 
of $\mathbb{Z}/2$ such that integral elements are even and non-integral elements are odd. 
We require that if $\Lambda_k$ is odd, the inequality is strict $\Lambda_{k-1} > \Lambda_k > \Lambda_{k+1}$. 
We denote $|\Lambda| = \displaystyle{\sum_{i=1}^{\ell(\Lambda)}} \Lambda_i$ and call it level of $\Lambda$. 
As in the case of partitions it is convenient to identify a super partition with the super Young diagram,
which consists of full boxes $\mathsf{s}=\square$ and upper half boxes (upper triangles)\footnote{In \cite{Blondeau-Fournier:2011sft}
and \cite{Blondeau-Fournier:2012exj} it is represented by a circle.}
$\mathsf{t}=
\begin{picture}(5,4)
\setlength{\unitlength}{0.8mm}
\thicklines
\put(0,4){\line(1,0){5}}
\put(0,-1){\line(0,1){5}}
\put(0,-1){\line(1,1){5}}
\end{picture}$~~. 
If the component $\Lambda_k$ of a super partition is even (bosonic) or odd (fermionic), 
we call the corresponding row of the super Young diagram even or odd, accordingly. 
If a row is even, the end of the row is a full box and if it is odd, the end of the row is an upper half box. 
It is convenient to think of a full box as a combination of an upper triangle and a lower triangle.

The generating function of the number of super partitions is 
\begin{equation}
\sum_{\Lambda} x^n y^m = \prod_{k=1}^\infty \frac{1+y x^{k-1}}{1-x^{k}},
\end{equation}
where $n$ is the number of full boxes and $m$ is the number of upper half-triangles. 
Since $|\Lambda| = n + \frac{m}{2}$, we have
\begin{equation}
\sum_{\Lambda} q^{2|\Lambda|} =  \prod_{k=1}^\infty \frac{1+ q^{2k-1}}{1-q^{2k}}.
\end{equation}
Note that this agrees with the character of the tensor product of the Fock spaces of a free boson $a_n~(n \in \mathbb{Z})$ 
with the commutation relation
\begin{equation}
\left[ a_n, a_m \right] = \delta_{n+m,0},
\end{equation}
and a free (NS) fermions $\psi_r~(r \in \mathbb{Z} + \frac{1}{2})$ with the anti-commutation relation
\begin{equation}
\left\{ \psi_{r}, \psi_{s} \right\} = \delta_{r+s, 0},
\end{equation}
The bosonic and the fermionic creation operators $a_{-n}~(n \in \mathbb{Z}_{>0})$ and $\psi_{-r}~(r \in \mathbb{Z}_{\geq 0} + \frac{1}{2})$
correspond to a bosonic row of length $n$ and a fermionic row of length $r$, respectively. 

%%%%%%%%%%%%%%%%%%%%%%%%%%%%%%%%%

We can express a super partition $\Lambda$ by a pair $(\lambda, \barsigma)$, where $\lambda$ is a partition
and $\barsigma = (\sigma_i),~ \sigma_i \in \{0, 1\}$. The components of $\Lambda$ is given by
\begin{equation}
\Lambda_ i = \lambda_i - \frac{1}{2} \sigma_i.
\end{equation}
The $k$-th row is bosonic (even) or fermionic (odd) according to $\sigma_k = 0$ or $\sigma_k =1$. 
This way of describing super partitions is convenient, when we construct the vector representation
of the quantum toroidal algebra of type $\mathfrak{gl}_{1|1}$ in the next section. 

\begin{figure}[t]
$$
\ytableaushort{{}{}{}{}{}{\times},{}{}{},{}{\times},{},{\times}} \qquad \qquad \qquad
\ytableaushort{{\bullet}{\bullet}{}{}{}{\times},{}{}{},{\bullet}{\times},{},{\times}}
$$
\caption{Left: Super partition $\Lambda = (\frac{11}{2}, 3, \frac{3}{2}, 1, \frac{1}{2})$ with fermion number $3$. 
The marked boxes are counted as $\frac{1}{2}$. 
For each row and each column the number of the marked box is at most one. \\
Right: For each pair of marked boxes we associate a box with $\bullet$. 
The irrelevant boxes are boxes with $\times$ or $\bullet$. The number of irrelevant boxes is
$3 + \frac{3\cdot 2}{2}= 6$.}\label{super-Young}
\end{figure}

%%%%%%%%%%%%%%%%%%%%%%%%%%%%%%%%%%%%%%%%%%%%%%%%

There is yet another way of representing super partitions, which was employed to identify 
the fixed points of the torus action on the moduli space of perverse coherent sheaves 
on the blow-up of $\mathbb{P}^2$ in \cite{Nakajima:2008ss}.
This is a generalization of the fact that the fixed points of the torus action on the moduli space of framed instantons 
on $\mathbb{P}^2$ are labelled by $r$-tuples of Young diagrams. This fact plays a key role in the localization computation of the instanton
partition function of the supersymmetric gauge theories. In \cite{Nakajima:2008ss} 
the fixed points are labeled by pairs $(Y,S)$ of Young diagrams $Y$ and a subset $S$ of removable boxes in $Y$. 
We call a box in $S$ marked. 
In each row of $Y$ the rightmost box $\mathsf{s}= (k, \lambda_k)$ is removable, if $\lambda_{k-1} \geq \lambda_k > \lambda_{k+1}$. 
When $\mathsf{s}$ is marked, or $\mathsf{s} \in S$, we replace $\lambda_k$ by $\lambda_k -\frac{1}{2}$ to obtain a super partition. 
Then we have $\lambda_{k-1} > \lambda_k -\frac{1}{2} > \lambda_{k+1}$ and  the $k$-th row is regarded as fermionic.
Conversely given a super partition $\Lambda$ we obtain the partition for $Y$ by $\lambda_k \to \lambda_k + \frac{1}{2}$ 
for each fermionic row, namely we replace all the upper half-triangles by the full boxes. 
We put a marking $\times$ in the boxes obtained in this way.
Let $m=|S|$, then the super partition $(Y,S)$ has $m$ fermionic rows and $m$ fermionic columns.
For this reason, we call $m$ fermionic number of the super partition. 
For a pair of two marked boxes $\mathsf{s} \neq \mathsf{s}'$, 
we associate an irrelevant box in $Y$, which is represented by a box with $\bullet$ in Figure \ref{super-Young}.
Namely two end points of the hook of the irrelevant box are given two marked points.
There are $\frac{1}{2}m(m-1)$ such boxes. We will include the marked points to the set of irrelevant boxes of $(Y,S)$
and define the set of relevant boxes as the compliment of the set of irrelevant boxes. 
Note that the number of the relevant boxes is $|Y|- \frac{1}{2}m(m+1)$. 

In the following we use the notation $\Lambda$ for the super partition and $(Y,S)$ for the Young diagram $Y$ 
with the set $S$ of marked points in an interchangeable manner. 
Following \cite{Blondeau-Fournier:2012exj}, 
it is convenient to employ the notations $Y^{\circledast} = Y$ and  $Y^{*} = Y \setminus S$. 
We also use the corresponding notations $\Lambda^{\circledast}$ and $\Lambda^{*}$. Note that both 
$\Lambda^{\circledast}$ and $\Lambda^{*}$ are ordinary Young diagrams such that the skew diagram 
$\Lambda^{\circledast} \setminus \Lambda^{*}$ is the set of marked boxes $S$. 

%%%%%%%%%%%%%%%%%%%%%%%%%%%%%%%%%%%%%%%%%%%%%%%%%%%%%%%%%%%%%%%%%%%%%%%%%%%%%%%%%%%%%%%%%%%%%%%%%%%%%%%%%%%%%%%%%%%%%%%%%%%%%%%%%%%%%%%% 

\subsection{Correspondence of super partitions and pairs of partitions}
\label{sec:super=pair}

Originally in \cite{Nakajima:2003pg} the fixed points of the torus action on the moduli space of
framed torsion free sheaves on the blow-up are indexed by pairs of partitions. 
After the blow-up there is the exceptional curve $C$ which is topologically a rational curve $\mathbb{P}^1$.
The torus action on $C \simeq \mathbb{C}^2 / \sim$ which is induced by the natural torus action 
on $\mathbb{C}^2$ has two fixed points $[(1,0)]$ and $[(0,1)]$, which leads to a pair of partitions  $(\lambda, \mu)$. 
In fact, if we fix $m \in \mathbb{Z}_{\geq 0}$,
there is a bijection between the following two sets \cite{Nakajima:2008ss};
\begin{enumerate}
\item
Pairs $Y \supset S$ of Young diagrams $Y$ and sets $S$ of marked removable boxes with $|S|=m$ and
$n= |Y| - \frac{1}{2}m(m+1)$. 
\item
Pairs of Young diagrams $(\lambda, \mu)$ such that $\mu$ has at most $m$ rows 
and $n= |\lambda| + |\mu|$. 
\end{enumerate}

\ytableausetup{aligntableaux=top,mathmode,boxsize=1.0em}
\begin{figure}[h]
$$
\Lambda = \ytableaushort{{}{\bullet}{}{}{\times},{}{}{},{}{\times},{},{}}~; \quad
\Biggl( \ytableaushort{{}{}{},{},{}},~~
\ytableaushort{{\bigcirc}{\bigcirc}{}{}{},{\bigcirc}{}} \Biggr)~; \quad
(\lambda, \mu) = \Biggl( \ytableaushort{{}{}{},{},{}},~~
\ytableaushort{{}{}{},{}},~~m=2 \Biggr)
$$
\caption{One to one correspondence between the super partitions and pairs of the partitions with fermion number $m$.}
\label{Fig:super=pair}
\end{figure}

Given a pair $(Y,S)$ from the first set, by decomposing $Y$ into the bosonic and the fermionic rows,
we obtain a pair of partitions $(Y^{\mathrm{B}}, Y^{\mathrm{F}})$.
We identify $\lambda = Y^{\mathrm{B}}$. Namely, $\lambda$ is obtained by removing 
all the fermionic rows from $Y$ and shifting the rows up to fill out empty rows. 
The second partition $\mu$ is obtained from $Y^{\mathrm{F}}$ as follows;
Note that $Y^{\mathrm{F}}$ has length $m$ and its components are strictly decreasing. 
Then by removing all the irrelevant boxes from $Y^{\mathrm{F}}$ and shifting the remaining boxes to the left to fill out empty slots,
we obtain $\mu$. Since there are $m-i+1$ irrelevant boxes in the $i$-th fermionic row of $Y^{\mathrm{F}}$, 
the resultant Young diagram is $\mu= Y^{\mathrm{F}} - \delta_m$, where $\delta_m=(m, m-1, \ldots ,1)$ is 
the staircase partition of size $m$. See Figure \ref{Fig:super=pair}, where $\delta_m$ is indicated by the boxes with big circle. 
Conversely from a pair of partitions $(\lambda, \mu)$ such that $\mu$ has at most $m$ rows, 
we first enlarge $\mu$ to $Y^{\mathrm{F}}$ by adding a staircase diagram $\delta_m$ and put markings on the end box of each row.
Note that the corresponding partition $Y^{\mathrm{F}}$ is strictly decreasing, since we add $\delta_m$. 
Then by combining $Y^{\mathrm{B}}=\lambda$ and $Y^{\mathrm{F}}$ as the bosonic and the fermionic rows 
we obtain a super Young diagram $(Y,S)$ with marked removable boxes $S$.

\begin{figure}[h]
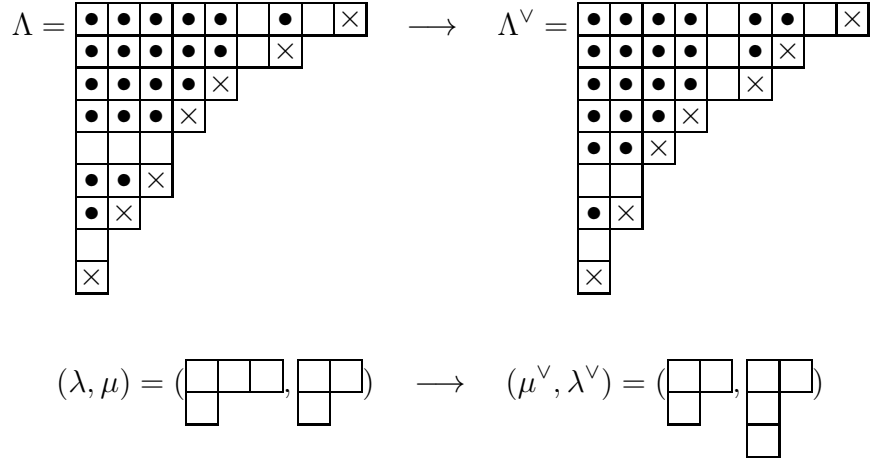

$$
\Lambda = \ytableaushort{{\bullet}{\bullet}{\bullet}{\bullet}{\bullet}{}{\bullet}{}{\times},
{\bullet}{\bullet}{\bullet}{\bullet}{\bullet}{}{\times},{\bullet}{\bullet}{\bullet}{\bullet}{\times},
{\bullet}{\bullet}{\bullet}{\times}, {}{}{},{\bullet}{\bullet}{\times},{\bullet}{\times},{},{\times}}
\quad \longrightarrow \quad \Lambda^\vee =
\ytableaushort{{\bullet}{\bullet}{\bullet}{\bullet}{}{\bullet}{\bullet}{}{\times},
{\bullet}{\bullet}{\bullet}{\bullet}{}{\bullet}{\times},{\bullet}{\bullet}{\bullet}{\bullet}{}{\times},
{\bullet}{\bullet}{\bullet}{\times},{\bullet}{\bullet}{\times},{}{},{\bullet}{\times},{},{\times}}
$$
\bigskip
$$
(\lambda, \mu) = (\ytableaushort{{}{}{},{}}, \ytableaushort{{}{},{}})
\quad \longrightarrow \quad 
(\mu^\vee, \lambda^\vee) =  (\ytableaushort{{}{},{}}, \ytableaushort{{}{},{},{}})
$$
\caption{Transpose of the super partition and the corresponding pair of partitions ($n=m=7$).}
\label{Fig:BF-exchange}
\end{figure}

By looking at the generating function of the numbers of super partitions of a fixed level,
we have argued that there is one to one correspondence between the super partition and the basis of 
the tensor product of the bosonic and the fermionic Fock spaces. From such a viewpoint the first partition $\lambda$
corresponds to a state in the bosonic Fock space. 
With a marking on the end box in each row, the staircase partition $\delta_m$ represents the fermion Fock vacuum 
$\vert m \rangle :=\psi_{-m+\frac{1}{2}}\psi_{-m+\frac{3}{2}}\cdots \psi_{-\frac{3}{2}}
\psi_{-\frac{1}{2}}\vert 0 \rangle$ of charge $m$. Hence the partition $\mu$ is regarded as an excitation 
from the Fock vacuum $\vert m \rangle$ in the fermionic Fock space. 
Note that we need a data of the fermion number $m$ to recover $Y^{\mathrm{F}}$ from $\mu$. 
See Figure \ref{Fig:embedding} for an example of the dependence on $m$ of the super partition $(Y,S)$.
In a sense we can \lq\lq embed\rq\rq\ a pair of partitions $(\lambda, \mu)$ with $|\lambda| + |\mu| = n$
to a sufficiently large super partition $\Lambda$ with $m \geq n$. If such an $m$ is fixed, the embedding is unique. 
Such an $m$ dependence of the super partition leads to the idea of the stable sector.
In \cite{Blondeau-Fournier:2012jca} the region with $m \geq n$ in $\mathbb{Z}_{\geq 0} \times \mathbb{Z}_{\geq 0}$
is called stable sector. In the stable sector the size of the super partition increased quadratically in $n$. 
One of the nice features of the stable sector is the following result (see Figure \ref{Fig:BF-exchange}
for an explicit example);

\begin{lem}[\cite{Blondeau-Fournier:2012jca}, Lemma 30]\label{transpose}
Under the correspondence $\Lambda \leftrightarrow (\lambda, \mu)= (\Lambda^{\mathrm{B}}, \Lambda^{\mathrm{F}} - \delta_m)$
with $m \geq n= |\lambda| + |\mu|$, the correspondence of the transposed super partition is
$\Lambda^\vee \leftrightarrow (\mu^\vee, \lambda^\vee)$. 
\end{lem}

If the super partition is not in the stable sector Lemma \ref{transpose} does not hold in general.
See Figure \ref{Fig:embedding} where the stable sector is $m \geq 3$. 
But we note the condition of the stable sector is only a sufficient condition for the validity 
of Lemma \ref{transpose}. There are super partitions which are not in the stable sector, but Lemma \ref{transpose} holds.

\begin{figure}[h]
$$
\ytableaushort{{}{}{}}~, \qquad
\ytableaushort{{}{}{}, {\times}}~, \qquad
\ytableaushort{{}{}{}, {\bullet}{\times}, {\times}}~, \qquad
\ytableaushort{{}{}{}, {\bullet}{\bullet}{\times},{\bullet}{\times}, {\times}}~, \qquad
\ytableaushort{{\bullet}{\bullet}{\bullet}{\times},{}{}{},{\bullet}{\bullet}{\times},{\bullet}{\times}, {\times}}~,  \qquad
\ytableaushort{{\bullet}{\bullet}{\bullet}{\bullet}{\times},{\bullet}{\bullet}{\bullet}{\times},{}{}{},
{\bullet}{\bullet}{\times},{\bullet}{\times}, {\times}}~.
$$
\caption{The embeddings of $\lambda = \ytableaushort{{}{}{}} = (\lambda, \varnothing)$ to the super partitions
with $m=0,1, \ldots, 5$. The pair of partitions corresponding to the transpose of the super partitions 
is $(\varnothing, \lambda^\vee)$ for $m \geq 3=|\lambda|$, But it is different from $(\varnothing, \lambda^\vee)$ for $m=0,1,2$. }
\label{Fig:embedding}
\end{figure}

\ytableausetup{aligntableaux=top,mathmode,boxsize=1.2em}
%%%%%%%%%%%%%%%%%%%%%%%%%%%%%%%%%%%%%%%%%%%%%%%%%%%%%%%%%%%%%%%%%%%%%%%%%%%%
%\newpage
\section{Quiver quantum toroidal algebra of type $\mathfrak{gl}_{1\vert 1}$}

The vertical (level zero) representations of the super DIM algebra are constructed in \cite{Noshita:2021dgj}
by following the method in \cite{FFJMM} for the DIM algebra.
In \cite{Noshita:2021dgj} the super DIM algebra is defined as the quiver quantum toroidal algebra
based on the quiver coming from the resolved conifold geometry (a resolution of the conifold singularity $xy=uv$ by the blow-up). 
In general the quiver quantum toroidal algebra of type $\mathfrak{gl}_{m\vert n}$ can be defined similarly 
based on the quiver coming from an appropriate toric geometry. 
For example the bosonic quantum toroidal algebra of type $\mathfrak{gl}_{n}$ can be associated with the quiver of
the resolution of $A_{n-1}$ singularity. 
%The quantum toroidal algebra of type $\mathfrak{gl}_{2|1}$ seems to be 
%related to the one-point blow-up of the resolution of $A_1$ singularity (the Eguchi-Hanson space). 

The level zero representation of the super DIM algebra associated with the super Macdonald polynomials is
related to the two dimensional crystal model introduced in \cite{Nishinaka:2011sv,Nishinaka:2011is,Nishinaka:2013mba}, 
which is a reduction of the three dimensional melting crystal model \cite{Ooguri:2009ijd}. 
Originally the three dimensional crystal model was proposed to describe $D2$-$D0$ system bound 
to $D6$-brane on a toric Calabi-Yau threefold. On the other hand the two dimensional reduction corresponds to 
$D2$-$D0$ system bound to a $D4$-brane wrapping on a toric divisor of the Calabi-Yau threefold. 
It was shown that the two dimensional crystal model reproduced the BPS index of the $D4$-$D2$-$D0$ system. 
When the toric Calabi-Yau threefold is $\mathbb{C}^3$, the crystal model is defined in terms of the plane partition.
In the case of the resolved conifold, the plane partition is replaced with the pyramid partition
\cite{Szendroi:2007nu,Jafferis:2008uf,Chuang:2009crq}. 

%%%%%%%%%%%%%%%%%%%%%%%%%%%%%%%%%%%%%%%%%%%%%%%%%%%%%%%%%%%%%%%%%%%%%

\subsection{Quiver quantum toroidal algebra}

We review the quantum toroidal algebra following \cite{Noshita:2021dgj} (see also \cite{Galakhov:2021vbo}
and references therein).
One of the canonical ways to derive the commutation relations of the quantum toroidal algebra is
the affinization of the procedure;
$$
\hbox{Dynkin diagram} \longrightarrow \hbox{Deformed Cartan matrix} \longrightarrow \hbox{Commutation relations}.
$$
For the Lie superalgebra $\mathfrak{gl}_{1\vert 1}$ this method does not work, since the Cartan matrix vanishes 
and its deformation looks quite subtle. 
Instead of the above procedure one can start with a quiver data;
$$
\hbox{Quiver diagram} \longrightarrow \hbox{Structure function} \longrightarrow \hbox{Commutation relations}.
$$

Denote a quiver diagram as $Q= \{ Q_0, Q_1 \}$, where $Q_0$ is the set of vertices and $Q_1$ is the set of
arrows between vertices. We use $i,j, \ldots \in Q_0$ to label vertices and $I, J, \ldots \in Q_1$ to label arrows. 
We assume that the quiver is symmetric;
\begin{equation}
| i \to j | = | j \to i |, \qquad i, j \in Q_0.
\end{equation}
To each arrow $I \in Q_1$, we assign a parameter $q_I$. The parameters $q_I,~(I \in Q_1)$ satisfy 
the loop constraints and the vertex constraints (see \cite{Noshita:2021dgj} for details). Consequently 
we have two independent parameters, which are regarded as deformation parameters of the quiver quantum toroidal algebra.

The quiver toroidal algebra $\mathcal{U}_Q$ is generated by $E_{i,k}, F_{i,k}, K_{i, \pm r}^{\pm}$
and a central element $C$, where $i \in Q_0, k \in \mathbb{Z}$ and $r \in \mathbb{Z}_{\geq 0}$. 
To write down the defining relations of the algebra it is convenient to introduce the generating currents;\footnote{In the case of 
super DIM algebra we have to consider the shifted algebra where the modes of $K_i^{\pm}(z)$ are shifted
(See section \ref{sec:super-Fock}).}
\begin{equation}
E_i(z) = \sum_{k \in \mathbb{Z}} E_{i,k} z^{-k}, \qquad F_i(z) = \sum_{k \in \mathbb{Z}} F_{i,k} z^{-k},
\qquad K_i^{\pm}(z) = \sum_{r \geq 0} K_{i, \pm r}^{\pm} z^{\mp r}.
\end{equation}
The generators have $\mathbb{Z}_2$ grading. When a vertex $i \in Q_0$ has no loop, we call the vertex fermionic
and denote $|i|=1$. Otherwise $i \in Q_0$ is bosonic and $|i|=0$.
The grading of $E_{i,k}$ and $F_{i,k}$ is defined by $|E_{i,k}| = |F_{i,k}| = |i|$. 
The Cartan generators $ K_{i, \pm r}^{\pm}$ are always bosonic. 
The defining relations are given by;
\begin{align}\label{K-zero-mode}
K_{i,0}^{+} K_{i,0}^{-} &= K_{i,0}^{-} K_{i,0}^{+} =1, 
\\
C^{-1} C &= C C^{-1}=1, 
\\
K_i^{\pm}(z) K_j^{\pm}(w) &= K_j^{\pm}(w) K_i^{\pm}(z),
\\
K_i^{-}(z) K_j^{+}(w) &= \frac{\varphi^{j \Rightarrow i}(z,Cw)}{\varphi^{j \Rightarrow i}(Cz,w)}
K_j^{+}(w) K_i^{-}(z),
\\
K_i^{\pm}(C^{\frac{1\mp 1}{2}}z) E_j(w) 
&= \varphi^{j \Rightarrow i}(z,w) E_j(w) K_i^{\pm}(C^{\frac{1\mp 1}{2}}z), 
\\
K_i^{\pm}(C^{\frac{1\pm 1}{2}}z) F_j(w) &= \varphi^{j \Rightarrow i}(z,w)^{-1}  
F_j(w) K_i^{\pm}(C^{\frac{1\pm 1}{2}}z),
\\
\left[ E_i(z), F_j(w) \right]_{\pm} &= \delta_{ij} \left[ \delta \left( \frac{Cw}{z} \right) K_i^{+}(z) - 
\delta \left( \frac{Cz}{w} \right)K_i^{-}(w) \right], 
\\
E_i(z) E_j(w) &= (-1)^{|i||j|} \cdot \varphi^{j \Rightarrow i}(z,w)  E_j(w) E_i(z),  
\\
F_i(z) F_j(w) &= (-1)^{|i||j|} \cdot \varphi^{j \Rightarrow i}(z,w)^{-1}  F_j(w) F_i(z), 
\end{align}
where the structure function $\varphi^{i \Rightarrow j}(z,w)$ is determined by the quiver data as follows;
\begin{equation}
\varphi^{i \Rightarrow j}(z,w) = \frac{\prod_{I \in \{j \to i\}} (q_I^{\frac{1}{2}} z - q_I^{-\frac{1}{2}}w)}
{\prod_{I \in \{i \to j\}} (q_I^{-\frac{1}{2}} z - q_I^{\frac{1}{2}}w)} 
= \frac{\prod_{I \in \{j \to i\}} \phi(q_I; z,w)}{\prod_{I \in \{i \to j\}} \phi(q_I^{-1}; z,w)},
\end{equation}
where
\beq
\phi(p; z,w) := p^{\frac{1}{2}}z -  p^{-\frac{1}{2}}w =  p^{\frac{1}{2}} (z - p^{-1} w).
\eeq

The quiver quantum toroidal algebra is a Hopf superalgebra with the following coproduct;\footnote{For the definition of
the counit and the antipode see \cite{Noshita:2021dgj}.}
\begin{align}
\Delta E_i(z) &= E_i(z) \otimes 1 + K_i^{-}(C_1 z ) \otimes E_i(C_1 z), 
\\
\Delta F_i(z) &= F_i(C_2z) \otimes K_i^{+}(C_2 z) + 1 \otimes F_i(z), 
\\
\Delta K_i^{+} &= K_i^{+}(z) \otimes K_i^{+}(C_1^{-1}z),
\\
\Delta K_i^{-} &= K_i^{-}(C_2^{-1} z) \otimes K_i^{-}(z),
\\
\Delta C &= C \otimes C,
\end{align}
where $C_1 = C \otimes 1$ and $C_2 = 1 \otimes C$. 

When $C=1$, we say the representation has level zero. In this case all the Cartan modes $K_{i,\pm r}^\pm$ are
commuting and there is a basis consisting of simultaneous eigenvectors of $K_{i,\pm r}^\pm$.

%%%%%%%%%%%%%%%%%%%%%%%%%%%%%%%%%%%%%%%%%%%%%%%%%%%%%%%%%%%%%%%%%%%%%%%%%%%%

%%%%%%%%%%%%%%%%%%%%%%%%%%%%%%%%%%%%%%%%%%%%%%%%%%%%%%%%%%%%%%%%%%%%%%%%%%%%%%%%%

\begin{figure}[t]
\begin{center}
\begin{picture}(30,15)
 \setlength{\unitlength}{1.4mm}
\thicklines
\put(-10,0){\circle{8}}
\put(20,0){\circle{8}}
\put(-7.3,3){\line(1,0){25}}
\put(-6.1,2){\line(1,0){22.2}}
\put(-6.1,-2){\line(1,0){22.2}}
\put(-7.3,-3){\line(1,0){25}}
\put(5,2.5){\line(-1,1){3}}
\put(5,2.5){\line(-1,-1){3}}
\put(6,-2.5){\line(1,1){3}}
\put(6,-2.5){\line(1,-1){3}}
\put(-11,-1){$1$}
\put(19.5,-1){$2$}
\put(1, 7){$(\epsilon_1, -\epsilon_1)$}
\put(3, -7){$(\epsilon_2, -\epsilon_2)$}
\end{picture}
\end{center}
\vspace{3mm}
\caption{The quiver for the super DIM algebra.
The edges from the vertex $1$ to the vertex $2$ have parameters $(\epsilon_1, -\epsilon_1)$,
while the edges with the reversed direction have $(\epsilon_2, -\epsilon_2)$.
We define $(q_1,q_2)= (e^{\epsilon_1}, e^{\epsilon_2})$.}\label{conifold-quiver}
\end{figure}
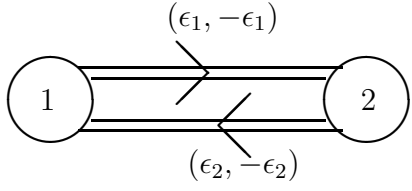
%%%%%%%%%%%%%%%%%%%%%%%%%%%%%%%%%%%%%%%%%%%%%%%%%%%%%%%%%%%%%%%%%%%%

\subsection{Vector representation}

Let us introduce the vector space spanned by $[u]_{j, \barsigma}$, where $ j \in \mathbb{Z}, \barsigma \in \mathbb{Z}_2 = \{ 0, 1\}$
and $u$ is the spectral parameter. 
We define the following action of the generating currents $E_i(z), F_i(z)$ and $K_i^{\pm}(z), i=1,2$ on $V$;
\begin{align}
& E_1(z) [u]_{k, 0} = \mathcal{E}_1(k) \delta \left( \frac{z}{u (q_1q_2)^{k+1}} \right) [u]_{k+1, 1},
\qquad
E_1(z) [u]_{k, 1} = 0,
\\
& E_2(z) [u]_{k, 0} = 0,
\qquad
 E_2(z)[u]_{k, 1} =  \mathcal{E}_2(k) \delta \left( \frac{z}{u q_1(q_1q_2)^k} \right) [u]_{k, 0},
\\
& F_1(z) [u]_{k, 0} = 0,
\qquad
 F_1(z) [u]_{k, 1} = \mathcal{F}_1(k) \delta \left( \frac{z}{u (q_1q_2)^k} \right) [u]_{k-1, 0},
\\
& F_2(z) [u]_{k, 0} = \mathcal{F}_2(k) \delta \left( \frac{z}{u q_1(q_1q_2)^k} \right) [u]_{k, 1},
\qquad
F_2(z) [u]_{k, 1} =  0,
\\
& K_i^{\pm}(z)  [u]_{k,0} = \left[ \Psi^{(i)}_{ [u]_{k,0}} (z)\right]_\pm [u]_{k,0},
 \qquad
K_i^{\pm}(z)   [u]_{k,1} =  \left[ \Psi^{(i)}_{ [u]_{k,1}}  (z) \right]_\pm  [u]_{k,1},
\end{align}
where $\mathcal{E}_i(k)$ and $\mathcal{F}_i(k)$ are normalization constants 
and $[ \bullet ]_\pm$ means the expansion around $z=\infty, 0$ of the rational function $\bullet$.
Diagrammatically the non-trivial actions of $E_i$ and $F_j$ are described as follows;
\vspace{5mm}
\begin{align*}
E_1 &:  \qquad
\begin{picture}(200,20)
 \setlength{\unitlength}{1.5mm}
\thicklines
\put(0,-1){\line(1,0){16}}
\put(0,3){\line(1,0){16}}
\put(16,-1){\line(0,1){4}}
\put(12,-1){\line(0,1){4}}
\put(8,-1){\line(0,1){4}}
\put(0,0){$\cdots\cdots$}
\put(22,0){$\longrightarrow $}
\put(31,-1){\line(1,0){20}}
\put(31,3){\line(1,0){20}}
\put(51,-1){\line(0,1){4}}
\put(47,-1){\line(0,1){4}}
\put(43,-1){\line(0,1){4}}
\put(39,-1){\line(0,1){4}}
\put(48,0){$\times$}
%\put(60,3){\line(1,-1){4}}
\put(31,0){$\cdots\cdots$}
\end{picture}
\\
E_2 &:\qquad
\begin{picture}(200,20)
 \setlength{\unitlength}{1.5mm}
\thicklines
\put(0,-1){\line(1,0){16}}
\put(0,3){\line(1,0){16}}
\put(16,-1){\line(0,1){4}}
\put(12,-1){\line(0,1){4}}
\put(8,-1){\line(0,1){4}}
\put(13,0){$\times$}
%\put(16,3){\line(1,-1){4}}
\put(0,0){$\cdots\cdots$}
\put(22,0){$\longrightarrow $}
\put(31,-1){\line(1,0){16}}
\put(31,3){\line(1,0){16}}
\put(47,-1){\line(0,1){4}}
\put(43,-1){\line(0,1){4}}
\put(39,-1){\line(0,1){4}}
\put(31,0){$\cdots\cdots$}
\end{picture}
\\
F_1 &: \qquad
\begin{picture}(200,20)
 \setlength{\unitlength}{1.5mm}
\thicklines
\put(0,-1){\line(1,0){16}}
\put(0,3){\line(1,0){16}}
\put(16,-1){\line(0,1){4}}
\put(12,-1){\line(0,1){4}}
\put(8,-1){\line(0,1){4}}
\put(13,0){$\times$}
%\put(16,3){\line(1,-1){4}}
\put(0,0){$\cdots\cdots$}
\put(22,0){$\longrightarrow $}
\put(31,-1){\line(1,0){12}}
\put(31,3){\line(1,0){12}}
\put(43,-1){\line(0,1){4}}
\put(39,-1){\line(0,1){4}}
\put(31,0){$\cdots\cdots$}
\end{picture}
\\
F_2 &: \qquad
\begin{picture}(200,20)
 \setlength{\unitlength}{1.5mm}
\thicklines
\put(0,-1){\line(1,0){16}}
\put(0,3){\line(1,0){16}}
\put(16,-1){\line(0,1){4}}
\put(12,-1){\line(0,1){4}}
\put(8,-1){\line(0,1){4}}
\put(0,0){$\cdots\cdots$}
\put(22,0){$\longrightarrow $}
\put(31,-1){\line(1,0){16}}
\put(31,3){\line(1,0){16}}
\put(47,-1){\line(0,1){4}}
\put(43,-1){\line(0,1){4}}
\put(39,-1){\line(0,1){4}}
%\put(56,3){\line(1,-1){4}}
\put(31,0){$\cdots\cdots$}
\put(44,0){$\times$}
\end{picture}
\end{align*}
In the above diagrams the last boxes on the left side of the arrow have the coordinate $k$. 
When the state is $\mathbb{Z}_2$ odd or fermionic, we put the marking $\times$ in the last box. 
In \cite{Galakhov:2024cry} and \cite{Galakhov:2024zqn} a box with the marking is described as a half tile (triangle)
and it contributes to the length of the row as $\frac{1}{2}$. In such a convention $E_i$ increases the length of the row
by $\frac{1}{2}$ and  $F_i$ decreases the length by $\frac{1}{2}$.

The generating functions of the eigenvalues of mutually commuting Cartan generators are
\beqa\label{Cartan-eigen1}
 \Psi^{(1)}_{ [u]_{k,0}} (z)  &=&    \Psi^{(1)}_{ [u]_{k+1,1}} (z)= \frac{\phi(q_1^{-k-2} q_2^{-k};z,u)}{\phi(q_1^{-k-1} q_2^{-k-1};z,u)}
 = \left(\frac{q_2}{q_1}\right)^{\frac{1}{2}} \frac{z-q_1^{k+2}q_2^{k}u}{z-q_1^{k+1}q_2^{k+1}u}, 
 \\
%  \Psi^{(1)}_{ [u]_{k,1}} (z) &=& \frac{\phi(q_1^{-k-1} q_2^{-k+1};z,u)}{\phi(q_1^{-k} q_2^{-k};z,u)},
%  \\
\label{Cartan-eigen2}
  \Psi^{(2)}_{ [u]_{k,0}} (z)  &=&   \Psi^{(2)}_{ [u]_{k,1}} (z)  =  \frac{\phi(q_1^{-k} q_2^{-k-1};z,u)}{\phi(q_1^{-k-1} q_2^{-k};z,u)}
   = \left(\frac{q_1}{q_2}\right)^{\frac{1}{2}} \frac{z-q_1^{k}q_2^{k+1}u}{z-q_1^{k+1}q_2^{k}u}.
% \\
%  \Psi^{(2)}_{ [u]_{k,1}} (z) &=&  \frac{\phi(q_1^{-k} q_2^{-k-1};z,u)}{\phi(q_1^{-k-1} q_2^{-k};z,u)},
\eeqa

With a suitable choice of the normalizations, this gives a level zero representation of the super DIM algebra. 
Namely $K_i^{\pm}(z)$ and $K_j^{\pm}(z)$ are mutually commuting 
and we have the following commutation relations;\footnote{$E_i(z)$ and $F_i(z)$ are regarded as fermionic currents.
Hence, we have the minus sign in \eqref{commu-4} and \eqref{commu-5} and the commutator in \eqref{commu-3} is actually
the anti-commutator.}
\begin{align}
K_i^{\pm}(z) E_j(w) &= \varphi^{j \Rightarrow i}(z,w)  E_j(w) K_i^{\pm}(z), \label{commu-1}
\\
K_i^{\pm}(z) F_j(w) &= \varphi^{j \Rightarrow i}(z,w)^{-1}  F_j(w) K_i^{\pm}(z), \label{commu-2}
\\
\left[ E_i(z), F_j(w) \right] &= \delta_{ij} \delta \left( \frac{w}{z} \right) \left(K_i^{+}(z) - K_i^{-}(w)  \right), \label{commu-3}
\\
E_i(z) E_j(w) &= (-1)  \cdot \varphi^{j \Rightarrow i}(z,w)  E_j(w) E_i(z),  \label{commu-4}
\\
F_i(z) F_j(w) &= (-1)  \cdot \varphi^{j \Rightarrow i}(z,w)^{-1}  F_j(w) F_i(z), \label{commu-5}
\end{align}
where the structure function (the bond factor) of the super DIM algebra is given by
\beqa\label{bond1}
 \varphi^{1 \Rightarrow 1}(z,w)  &=&  \varphi^{2\Rightarrow 2}(z,w)  = 1, 
 \\
 \varphi^{1 \Rightarrow 2}(z,w) &=&   \varphi^{2 \Rightarrow 1}(z,w)^{-1} \CR
 &=& \frac{\phi(q_2;z,w)  \phi(q_2^{-1};z,w)}{\phi(q_1;z,w)  \phi(q_1^{-1};z,w)}
 = \frac{(z-q_2 w)(z - q_2^{-1}w)}{(z-q_1 w)(z - q_1^{-1}w)}. \label{bond2}
 \eeqa
 Note that the structure function has $\mathbb{Z}_2 \times \mathbb{Z}_2$ symmetry; $q_1 \to q_1^{-1},~q_2 \to q_2^{-1}$.

 \subsubsection{Recursion relations from \eqref{commu-1}}

Given the structure functions \eqref{bond1} and  \eqref{bond2}, we can evaluate $ \Psi^{(i)}_{ [u]_{k,\overline{\sigma}}}(z)$ by
solving the following recursion relations derived from  \eqref{commu-1};
\beq\label{EK-recursion}
\frac{\Psi^{(i)}_{[u]_{k+1,1}}(z)} {\Psi^{(i)}_{[u]_{k,0}}(z)} = \varphi^{1 \Rightarrow i} (z; q_1^{k+1} q_2^{k+1} u),
\qquad
\frac{\Psi^{(i)}_{[u]_{k,0}}(z)} {\Psi^{(i)}_{[u]_{k,1}}(z)}  = \varphi^{2 \Rightarrow i} (z; q_1^{k+1} q_2^{k}u).
\eeq
The first relation for $i=1$ and the second relation for $i=2$ are trivially satisfied by definition.
By using
\beq\label{s-shift}
\phi(st; z,u) = s^{\frac{1}{2}} \phi(t; z, s^{-1} u),
\eeq
one can check the remaining relations;
\begin{align*}
%\frac{\Psi^{(1)}_{[u]_{k+1,1}}(z)} {\Psi^{(1)}_{[u]_{k,0}}(z)} 
%&= \frac{\phi(q_1^{-k-2} q_2^{-k}; z,u) \phi(q_1^{-k-1} q_2^{-k-1};z, u)}
%{\phi(q_1^{-k-1} q_2^{-k-1}; z,u) \phi(q_1^{-k-2} q_2^{-k};z, u)} =1,
%\\
\frac{\Psi^{(2)}_{[u]_{k+1,1}}(z)} {\Psi^{(2)}_{[u]_{k,0}}(z)} 
&=  \frac{\phi(q_1^{-k-1} q_2^{-k-2}; z,u) \phi(q_1^{-k-1} q_2^{-k};z, u)}
{\phi(q_1^{-k-2} q_2^{-k-1}; z,u) \phi(q_1^{-k} q_2^{-k-1};z, u)}  \\
&=  \frac{\phi(q_2; z,q_1^{k+1} q_2^{k+1} u) \phi(q_2^{-1};z, q_1^{k+1} q_2^{k+1}u)}
{\phi(q_1; z, q_1^{k+1} q_2^{k+1} u) \phi(q_1^{-1} ;z, q_1^{k+1} q_2^{k+1}u)}  
= \varphi^{1 \Rightarrow 2}(z, q_1^{k+1} q_2^{k+1}u), 
\\
\frac{\Psi^{(1)}_{[u]_{k,0}}(z)} {\Psi^{(1)}_{[u]_{k,1}}(z)} 
&=  \frac{\phi(q_1^{-k-2} q_2^{-k}; z,u) \phi(q_1^{-k} q_2^{-k};z, u)}
{\phi(q_1^{-k-1} q_2^{-k-1}; z,u) \phi(q_1^{-k-1} q_2^{-k+1};z, u)}  \\
&=  \frac{\phi(q_1; z,q_1^{k+1} q_2^{k} u) \phi(q_1^{-1};z, q_1^{k+1} q_2^{k}u)}
{\phi(q_2; z, q_1^{k+1} q_2^{k} u) \phi(q_2^{-1} ;z, q_1^{k+1} q_2^{k}u)}  
= \varphi^{2 \Rightarrow 1}(z, q_1^{k+1} q_2^{k} u).
%\\
%\frac{\Psi^{(2)}_{[u]_{k,0}}(z)} {\Psi^{(2)}_{[u]_{k,1}}(z)} 
%&= \frac{\phi(q_1^{-k} q_2^{-k-1}; z,u) \phi(q_1^{-k-1} q_2^{-k};z, u)}
%{\phi(q_1^{-k-1} q_2^{-k}; z,u) \phi(q_1^{-k} q_2^{-k-1};z, u)} =1.
\end{align*}

The recursion relations \eqref{EK-recursion} allow an ambiguity of the normalization factor which is independent of $k$. 
In \eqref{Cartan-eigen1} and \eqref{Cartan-eigen2} we fix the normalization by requiring the condition \eqref{K-zero-mode}
on the zero modes of the Cartan currents $K_i^{\pm}(z)$.
In section \ref{sec:Pieri} we are going to derive the Pieri rule of the super Macdonald polynomials $\mathcal{M}_{\Lambda}(x, \theta;q,t)$
based on the representation of the super DIM algebra. 
Due to the regularization of the Cartan currents which is necessary to construct the super Fock representation
the representation becomes a shifted representation. In this case the condition \eqref{K-zero-mode} loses the meaning 
and the normalization should be fixed by another way (See section \ref{sec:super-Fock}).
However, it breaks the symmetry of the Pieri rule under $(q_1, q_2) \to (q_1^{-1}, q_2^{-1})$,
which seems to be related to the fact that $\mathcal{M}_{\Lambda}(x, \theta;q,t)$ are not 
invariant under $(q, t) \to (q^{-1}, t^{-1})$ in general.

\subsubsection{Another formulation}

 In order to investigate the spectral duality of the super DIM algebra, it may be instructive to recast
 \eqref{Cartan-eigen1} and \eqref{Cartan-eigen2} as follows;
 \begin{align}\label{K-exp1}
  \Psi^{(1)}_{ [u]_{k,0}} (z)  &=    \Psi^{(1)}_{ [u]_{k+1,1}} (z) 
  = \left(\frac{q_2}{q_1}\right)^{\frac{1}{2}}\frac{z- q_1^{k+2}q_2^{k} u}{z- q_1^{k+1}q_2^{k+1}u}
  \simeq \exp \left( \sum_{n=1}^\infty \frac{1}{n} (1- (q_1q_2^{-1})^n) \left(\frac{y[k]}{z} \right)^n \right), \\
\label{K-exp2}
   \Psi^{(2)}_{ [u]_{k,0}} (z)  &=   \Psi^{(2)}_{ [u]_{k,1}} (z)  
   =  \left(\frac{q_1}{q_2}\right)^{\frac{1}{2}}\frac{z- q_1^{k}q_2^{k+1} u}{z- q_1^{k+1}q_2^{k}u}
  \simeq \exp \left( \sum_{n=1}^\infty \frac{1}{n} (1- (q_1^{-1}q_2)^n)\left(\frac{y[k]}{q_2 z} \right)^n \right),
\end{align}
where we have introduced the auxiliary variable $y[k] := (q_1q_2)^{k+1}u$.
As we will see below the natural parameters for the super Macdonald polynomials are 
$t^2 = q_1^{-1} q_2, q^2 = q_1q_2$ (See \eqref{dictionary}).
The advantage of the auxiliary variable $y[k]$ is that the operation $k \to k \pm 1$ is represented by
the $q^2$-shift operator $e^{\pm q^2 \partial_y}$, which is dual to the multiplication of $y$. 
Moreover by comparing \eqref{K-exp1} and \eqref{K-exp2} with 
\begin{equation}
K_1^{+}(z) = \exp \left( \sum_{n=1}^\infty \frac{1}{n} (h_1)_n z^{-n} \right), \qquad
K_2^{+}(z) = \exp \left( \sum_{n=1}^\infty \frac{1}{n} (h_2)_n z^{-n} \right),
\end{equation}
we find
\begin{equation}
(h_1)_n \sim (1-t^{-n}) y[k]^n, \qquad
(h_2)_n \sim  (1-t^{n}) (y[k]/q t)^n,
\end{equation}
which is going to be promoted to the power sum, after we construct the super Fock representation. 
Let us look at the action of $(E_i(z), F_i(z))$. Since we are interested in the zero modes,
we can neglect the delta function factor. We find a problem. Namely $E_1$ and $F_1$ are represented by
the shift of $k$, but $E_2$ and $F_2$ are not in the above formulation.
To remedy it, let us use the notation $(k, \barsigma) \to (k_1, k_2)= (k, k-\barsigma) \in \mathbb{Z} \times \mathbb{Z}$
with the constraint $k_1 - k_2 = \barsigma \in \{0,1 \}$. Then all of $E_i$ and $F_i$ are represented by the shift operator;
\begin{align}
E_1 : (k_1, k_2) \longrightarrow (k_1+1, k_2), \qquad
E_2 : (k_1, k_2) \longrightarrow (k_1, k_2+1), \\
F_1 : (k_1, k_2) \longrightarrow (k_1-1, k_2), \qquad
F_2 : (k_1, k_2) \longrightarrow (k_1, k_2-1).
\end{align}
Note that due to the constraint on $(k_1,k_2)$, we must have $E_1^2 = E_2^2 = F_1^2 = F_2^2 =0$. 
Anti-commutativity of $E_1$ and $E_2$ is trivially satisfied by $E_1E_2= E_2E_1=0$ from the delta function. 

In fact such a change of the notation seems to be in accord with the formulation of \cite{Blondeau-Fournier:2012exj},
where they associated a pair of Young diagrams $(\Lambda^{\circledast}, \Lambda^{\ast})$ to a super partition $\Lambda$. 
By looking at their rule at each row, we see that that the larger Young diagram $\Lambda^{\circledast}$ 
is obtained from $k_1$, while the smaller Young diagram $\Lambda^{\ast}$ is from $k_2$. 
Hence, in this language the action of $E_1$ and $F_1$ add or remove a box in $\Lambda^{\circledast}$, 
while the action of $E_2$ and $F_2$ add or remove a box in $\Lambda^{\ast}$. 
On the other hand the generating functions of the eigenvalues $\Psi^{(1)}$ refers to $k_2$,
while $\Psi^{(2)}$ does to $k_1$. Therefore for the Cartan currents the index is swapped. 
This phenomena should be related to the fact that $\varphi^{1 \Rightarrow 1}(z,w)$ and $\varphi^{2\Rightarrow 2}(z,w)$ are
trivial, while $\varphi^{1 \Rightarrow 2}(z,w)$ and $\varphi^{2 \Rightarrow 1}(z,w)$ are non-trivial. 

%%%%%%%%%%%%%%%%%%%%%%%%%%%%%%%%%%%%%%%%%%%%%%%%%%%%%%%%%%%%%%%%%%%%%%%%%%%%%%%

\subsection{Construction of super Fock representation}
\label{sec:super-Fock}

By a similar way to the DIM algebra, we can construct the super Fock representation 
by taking an infinite tensor product of the vector 
representations whose adjacent spectral parameters are shifted appropriately. 
A basis of the representation space is given by the set of super Young diagrams.
By the shift of the spectral parameters the empty partition becomes the highest weight vector, 
which is annihilated by $F_i(z)$. 
 
 \begin{figure}[t]
 \begin{center}
 \begin{picture}(180,220)
 \setlength{\unitlength}{0.95mm}
\thicklines
\put(-10,0){\vector(1,0){100}}
 \put(0,-10){\vector(0,1){100}}
 \put(80,-5){$q_1q_2=q^2$}
  \put(3,84){$q_1q_2^{-1}=t^{-2}$}
 \put(0,20){\line(1,0){80}}
  \put(0,40){\line(1,0){60}}
    \put(0,60){\line(1,0){40}}
     \put(0,80){\line(1,0){20}}
  \put(20,0){\line(0,1){80}}
  \put(40,0){\line(0,1){60}}
    \put(60,0){\line(0,1){40}}
     \put(80,0){\line(0,1){20}}
 \put(0,20){\line(1,-1){20}}
  \put(0,40){\line(1,-1){40}}
    \put(0,60){\line(1,-1){60}}
       \put(0,80){\line(1,-1){80}}
\put(45,45){\vector(1,1){10}} 
\put(50,47){$q_1$}

\put(55,65){\vector(1,-1){10}} 
\put(60,62){$q_2$}

\put(5,5){$1$}
\put(23,5){$q_1q_2$}
\put(42,5){$q_1^2q_2^2$}
\put(62,5){$q_1^3q_2^3$}

\put(12,12){$q_1$}
\put(30,12){$q_1^2 q_2$}
\put(50,12){$q_1^3 q_2^2$}

\put(1,25){$q_1q_2^{-1}$}
\put(24,25){$q_1^2$}
\put(42,25){$q_1^3q_2$}

\put(9,33){$q_1^2q_2^{-1}$}
\put(31,33){$q_1^3$}

\put(1,45){$q_1^2q_2^{-2}$}
\put(20,45){$q_1^3q_2^{-1}$}
\put(9,52){$q_1^3q_2^{-2}$}
\put(1,65){$q_1^3q_2^{-3}$}
 
 \end{picture}
 \end{center}
 \caption{\lq\lq Coordinates\rq\rq\ of a triangle (= a half box):
 the lower triangle at $(i,j),~ i,j \in \mathbb{Z}_{\geq 0}$ is assigned to  the coordinates $q_1^{i+j}q_2^{i-j}$.
 the upper one at the same box is to $q_1^{i+j+1}q_2^{i-j}$}\label{Fig:Coordinates}
 \end{figure}
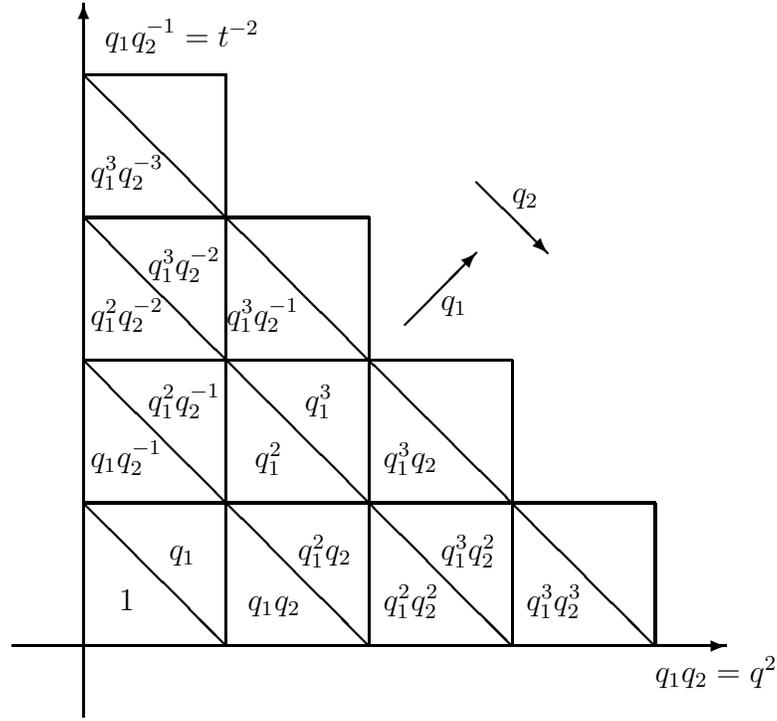

The (Drinfeld) coproduct of the super DIM algebra is
\beqa
\Delta(E_i(z)) &=& E_i(z) \otimes 1 + K_i^{-}(z) \otimes E_i(z), \\ 
\Delta(F_i(z)) &=& 1 \otimes F_i(z)  + F_i(z) \otimes K_i^{+}(z) , \\ 
\Delta(K_i^{\pm}(z)) &=& K_i^{\pm}(z)  \otimes K_i^{\pm}(z).
\eeqa
Note that we have
\beq
(x \otimes y)(z \otimes w) = (-1)^{|y|\cdot|z|} xz \otimes yw.
\eeq
For example
\beqa
(K_i^{-}(z) \otimes E_i(z))(E_j(w) \otimes 1) &=& (-1) (K_i^{-}(z) E_j(w) \otimes E_i(z)), \\
(E_i(z) \otimes 1)(K_j^{-}(w) \otimes E_j(w)) &=& E_i(z) K_j^{-}(w) \otimes E_j(w). 
\eeqa
Hence,
\begin{align*}
\Delta(E_i(z))\Delta(E_j(w)) &= (E_i(z) \otimes 1 + K_i^{-}(z) \otimes E_i(z)) (E_j(w) \otimes 1 + K_j^{-}(w) \otimes E_j(z)) \\
&= E_i(z) E_j(w) \otimes 1 -  K_i^{-}(z) E_j(w) \otimes E_i(z)  \\
& \qquad + E_i(z) K_j^{-}(w) \otimes E_j(z) + K_i^{-}(z)K_j^{-}(w) \otimes E_i(z) E_j(w) \\
&= -\varphi^{j \Rightarrow i}(z,w)  E_j(w) E_i(z)  \otimes 1 -  \varphi^{j \Rightarrow i}(z,w) E_j(w) K_i^{-}(z) \otimes E_i(z) \\
& \qquad + \varphi^{i \Rightarrow j}(w,z)^{-1} K_j^{-}(w) E_i(z)  \otimes E_j(z) 
- \varphi^{j \Rightarrow i}(z,w)K_j^{-}(w) K_i^{-}(z)\otimes  E_j(w) E_i(z) \\
&= - \varphi^{j \Rightarrow i}(z,w) \Delta(E_j(w))\Delta(E_i(z)),
\end{align*}
where we have used $\varphi^{i \Rightarrow j}(w,z)^{-1}= \varphi^{j \Rightarrow i}(w,z)=  \varphi^{j \Rightarrow i}(z,w)$. 

As preliminary computations, let us take the tensor product of 
two vector representations $V(u) \otimes V(v)$ with spectral parameters $u$ and $v=uq_1q_2^{-1}$.
Look at Figure \ref{Fig:Coordinates} to motivate the shift of adjacent spectral parameters.\footnote{As we will see below,
$q_1q_2^{-1}= t^{-2}$.}
The tensor product representation is spanned by four types of vectors;
$$
[u]_{k,0} \otimes [v]_{\ell,0}, \qquad [u]_{k,0} \otimes [v]_{\ell,1}, \qquad
[u]_{k,1} \otimes [v]_{\ell,0}, \qquad [u]_{k,1} \otimes [v]_{\ell,1}.
$$
By computing  the action of $\Delta(E_s(z)) = E_s \otimes 1 + K_s^{-}(z) \otimes E_s(z)$ on these tensor product,
we can check that the action is well-defined and that the set of the tensor products corresponding to the super partitions 
is an invariant subspace. In fact the action of $\Delta(E_s(z))$ agrees with the melting rule proposed  
in \cite{Nishinaka:2011sv,Nishinaka:2011is,Nishinaka:2013mba}.

Now let us consider $N$-times tensor product $V(u_1) \otimes V(u_2) \otimes \cdots \otimes V(u_N)$
with $u_r = (q_1 q_2^{-1})^{r-1} u$. We assign $(\lambda_i, \overline{\sigma}_i),~\lambda_i \in \mathbb{Z}_{\geq 0},
\overline{\sigma}_i \in \mathbb{Z}_2$ for the $i$-th component.
The states in the tensor product are denoted 
\beq
\vert \lambda, \overline{\sigma} \rangle := \prod_{i=1}^N [(q_1q_2^{-1})^{i-1} u ]_{\lambda_i -1, \overline{\sigma}_i}
\in V(u_1) \otimes V(u_2) \otimes \cdots \otimes V(u_N). 
\eeq
It is convenient to identify $E_2(z) \equiv E_0(z)$ and similarly for other currents
and to use the notation $\overline{s}$ when $s$ is regarded as an element in $\mathbb{Z}_2$. 
With this convention the vector representation is defined uniformly as follows;
\begin{align}
E_{\overline{s}}(z) \cdot [u]_{k, \barsigma} &= \mathcal{E}_{\overline{s}}~
\delta \left( \frac{z}{u q_1^{k+1} q_2^{k+1 - \barsigma}} \right) \overline{\delta}_{s+\sigma, 1} 
[u]_{k+\overline{s},1-\barsigma},
\\
F_{\overline{s}}(z) \cdot [u]_{k, \overline{\sigma}} &= \mathcal{F}_{\overline{s}}~
\delta \left( \frac{z}{u q_1^{k+1-\overline{s}} q_2^{k}} \right) \overline{\delta}_{s+\sigma, 0}
[u]_{k-\overline{s},1-\barsigma},
\\
\label{Cartan-eigen1a} 
\Psi^{(1)}_{[u]_{k, \barsigma}}(z) &= \frac{\phi(q_1^{-k-2+\barsigma} q_2^{-k+\barsigma}; z,u)}
{\phi(q_1^{-k-1+\barsigma} q_2^{-k-1+\barsigma}; z,u)},
\\
\label{Cartan-eigen2a} 
\Psi^{(2)}_{[u]_{k, \barsigma}}(z) &= \frac{\phi(q_1^{-k} q_2^{-k-1};z,u)}{\phi(q_1^{-k-1} q_2^{-k};z,u)}.
\end{align}

Then the action of $E_{\overline{s}}(z)$ is 
\begin{align}\label{E-action}
E_{\overline{s}}(z) \cdot \vert \lambda, \barsigma \rangle 
&= \mathcal{E}_{\overline{s}} \sum_{k=1}^{\ell(\lambda)+1}
 (-1)^{F(k)} \cdot \overline{\delta}_{s + \barsigma_k,1} 
\prod_{i=1}^{k-1} \left[  \Psi^{(s)}_{[u (q_1 q_2^{-1})^{i-1}]_{\lambda_i -1, \overline{\sigma_i}}}(z) \right]_{-}
\CR
& \qquad \times 
 \delta\left(\frac{z}{u (q_1q_2^{-1})^{k-1} q_1^{\lambda_k} q_2^{\lambda_k - \overline{\sigma_k}}} \right) 
\vert \lambda + 1_{k} \cdot \delta_{\overline{s},1}, \barsigma + \overline{1_k} \rangle,
\end{align}
where $F(k):= \displaystyle{\sum_{i=1}^{k-1}} \barsigma_i $ is the number of fermionic rows above the $k$-th row and 
the sign factor $ (-1)^{F(k)}$ comes from the fermionic nature of $E_{\overline{s}}(z)$. 

\subsubsection{Regularization of the Cartan currents}

As in the case of DIM algebra, the $N \to \infty$ limit of the $E_{\overline{s}}(z)$ action is well-defined,
since it is bounded by $\ell(\lambda)+1$. On the other hand we have to regularize the action of the Cartan 
currents $K^\pm_{\overline{s}}(z)$, since it involves the infinite product. 
We define the regularization by specifying the order of taking the product, such that the cancellation of 
infinitely many factors takes place. 
Namely we define
\begin{align}
\prod_{i=\ell(\lambda) +1}^\infty \Psi^{(1)}_{[u(q_1q_2^{-1})^{i-1}]_{-1,0}}
&= \prod_{i=\ell(\lambda) +1}^\infty \frac{\phi(q_1^{-1} q_2; z, u(q_1q_2^{-1})^{i-1})}
{\phi(1; z, u(q_1q_2^{-1})^{i-1})}
\CR
&= \frac {q_1^{-\frac{1}{2}}q_2^{\frac{1}{2}}}{z - u (q_1q_2^{-1})^{\ell(\lambda)}}
= \frac{q_1^{-\frac{1}{2}(\ell(\lambda)+1)} q_2^{\frac{1}{2}(\ell(\lambda)+1)}}
{\phi(q_1^{-\ell(\lambda)}q_2^{\ell(\lambda)}; z,u)}.
\end{align}
Hence.
\begin{equation}
K_1(z) \vert \lambda, \barsigma \rangle = \frac{q_1^{-\frac{1}{2}(\ell(\lambda)+1)} q_2^{\frac{1}{2}(\ell(\lambda)+1)}}
{\phi(q_1^{-\ell(\lambda)}q_2^{\ell(\lambda)}; z,u)} \prod_{i=1}^{\ell(\lambda)}
\Psi^{(1)}_{[u(q_1q_2^{-1})^{i-1}]_{\lambda_i -1, \barsigma_i}} (z) \vert \lambda, \barsigma \rangle. 
\end{equation}
Note that the number of $\phi$-factor is {\it not} balanced, which is a significant difference from DIM algebra.
This means the super Fock representation is a shifted representation of the quantum toroidal algebra. 

Similarly we define
\begin{align}
\prod_{i=\ell(\lambda) +1}^\infty \Psi^{(2)}_{[u(q_1q_2^{-1})^{i-1}]_{-1,0}}
&=  \prod_{i=\ell(\lambda) +1}^\infty \frac{\phi(q_1; z, u(q_1q_2^{-1})^{i-1})}
{\phi(q_2; z, u(q_1q_2^{-1})^{i-1})} \CR
&= q_1^{\frac{1}{2}} q_2^{-\frac{1}{2}} 
\left( q_1^{\frac{1}{2}} z - q_1^{-\frac{1}{2}} q_1^{\ell(\lambda)} q_2^{-\ell(\lambda)}  u \right) \CR
&= q_1^{-\frac{1}{2}(\ell(\lambda)-1} q_2^{\frac{1}{2}(\ell(\lambda)-1)} \phi(q_1^{1- \ell(\lambda)}q_2^{\ell(\lambda)}; z,u).
\end{align}
Hence,
\begin{equation}
K_2(z) \vert \lambda, \barsigma \rangle = 
q_1^{-\frac{1}{2}(\ell(\lambda)-1)} q_2^{\frac{1}{2}(\ell(\lambda)-1)}
\phi(q_1^{1-\ell(\lambda)}q_2^{\ell(\lambda)}; z,u)
\cdot \prod_{i=1}^{\ell(\lambda)}
\Psi^{(2)}_{[u(q_1q_2^{-1})^{i-1}]_{\lambda_i -1, \barsigma_i}} (z) \vert \lambda, \barsigma \rangle. 
\end{equation}
In contrast to $K_1(z)$, the unbalanced $\phi$-factor appears in the numerator for $K_2(z)$.
Substituting the definition \eqref{Cartan-eigen1a} and \eqref{Cartan-eigen2a}, 
we obtain the following generating functions of the eigenvalues of the Cartan modes;
\begin{align}\label{K1-eigen}
K_1(z) &= \frac{q_1^{-\frac{1}{2}(\ell(\lambda)+1)} q_2^{\frac{1}{2}(\ell(\lambda)+1)}}
{\phi(q_1^{-\ell(\lambda)}q_2^{\ell(\lambda)}; z,u)} \prod_{i=1}^{\ell(\lambda)}
\frac{\phi(q_1^{-\lambda_i+\barsigma_i -1} q_2^{-\lambda_i +\barsigma_i +1}(q_1q_2^{-1})^{1-i}; z,u)}
{\phi(q_1^{-\lambda_i+\barsigma_i} q_2^{-\lambda_i +\barsigma_i}(q_1q_2^{-1})^{1-i}; z,u)} \CR
&=\frac{t^{\ell(\lambda)+1}}{\phi(t^{2\ell(\lambda)}; z,u)} \prod_{i=1}^{\ell(\lambda)}
\frac{\phi(q^{-2\lambda_i+2\barsigma_i}t^{2i}; z,u)}
{\phi(q^{-2\lambda_i+2\barsigma_i}t^{2i-2}; z,u)} \CR
&= \frac{t^{\ell(\lambda)+1}}{z - t^{-2\ell(\lambda)}u} \prod_{i=1}^{\ell(\lambda)}
\frac{z-q^{2\lambda_i - 2 \barsigma_i}t^{-2i}u}{z-q^{2\lambda_i - 2 \barsigma_i}t^{-2i+2}u},
\\
\label{K2-eigen}
K_2(z) &= 
q_1^{-\frac{1}{2}(\ell(\lambda)-1)} q_2^{\frac{1}{2}(\ell(\lambda)-1)}
\phi(q_1^{1-\ell(\lambda)}q_2^{\ell(\lambda)}; z,u) \CR
& \qquad\times \prod_{i=1}^{\ell(\lambda)}
\frac{\phi(q_1^{-\lambda_i+1} q_2^{-\lambda_i}(q_1q_2^{-1})^{1-i}; z,u)}
{\phi(q_1^{-\lambda_i} q_2^{\lambda_i+1}(q_1q_2^{-1})^{1-i}; z.u)} \CR
&= t^{\ell(\lambda)-1}\phi(q t^{2\ell(\lambda)-1}; z,u)\prod_{i=1}^{\ell(\lambda)}
\frac{\phi(q^{-2\lambda_i+1}t^{2i-3}; z,u)}{\phi(q^{-2\lambda_i+1}t^{2i-1}; z,u)} \CR
&= t^{\ell(\lambda)-1}(q/t)^{\frac{1}{2}} ( z- (t/q)t^{-2\ell(\lambda)}u)
\prod_{i=1}^{\ell(\lambda)} \frac{z-q^{2\lambda_i -1}t^{-2i+3}u}{z-q^{2\lambda_i -1}t^{-2i+1}u},
\end{align}
where we have defined $q_1q_2^{-1}=t^{-2}$ and $q_1q_2= q^2$ (See Section \ref{Eigenvalues}).\footnote{
We can exchange $q$ and $t$ by the involution $q_1 \to q_1^{-1}$, which is a symmetry of the structure function of super DIM algebra.}

Since the representation is shifted, we cannot impose $(K_1^{+})_0(K_1^{-})_0 = (K_2^{+})_0(K_2^{-})_0 =1$ any more. 
The leading coefficients of the Cartan currents $K_{i}^\pm(z)$ are
\begin{equation}
 (K_1^{+})_{1} = t^{\ell(\lambda)+1}, \qquad (K_1^{-})_0 = - u^{-1} t^{\ell(\lambda) +1}, 
\end{equation}
and 
\begin{equation}
(K_2^{+})_{-1} = t^{\ell(\lambda) -1} (q/t)^{\frac{1}{2}}, \qquad (K_2^{-})_0 = t^{\ell(\lambda)-1} (t/q)^{\frac{1}{2}} (-u),
\end{equation}
which depend on $\ell(\lambda)$. Hence, we propose to change the normalization of the Cartan currents $K_{i}^\pm(z)$
so that they are independent of $\ell(\lambda)$.
Namely we multiply \eqref{Cartan-eigen1} and \eqref{Cartan-eigen2} with the additional factor $(q_1/q_2)^{\frac{1}{2}} = t^{-1}$.
Then the generating functions of eigenvalues of the Cartan generators become
\beqa\label{Cartan-eigen11}
 \widetilde{\Psi}^{(1)}_{ [u]_{k,0}} (z)  &=& \widetilde{\Psi}^{(1)}_{ [u]_{k+1,1}} (z)
 = \frac{z-q^{2k+2}t^{-2}u}{z-q^{2k+2}u}, 
 \\
\label{Cartan-eigen12}
  \widetilde{\Psi}^{(2)}_{ [u]_{k,0}} (z)  &=&  \widetilde{\Psi}^{(2)}_{ [u]_{k,1}} (z) 
   = t^{-2} \cdot \frac{z-q^{2k+1}tu}{z-q^{2k+1}t^{-1}u}.
\eeqa

The super DIM algebra is symmetric under the involution of parameters $(q,t) \to (q^{-1}, t^{-1})$. 
However, through the construction of the super Fock representation the regularization of the Cartan currents breaks the symmetry.
In fact, in contrast to the Macdonald polynomials the super Macdonald polynomials are not invariant under $(q,t) \to (q^{-1}, t^{-1})$
\cite{Alarie-Vezina:2019ohz,Galakhov:2025phf}. It implies there are two families of the super Macdonald polynomials
which are related by the involution $(q,t) \to (q^{-1}, t^{-1})$. It is an interesting problem to understand the symmetry braeking
by the statistical model of crystal melting.

%%%%%%%%%%%%%%%%%%%%%%%%%%%%%%%%%%%%%%%%%%%%%%%%%%%%%%%%%%%%%%%%%%%%%%%%%

\subsubsection{Action of $F_{\overline{s}}(z)$ and highest weight state}

The action of $F_{\overline{s}}(z)$ is
\begin{align}\label{F-action}
F_{\overline{s}}(z) \vert \lambda, \barsigma \rangle &= \mathcal{F}_{\overline{s}}
 \sum_{k=1}^{\ell(\lambda)} (-1)^{F(k)} \cdot \overline{\delta}_{\overline{s},\sigma_k}
\prod_{i=k+1}^{\ell(\lambda)} \left[ \Psi^{(s)}_{[u(q_1q_2^{-1})^{i-1}]_{\lambda_i-1, \barsigma_i}}(z) \right]_{+} \CR
& \times  \prod_{i=\ell(\lambda)+1}^{\infty} \left[ \Psi^{(s)}_{[u(q_1q_2^{-1})^{i-1}]_{-1, 0}}(z) \right]_{+}  
\delta\left( \frac{z}{uq_1^{\lambda_k +k -1 -\overline{s}}q_2^{\lambda_k -k}} \right) 
\vert \lambda - 1_k \cdot \overline{\delta}_{s,1}, \barsigma - \overline{1_k} \rangle.
\end{align}
The infinite product is regularized in the same manner as $K_s(z)$;
\begin{equation}\label{F-regularization}
\prod_{i=\ell(\lambda)+1}^{\infty} \left[ \Psi^{(s)}_{[u(q_1q_2^{-1})^{i-1}]_{-1, 0}}(z) \right]_{+} 
=
\begin{cases}
\phi(q_1^{-\ell(\lambda)} q_2^{\ell(\lambda)};z,u)^{-1}, \qquad s=1,
\\
\phi(q_1^{-\ell(\lambda)+1} q_2^{\ell(\lambda)};z,u), \qquad s=2.
\end{cases}
\end{equation}

The vector representation is not a highest weight module. But the super Fock representation has a highest weight state
$\vert 0 \rangle := \otimes_{i=1}^\infty [u(q_1q_2^{-1})^{i-1}]_{-1,0}$, which satisfies $F_s(z) \vert 0 \rangle =0$. 
This is trivial for $s=1$. We also have
\begin{equation}
F_2(z) \vert 0 \rangle \sim \phi(q_2, z,u) \delta\left( \frac{z}{uq_2^{-1}} \right) \vert 0 \rangle =0.
\end{equation}
Note that the factor $\phi(q_2, z,u)$ comes from the regularization of the infinite product. 

%%%%%%%%%%%%%%%%%%%%%%%%%%%%%%%%%%%%%%%%%%%%%%%%%%%%%%%%%%%%%%%%%%%%%%%%%%%%%%%

\subsection{Eigenvalues of the Cartan modes}
\label{Eigenvalues}

For the purpose of comparing the formulas with those in \cite{Galakhov:2024zqn},
it is convenient to switch the parameter $(q_1, q_2)$ to $(q,t)$ with 
\begin{equation}\label{dictionary}
q_1=q/t, \qquad q_2=qt,
\end{equation}
namely $q_1q_2^{-1}=t^{-2}=e^{\epsilon_1 - \epsilon_2}$ and $q_1q_2= q^2 =e^{\epsilon_1 + \epsilon_2}$.
For simplicity let us introduce the notations\footnote{The orientation of the vertical direction is reversed. 
In the previous subsections it was upwards (see Figure 1). In the following it is changed downwards.}
$\mathsf{t} \equiv
\begin{picture}(5,4)
\setlength{\unitlength}{0.8mm}
\thicklines
\put(0,4){\line(1,0){5}}
\put(0,-1){\line(0,1){5}}
\put(0,-1){\line(1,1){5}}
\end{picture}$
~~and 
$\overline{\mathsf{t}} \equiv
\begin{picture}(5,4)
\setlength{\unitlength}{0.8mm}
\thicklines
\put(0,-1){\line(1,0){5}}
\put(5,-1){\line(0,1){5}}
\put(0,-1){\line(1,1){5}}
\end{picture}$
~~for a \lq\lq half-tile\rq\rq\ used in \cite{Galakhov:2024zqn}. 
We also use the notation $\mathsf{b}$ to represent either $\mathsf{t}$ or $\overline{\mathsf{t}}$. 
In \cite{Galakhov:2024zqn} the theory of super Macdonald polynomials was developed and they obtained
the following formula for the generating functions of the eigenvalues of the Cartan modes;
\begin{equation}\label{Eigen-factor}
\Psi^{(\mathsf{t})}_\lambda (u) = \frac{1}{1-u} \prod_{\mathsf{b} \in \lambda} \varphi_{\mathsf{t}, \mathsf{b}}(u/\omega_{\mathsf{b}}),
\qquad
\Psi^{(\overline{\mathsf{t}})}_\lambda (u) = \frac{1- (q/t)u}{(1-t^{-2})(1-q^2)} \prod_{\mathsf{b} \in \lambda}
\varphi_{\overline{\mathsf{t}}, \mathsf{b}}(u/\omega_{\mathsf{b}}),
\end{equation}
where 
$\omega_{\mathsf{b}}$ denotes the contents of the half-tile $\mathsf{b}$, which is defined by\footnote{
We have exchanged $q$ and $t$ in \cite{Galakhov:2024zqn} so that the row lengths of the partition $\lambda$ 
appears in the power of $q$, which is the standard convention in the literature. We also make shifts of
the powers of $q$ and $t$, since the convention in \cite{Galakhov:2024zqn} is 
that the first box in the first row has the coordinates $(0,0)$ and in our convention it has $(1,1)$.} 
\begin{equation}
\omega_{\mathsf{t}}= q^{2x_{\mathsf t}-2} t^{-2y_{\mathsf{t}}+2}, \qquad 
\omega_{\overline{\mathsf{t}}}= q^{2x_{\overline{\mathsf{t}}}-1} t^{-2y_{\overline{\mathsf{t}}}+1}
= q_1 \omega_{\mathsf{t} \to \overline{\mathsf{t}}},
\end{equation}
and $(x_{\mathsf{b}}, y_{\mathsf{b}})$ are the horizontal and vertical coordinates 
of the box which the half-tile $\mathsf{b}$ belongs to.\footnote{Compared with the convention of Figure \ref{Fig:Coordinates}, 
the sign of the coordinates are flipped.}
The function $\varphi(u)$ is nothing but the structure function of the super DIM algebra. 
Namely $\varphi_{\mathsf{t}, \mathsf{t}}(u) = \varphi_{\overline{\mathsf{t}}, \overline{\mathsf{t}}}(u)=1$
and\footnote{The structure functions are invariant under the exchange of $q$ and $t$.}
\begin{align}
\varphi_{\mathsf{t}, \overline{\mathsf{t}}}(u) &= \varphi^{2 \Rightarrow 1}(1,u)
= \frac{(1-(q/t)u)(1-(t/q)u)}{1- (1/qt)u)(1-qtu)}, \\
\varphi_{\overline{\mathsf{t}}, \mathsf{t}}(u) &= \varphi^{1 \Rightarrow 2}(1,u)
= \frac{1- (1/qt)u)(1-qtu)}{(1-(q/t)u)(1-(t/q)u)}.
\end{align} 
Substituting all the definitions, 
%we obtain
%\begin{align}
%\Psi^{(\mathsf{t})}_\lambda (u) 
%%&= \frac{1}{1-u} \prod_{i=0}^{\ell(\lambda)-1} 
%%\prod_{j=0}^{\lambda_i-1-\barsigma_i} \varphi_{\mathsf{t}, \overline{\mathsf{t}}}(u q^{-1-2j} t^{1+2i}) \CR
%&= \frac{1}{1-u} \prod_{i=1}^{\ell(\lambda)} 
%\prod_{j=1}^{\lambda_i-\barsigma_i} \frac{(1- q^{2-2j}t^{-2+2i}u)(1-q^{-2j}t^{2i}u)}
%{(1-q^{-2j}t^{-2+2i}u)(1-q^{2-2j}t^{2i}u)} \CR
%&= \frac{1}{1-u} \prod_{i=1}^{\ell(\lambda)} \frac{(1-t^{2i-2}u)(1-q^{-2(\lambda_i-\barsigma_i)}t^{2i}u)}
%{(1-t^{2i}u)(1-q^{-2(\lambda_i - \barsigma_i)}t^{2i-2}u)} \CR
%&= \frac{1}{1-t^{2\ell(\lambda)}u}
%\prod_{i=1}^{\ell(\lambda)} \frac{1-q^{-2(\lambda_i-\barsigma_i)}t^{2i}u}{1-q^{-2(\lambda_i - \barsigma_i)}t^{2i-2}u},
%\end{align}
%and
%\begin{align}
%\Psi^{(\overline{\mathsf{t}})}_\lambda (u) 
%%&= \frac{1- (q/t)u}{(1-t^{-2})(1-q^2)} \prod_{i=0}^{\ell(\lambda)-1} 
%%\prod_{j=0}^{\lambda_i-1} \varphi_{\overline{\mathsf{t}}, \mathsf{t}}(u q^{-2j} t^{2i}u) \CR
%&=\frac{1- (q/t)u}{(1-t^{-2})(1-q^2)} \prod_{i=1}^{\ell(\lambda)} 
%\prod_{j=1}^{\lambda_i} \frac{(1-q^{-2j+3}t^{2i-3}u)(1-q^{-2j+1}t^{2i-1}u)}{(1-q^{-2j+1}t^{2i-3}u)(1-q^{-2j+3}t^{2i-1}u)} \CR
%&=\frac{1- (q/t)u}{(1-t^{-2})(1-q^2)} \prod_{i=1}^{\ell(\lambda)} 
%\frac{(1-q^{-2\lambda_i+1}t^{2i-3}u)(1-qt^{2i-1}u)}{(1-qt^{2i-3}u)(1-q^{-2\lambda_i+1}t^{2i-1}u)} \CR
%&= \frac{1-qt^{2\ell(\lambda)-1}u}{(1-t^{-2})(1-q^2)}
%\prod_{i=1}^{\ell(\lambda)} \frac{(1-q^{-2\lambda_i+1}t^{2i-3}u)}{(1-q^{-2\lambda_i+1}t^{2i-1}u)}.
%\end{align}
we confirm that up to normalization, $\Psi^{(\mathsf{t})}(u)$ and $\Psi^{(\overline{\mathsf{t}})}(u)$ agrees with 
\eqref{K1-eigen} and \eqref{K2-eigen}, respectively.

It may be instructive to compare the above result with the case of the DIM algebra. 
For the Cartan current of the DIM algebra we have
\begin{equation}
K^{+}(z) \vert \lambda \rangle = \prod_{s=1}^{\ell(\lambda)} \psi(x_s u/z) \prod_{s=1}^{\ell(\lambda)+1}\psi(q_2^{-1}x_s u/z)^{-1}
 \vert \lambda \rangle.
\end{equation}
Using
\begin{equation}
\psi(z) = q_3^{1/2} \frac{1-q_3^{-1}z}{1-z}, \qquad x_s := q_1^{\lambda_s -1}q_2^{s-1} = q^{\lambda_s -1} t^{1-s},
\end{equation}
we find
\begin{equation}
K^{+}(z) \vert \lambda \rangle = q_3^{1/2} \frac{1- q^{-1}t^{1-\ell(\lambda)}}{1- t^{-\ell(\lambda)}}
\prod_{s=1}^{\ell(\lambda)} \frac{(1-q^{\lambda_s}t^{-s}(u/z))(1- q^{\lambda_s-1} t^{2-s}(u/z))}
{(1-q^{\lambda_s-1}t^{1-s}(u/z))(1- q^{\lambda_s} t^{1-s}(u/z))}.
\end{equation}
On the other hand, when $\barsigma_i =0$, the product of \eqref{K1-eigen} and \eqref{K2-eigen} gives
\begin{equation}
\frac{1}{(1-q^{-2})(1-t^2)}
 \frac{1-(t/q)q^2t^{2\ell(\lambda)-2}u}{1-t^{2\ell(\lambda)}u}
\prod_{i=1}^{\ell(\lambda)} \frac{(1-q^{-2\lambda_i}t^{2i}u)}{(1-q^{-2\lambda_i}t^{2i-2}u)}
\frac{(1-(t/q)q^{-2\lambda_i+2}t^{2i-4}u)}{(1-(t/q)q^{-2\lambda_i+2}t^{2i-2}u)}.
\end{equation}
where we have made a shift $i \to i+1$.
Thus, up to normalization and the additional factor $(t/q)=q_1^{-1}$,\footnote{This factor can be
eliminated by a redefinition of the spectral parameter $u$.} the eigenvalues of the product $K_1(z)K_2(z)$
agrees with those of $K(z)$ of DIM algebra by the rule $(q,t) \to (q^{-2}, t^{-2})$.

%%%%%%%%%%%%%%%%%%%%%%%%%%%%%%%%%%%%%%%%%%%%%%%%%%%%%%%%%%%%%%%%%%%%%%%%%%%%%%
\subsection{Pieri rule}
\label{sec:Pieri}

The Pieri rule coming from the super-Fock representation is obtained from the action of the zero modes of $E_s(z)$,
which is the constant part of \eqref{E-action}.  Eliminating $z$ from $\Psi^{(s)}(z)$ by using the delta function, we find
\begin{align}
E_{1,0} \vert \lambda, \barsigma \rangle &= \sum_{k=1}^{\ell(\lambda)+1} \delta_{\barsigma_k}  (-1)^{F(k)} 
\prod_{i=1}^{k-1} \widetilde{\Psi}^{(1)}_{[u t^{2-2i}]_{\lambda_i-1, \barsigma_i}}
\left( u t^{2-2k} q^{2\lambda_k}\right)  \vert \lambda + 1_k , \barsigma+\bar{1}_k \rangle,
\\
E_{2,0} \vert \lambda, \barsigma \rangle &= \sum_{k=1}^{\ell(\lambda)} \delta_{\barsigma_k-1}  (-1)^{F(k)} 
\prod_{i=1}^{k-1} \widetilde{\Psi}^{(2)}_{[u t^{2-2i}]_{\lambda_i-1, \barsigma_i}} 
\left(u t^{1-2k} q^{2\lambda_k-1}\right)  \vert \lambda, \barsigma+\bar{1}_k \rangle,
\end{align}
where the normalization factor $\mathcal{E}_{\overline{s}}$ is set to be trivial and we employ
the renormalized generating functions \eqref{Cartan-eigen11} and \eqref{Cartan-eigen12}.
The Kronecker delta means the addable half tiles for $E_{1,0}$ are in the row with $\barsigma_i=0$, namely they are $\mathsf{t}$,
while those for $E_{2,0}$ are in the row with $\barsigma_i=1$, namely they are $\overline{\mathsf{t}}$.
By the relation \eqref{dictionary}, for the addable tile $\mathsf{t}$ in the $k$-th row, the coefficient is
\begin{align}
\psi_{\lambda_k}^{(1)}(q,t) :=&~(-1)^{F(k)}  t^{1-k} \prod_{i=1}^{k-1} \frac{\phi(t^2 q^{-2(\lambda_i - \barsigma_i)}; t^{2-2k}q^{2\lambda_k}, t^{2-2i})}
{\phi(q^{-2(\lambda_i - \barsigma_i)}; t^{2-2k}q^{2\lambda_k}, t^{2-2i})} \CR
%=&~ (-1)^{F(k)}  t^{k-1} \prod_{i=1}^{k-1} \frac{\phi(1; t^{2-2k} q^{2\lambda_k}, t^{-2i}q^{2(\lambda_i -\barsigma_i)})}
%{\phi(1; t^{2-2k} q^{2\lambda_k}, t^{2-2i}q^{2(\lambda_i -\barsigma_i)})} \CR
=&~ (-1)^{F(k)} \prod_{i=1}^{k-1} \frac{1 - t^{2k-2i-2}q^{2(\lambda_i -\lambda_k -\barsigma_i)}}
{1- t^{2k-2i}q^{2(\lambda_i -\lambda_k - \barsigma_i)}}, \label{matrix-elem-super1}
%=&~ (-1)^{F(k)}  t^{1-k} \prod_{i=1}^{k-1} \frac{1 - t^{2i-2k+2}q^{2(\lambda_k -\lambda_i +\barsigma_i)}}
%{1- t^{2i-2k}q^{2(\lambda_k -\lambda_i + \barsigma_i)}}, 
\end{align}
and for the addable tile $\overline{\mathsf{t}}$ in the $k$-th row, it is
\begin{align}
\psi_{\lambda_k}^{(2)}(q,t):=&~(-1)^{F(k)} t^{1-k} \prod_{i=1}^{k-1} \frac{\phi(t^{-1} q^{-2\lambda_i +1}; t^{1-2k}q^{2\lambda_k-1}, t^{2-2i})}
{\phi(t q^{-2\lambda_i +1}; t^{1-2k}q^{2\lambda_k-1}, t^{2-2i})} \CR
%=&~ (-1)^{F(k)} t^{1-k} \prod_{i=1}^{k-1}\frac{\phi(1; t^{1-2k} q^{2\lambda_k-1}, t^{3-2i}q^{2\lambda_i -1})}
%{\phi(1; t^{1-2k} q^{2\lambda_k-1}, t^{1-2i}q^{2\lambda_i - 1})} \CR
%=&~ (-1)^{F(k)} t^{1-k} \prod_{i=1}^{k-1}\frac{1 - t^{2k -2i+2}q^{2\lambda_i -2\lambda_k}}
%{1- t^{2k-2i}q^{2\lambda_i - 2\lambda_k}} \CR
=&~  (-1)^{F(k)} \prod_{i=1}^{k-1}\frac{1 - t^{2i -2k-2}q^{2\lambda_k -2\lambda_i}}
{1- t^{2i-2k}q^{2\lambda_k - 2\lambda_i}}. \label{matrix-elem-super2}
\end{align}
These formula should be compared with the Pieri formula for the Macdonald polynomials (\cite{MacD},\cite{FT} Lemma 6.3);
\begin{equation}
P_\mu e_r = \sum_\lambda \psi_{\lambda/\mu} P_\lambda,
\qquad
\psi_{\lambda/\mu} := \prod \frac{(1- q^{\mu_i -\mu_j}t^{j-i-1})(1- q^{\lambda_i -\lambda_j}t^{j-i+1})}
{(1- q^{\mu_i -\mu_j}t^{j-i})(1- q^{\lambda_i -\lambda_j}t^{j-i})},
\end{equation}
where $\lambda/\mu$ is a vertical $r$-strip and the product is taken over all the pairs $(i,j)$ 
with $i<j$ and $\lambda_i=\mu_i$ and $\lambda_j= \mu_j+1$. In particular when $r=1$,
\begin{equation}\label{Pieri1}
\psi_{\mu + j/\mu} =  \prod_{i=1}^{j-1}  \frac{(1- q^{\mu_i -\mu_j}t^{j-i-1})(1- q^{\mu_i -\mu_j-1}t^{j-i+1})}
{(1- q^{\mu_i -\mu_j}t^{j-i})(1- q^{\mu_i -\mu_j-1}t^{j-i})}.
\end{equation}
To reproduce the Pieri coefficient \eqref{Pieri1}, let us set $\barsigma_i =0$ for any $i$.
We first add $\mathsf{t}$ in the $k$-th row and then add $\overline{\mathsf{t}}$ in the same row.
Note that we should shift $\lambda_k \to \lambda_k +1$ in the second step.
We find
\begin{equation}
\psi_{\lambda_k}^{(1)}(q,t)\cdot \psi_{\lambda_k+1}^{(2)}(q^{-1},t^{-1}) 
= \prod_{i=1}^{k-1} \frac{1 - t^{2k-2i-2}q^{2\lambda_i -2\lambda_k}}
{1- t^{2k-2i}q^{2\lambda_i -2\lambda_k}} 
\frac{1 - t^{2k -2i+2}q^{2\lambda_i -2\lambda_k-2}}
{1- t^{2k-2i}q^{2\lambda_i - 2\lambda_k-2}},
\end{equation}
which agrees with \eqref{Pieri1}, if we change $(q^2, t^2) \to (q,t)$. 
It is remarkable that the arguments of $\psi_{\lambda_k+1}^{(2)}$ are not $(q,t)$ but $(q^{-1}, t^{-1})$. 
Note that the Pieri formula for the Macdonald polynomials is invariant under $(q,t) \to (q^{-1}, t^{-1})$.
But in the case of the super Macdonald polynomials it is not invariant. The change $(q,t) \to (q^{-1}, t^{-1})$
in \eqref{matrix-elem-super1} and \eqref{matrix-elem-super2} produces an additional power $t^{\pm 2(k-1)}$.

Since the generating functions of the eigenvalues of the Cartan generator allow the product formula 
\eqref {Eigen-factor} over half-tiles in the super Young diagram, the coefficients $\mathbf{E}_{\lambda, \lambda + \mathsf{b}}$
in the Pieri formula can be also expressed as the product over half-tiles \cite{Galakhov:2024zqn};
\begin{equation}
E_{1,0} \cdot \mathcal{M}_\lambda = \sum_{\mathsf{t} \in \mathrm{Add}(\lambda)} 
\mathbf{E}_{\lambda, \lambda + \mathsf{t}}  \cdot \mathcal{M}_{\lambda + \mathsf{t}}, 
\qquad
E_{2,0} \cdot \mathcal{M}_\lambda = \sum_{\overline{\mathsf{t}} \in \mathrm{Add}(\lambda)} 
\mathbf{E}_{\lambda, \lambda + \overline{\mathsf{t}}}  \cdot \mathcal{M}_{\lambda + \overline{\mathsf{t}}}, 
\end{equation}
where 
\begin{equation}
\mathbf{E}_{\lambda, \lambda + \mathsf{b}} = \Gamma_{\mathsf{b}}(\lambda) \prod_{\mathsf{b}' \in \lambda} 
\Delta_{\mathsf{b}, \mathsf{b}'}(x_{\mathsf{b}}-x_{\mathsf{b}'},y_{\mathsf{b}}-y_{\mathsf{b}'}).
\end{equation}
and $(x_{\mathsf{b}}, y_{\mathsf{b}})$ denotes the position of the box which the half-tile belongs to.\footnote{
In \cite{Galakhov:2024zqn} the coordinates of the first box in the first row is defined to be $(0,0)$. 
But here we define its coordinates as $(1,1)$.}
Each factor is related to the structure function as follows\footnote{As before 
we exchange $q$ and $t$ in the original definition in \cite{Galakhov:2024zqn}.};
\begin{equation}
\Delta_{\mathsf{t}, \overline{\mathsf{t}}}(x,y) = \varphi_{\mathsf{t}, \overline{\mathsf{t}}}(q^{2x-1}t^{-2y+1}),
\qquad
\Delta_{\overline{\mathsf{t}}, \mathsf{t}}(x,y) = \varphi_{\overline{\mathsf{t}}, \mathsf{t}}(q^{2x+1}t^{-2y-1}),
\qquad y>0,
\end{equation}
and $\Delta_{\mathsf{t}, \mathsf{t}}(x,y) = \Delta_{\overline{\mathsf{t}}, \overline{\mathsf{t}}}(x,y) =1$. 
When $y \leq 0$, namely when $y_{\mathsf{b}} \leq y_{\mathsf{b}'}$, we define $\Delta_{\mathsf{b}, \mathsf{b}'}=1$.
This means that the half tiles below the addable half-tile do not contribute to the product. 
We have checked our Pieri formula is consistent with the list of lower super Macdonald polynomials 
in \cite{Galakhov:2024zqn} (See Appendix \ref{lower-super-Mac}).

%%%%%%%%%%%%%%%%%%%%%%%%%%%%%%%%%%%%%%%%%%%%%%%%%%%%%%%%%%%%%%%%%%%%%%%%%%%%%%%%%%%%%%%%%%%%%%%%%%%%%%%%%%%%%%%%%%%%%%%%%%%%%%%%%%%

\section{Moduli space of the framed sheaves on $\widehat{\mathbb{P}}^{2}$}
\label{sec:framed-sheaves}

The problem of instanton counting on the blow-up of $\mathbb{C}^2$ or $\mathbb{P}^2$ was originally
introduced in \cite{Nakajima:2003pg}, which led to the blow-up formula for the Nekrasov's instanton partition function.
By virtue of the blow-up formula it was shown that the Seiberg-Witten prepotential is reconstructed 
from the instanton partition function which allows an explicit combinatorial expression 
in terms of the Young diagrams \cite{Nekrasov:2002qd}. 

\subsection{Quiver description}
\label{subsec:quiver}

%%%%%%%%%%%%%%%%%%%%%%%%%%%%%%%
\begin{figure}[t]
\begin{center}
\begin{picture}(50,60)
 \setlength{\unitlength}{1.2mm}
\thicklines
\put(-10,20){\circle{8}}
\put(20,20){\circle{8}}
\put(-11.5,19){\Large$v_0$}
\put(18.5,19){\Large$v_1$}
\put(17,23){\vector(-1,0){24.3}}
\put(16.1,21){\vector(-1,0){22}}
\put(-6,18){\vector(1,0){22.2}}

\put(2,-3){\line(1,0){6}}
\put(2,-3){\line(0,1){6}}
\put(2,3){\line(1,0){6}}
\put(8,-3){\line(0,1){6}}
\put(4,-1){\Large$r$}
\put(-6.5,6){\large$i$}
\put(15.5,6){\large$j$}
\put(3,25){\large$B_{1,2}$}
\put(3,14){\large$d$}
%%
%\put(6,18){\line(1,1){2.5}}
%\put(6,18){\line(1,-1){2.5}}
%%
\put(2,3){\vector(-1,1){12.5}}
\put(20,15.5){\vector(-1,-1){12.5}}
%\put(-11,-1){$1$}
%\put(19.5,-1){$2$}
%%
\end{picture}
\end{center}
\caption{Quiver for the framed sheaves on the blow-up}
\label{blowupADHM}
\end{figure}
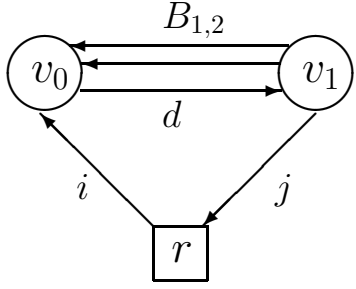
%%%%%%%%%%%%%%%%%%%%%%%%%%%%%%%%%%%

Let us consider the blow-up $p : \widehat{\mathbb{P}}^{2} \longrightarrow \mathbb{P}^2$ 
at a point $0$ with the exceptional curve $C$. 
Similarly to the famous ADHM construction of instantons on $\mathbb{P}^2$,
there is a quiver description of the framed locally free sheaves on $\widehat{\mathbb{P}}^{2}$
\cite{King,Nakajima:2008eq}.
In Figure \ref{blowupADHM} we have vector spaces $V_1, V_2$ denoted by circles and $W$ denoted by a square.
The morphisms $B_{1,2} \in \operatorname{Hom}(V_1, V_0)$, $d \in \operatorname{Hom}(V_0, V_1)$,
$i \in \operatorname{Hom}(W, V_0)$ and $j \in \operatorname{Hom}(V_1, W)$ satisfy
\begin{equation}\label{moment-map}
\mu(B_1, B_2, d, i,j) := B_1 d B_2 - B_2 d B_1 +ij =0.
\end{equation}
The rank of a framed sheaf $E$ is given by $r:= \dim W$. 
On the other hand the dimensions $v_1 := \dim V_1$ and  $v_2 := \dim V_2$ correspond to
the cohomological data;
\begin{equation}
k := - (c_1(E), C), \qquad
n := (c_2(E) - (r-1)c_1(E)^2/(2r), \widehat{\mathbb{P}}^{2}),
\end{equation}
where $c_1(E)$ and $c_2(E)$ are the first and the second Chen classes of $E$. Then we have
\begin{equation}
k = v_1 - v_0, \qquad n + \frac{k^2}{2r} = \frac{1}{2}(v_1 + v_0).
\end{equation}
In particular for the rank one case we find
\begin{equation}
%(c_1(E), C) = v_0 - v_1, \qquad
(c_2(E), \widehat{\mathbb{P}}^{2}) = v_0 - \frac{k(k-1)}{2} = v_1 - \frac{k(k+1)}{2},
\end{equation}
which implies a simple relation between the dimension vector $(v_0, v_1)$ and the super partition $(Y,S)$. 
Namely if $k \leq 0$,\footnote{According to \cite{Nakajima:2008eq}
the moduli space $\widehat{M}_{0}(r,k,n)$ of the framed locally sheaves is empty for $k>0$.}
 we can identify $-k$ with the number of marked boxes $|S|$ 
or the fermion number $m$ of $Y$ so that $v_0 = |Y^{\circledast}|$ and $v_1 = |Y^{*}|$. 
Then $n$ is nothing but the number of the relevant boxes of $(Y,S)$.
In the correspondence to a pair $(\lambda, \mu)$ of Young diagrams explained in section \ref{sec:super=pair},
$n = |\lambda| + |\mu|$.
Note that the action of the super DIM algebra allows a natural description in terms of 
the dimension vector of the quiver. 
Namely, $(E_1, F_1)$ makes the shift $v_0 \to v_0 \pm 1$ and $(E_2, F_2)$ makes the shift $v_1 \to v_1 \pm 1$.

By imposing the appropriate stability condition parametrized by $m \in \mathbb{Z}$ (see \cite{Nakajima:2008eq} for a precise definition), 
the moduli space $\widehat{M}_{0}(r,k,n)$ of the framed locally sheaves is embedded into the set of the ADHM-like data $(B_1, B_2, d, i,j)$
which satisfies \eqref{moment-map} and the stability condition modulo the action of the gauge group  $GL(v_0, \mathbb{C}) \times GL(v_1, \mathbb{C})$.
It is known that the description of the moduli space in terms of $(B_1, B_2, d, i,j)$ allows a generalization to the framed torsion free sheaves on $\widehat{\mathbb{P}}^{2}$.

Now we have two quivers in hand; one is given before in Figure \ref{conifold-quiver}.
It is for defining the commutation relations of the quantum toroidal algebra of type $\mathfrak{gl}_{1\vert 1}$ (the super DIM algebra).
The other is the quiver for the ADHM like description of the moduli space for the framed locally free sheaves 
on the blow-up given here in Figure \ref{blowupADHM}. It is interesting that 
we can understand the relation of the quivers by the dimensional reduction of the quiver with potential \cite{Rapcak:2020ueh}.
See Appendix \ref{App:quiver-reduction} for a brief review on this topic.

\subsection{Fixed points of the torus action}

To define the moduli space of the framed stable sheaves on the blow-up $\widehat{\mathbb{P}}^{2}$,
we need a stability parameter $m \in \mathbb{Z}$.
For the homological data $c = (r, c_1, \mathrm{ch}_2) \in H^{0}(\widehat{\mathbb{P}}^{2}) \oplus H^{2}(\widehat{\mathbb{P}}^{2})
 \oplus H^{4}(\widehat{\mathbb{P}}^{2})$, let $\widehat{M}^m(c)$ be the moduli space of coherent sheaves $E$ with $\operatorname{ch}(E) =c$ 
such that $E(-mC) := E \otimes \mathcal{O}_{\widehat{\mathbb{P}}^{2}}(-mC)$\footnote{$\mathcal{O}_{\widehat{\mathbb{P}}^{2}}(-mC)$ 
is the sheaf of holomorphic functions that has a pole of order at most $m$ along the divisor $C$.} 
is stable perverse coherent. 
In the rank one case, let $c=(1,0, -N \hbox{pt})$. Then $\widehat{M}^0(c)$ is the Hilbert 
scheme $(\mathbb{C}^2)^{[N]}$ of $N$ points on $\mathbb{C}^2$, while for sufficient large $m$
depending on $N$, $\widehat{M}^m(c)$ agrees with the Hilbert scheme $(\widehat{\mathbb{C}}^{2})^{[N]}$ 
of $N$ points on the blow-up $\widehat{\mathbb{C}}^{2}$.
It is known that $\widehat{M}^0(c)$ and $\widehat{M}^\infty(c) := \lim_{m \to \infty}\widehat{M}^m(c)$ are
related by the wall crossing.\footnote{
This is well described by the formula of the generating function of the Poincar\'e polynomials 
given by Corollary 5.14 of \cite{Nakajima:2008ss}.}
In \cite{Nakajima:2008ss} (see Lemma 5.1) it is shown that
the torus fixed points in $\widehat{M}^0(c)$ are in bijection to
$r$-tuples of pairs $(Y_\alpha, S_\alpha)$ of a Young diagram $Y_\alpha$ 
and a subset $S_\alpha$ of removable boxes such that
\begin{equation}
\sum_\alpha |S_\alpha| = (c_1, [C]), \qquad \sum_\alpha |Y_\alpha| 
= - \int_{\widehat{\mathbb{P}}^{2}}\operatorname{ch}_2 + \frac{1}{2} (c_1, [C]).
\end{equation}
As we explained in section \ref{sec:superY}, $(Y_\alpha, S_\alpha)$ can be identified with 
$r$-tuple of super partitions. A box in $S_\alpha \subset Y_\alpha$ is called marked. 
On the other hand, the fixed points of the torus action on the moduli space of framed torsion free sheaves on $\widehat{\mathbb{P}}^{2}$
are labeled by pair of partitions \cite{Nakajima:2003pg}. 
In section \ref{sec:super=pair} we have seen that {\it in the stable sector} the super partitions are
in one to one correspondence with pair of partitions. 
Let $\widehat{M}^\infty(r,k,n)$ be the moduli space of Gieseker stable framed torsion free sheaves on 
$\widehat{\mathbb{P}}^{2}$ with the rank $r$, $c_1 = k[C]$ and $c_2 =n$. It is known that
$\widehat{M}^m(r,k,n)$ is identical to $\widehat{M}^\infty(r,k,n)$ for $m \geq n$.
The condition $m \geq n$ is nothing but the definition of the stable sector in section \ref{sec:super=pair}. 
Note that by tensoring $\mathcal{O}_{\widehat{\mathbb{P}}^{2}}(-m C)$ we obtain an isomorphism 
$\widehat{M}^m(1,k,n) \simeq \widehat{M}^0(1,k+m,n)$.

\subsection{Nekrasov factor for super partitions}

In \cite{Nakajima:2008ss} the equivariant character for a pair of super partitions 
$\Lambda_\alpha =(Y_\alpha, S_\alpha)$ and $\Lambda_\beta =(Y_\beta, S_\beta)$
is computed as follows;
\begin{equation}\label{NY-1}
%\operatorname{ch} \operatorname{Ext}^1 (E_\alpha, E_\beta(-\ell_\infty))
\chi_{(\Lambda_\alpha, \Lambda_\beta)} (q,t) = {\sum_{\mathsf{s} \in Y_\alpha \setminus S_\alpha}}^{\hspace{-2mm}\prime}~t^{-\ell_{Y_\beta}(\mathsf{s})} 
q^{-a_{Y_\alpha\setminus S_\alpha}(\mathsf{s})-1} +
 {\sum_{\mathsf{t} \in Y_\beta}}^{\prime}  t^{\ell_{Y_\alpha\setminus S_\alpha}(\mathsf{t})+1} q^{a_{Y_\beta}(\mathsf{t})}.
\end{equation}
The arm length and the leg length at $\mathsf{s}=(i,j)$ are defined by $a_\lambda(i,j) = \lambda_i -j$ and $\ell_\lambda(i,j)= \lambda^\vee_j -i$.
The prime on the sum means that the irrelevant boxes determined by $(S_\alpha, S_\beta)$ are removed from the sum (see below for more details). 
For later convenience we have set $(t_1, t_2) = (t, q^{-1})$ in the original formula in \cite{Nakajima:2008ss}.
When $S_\alpha$ is empty, the corresponding super partition is an ordinary partition and the formula \eqref{NY-1} implies
the standard Nekrasov factor $\mathsf{N}_{Y_\alpha,Y_\beta}(q,t)$. 
In Appendix \ref{App:matching} we confirm that in the stable sector $\chi_{(\Lambda_\alpha, \Lambda_\beta)} (q,t)$
agrees with the equivariant character for the instanton counting on the blow-up,
where the corresponding pair of partitions is determined by the rule explained in section \ref{sec:super=pair}.

In the formula \eqref{NY-1}, the concept of the (ir)relevant boxes plays a crucial role. 
For a single super Young diagram we have already provided the definition of  the (ir)relevant boxes in section \ref{sec:superY}.
Now we define the irrelevant boxes for a pair of super partitions $\Lambda_\alpha$ and $\Lambda_\beta$. 
The relevant boxes are those which are not irrelevant. 
For a pair of the marked boxes $(\mathsf{s}, \mathsf{s}') \in S_\alpha \times S_\beta$, we define two boxes $\mathsf{u}$ and $\mathsf{u}'$.
The box $\mathsf{u}$ is in the same column as $\mathsf{s}$ and in the same row as $\mathsf{s}'$.
The other box $\mathsf{u}'$ is in the same column as $\mathsf{s}'$ and in the same row as $\mathsf{s}$.
Our convention of the Young diagram follows Macdonald \cite{MacD}. To move to the convention of \cite{Nakajima-AMS}, 
we should clockwise rotate the diagram by $\pi/2$. Then we can keep the same definitions of the arm length and the leg length. 
Recall the co-arm length is $a'(i,j) = j-1$.
The definition of $\mathsf{u}$ and $\mathsf{u}'$ implies that
\begin{enumerate}
\item
If $a'(\mathsf{s}) < a'(\mathsf{s}')$, $\mathsf{u} \in Y_\beta$ and $\mathsf{u}' \notin Y_\alpha\setminus S_\alpha$. 
\item
If $a'(\mathsf{s}) = a'(\mathsf{s}')$, $\mathsf{u} \in S_\beta \subset Y_\beta$ and $\mathsf{u}' \notin Y_\alpha\setminus S_\alpha$. 
\item
If $a'(\mathsf{s}) >  a'(\mathsf{s}')$, $\mathsf{u} \notin Y_\beta$ and $\mathsf{u}' \in Y_\alpha\setminus S_\alpha$. 
\end{enumerate}
We define $\mathsf{u} \in Y_\beta$ in the first two cases and $\mathsf{u}' \in Y_\alpha\setminus S_\alpha$ in the last case 
as the irrelevant box associated with the pair $(\mathsf{s},\mathsf{s}')$. 

%%%%%%%%%%%%%%%%%%%%%%%%%%%%%%%%%%%%%%%%%%%%%%%%%%%%%%%%%%%%%%%%%%%%%%%%%%%%%
\begin{figure}[h]
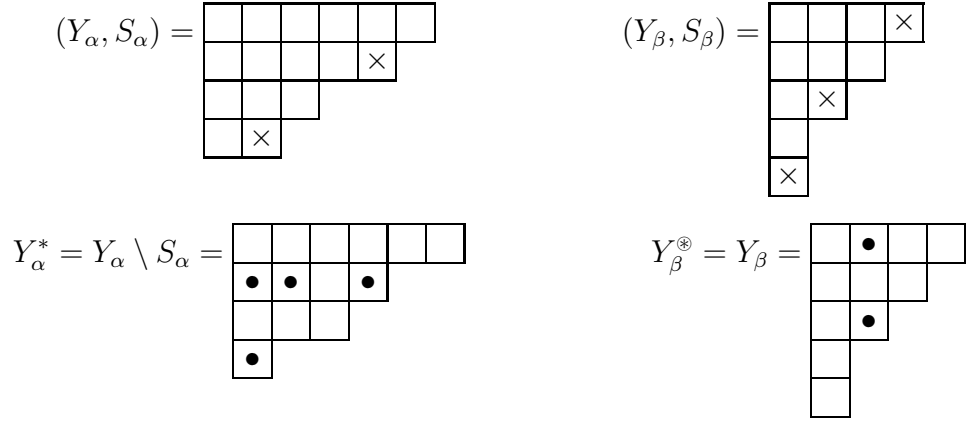

$$
(Y_\alpha, S_\alpha)=\ytableaushort{{}{}{}{}{}{},{}{}{}{}{\times},{}{}{},{}{\times}} \qquad \qquad \qquad
(Y_\beta, S_\beta)=\ytableaushort{{}{}{}{\times},{}{}{},{}{\times},{},{\times}}
$$

$$
Y_\alpha^{*} = Y_\alpha \setminus S_\alpha=\ytableaushort{{}{}{}{}{}{},{\bullet}{\bullet}{}{\bullet},{}{}{},{\bullet}} \qquad \qquad \qquad
Y_\beta^{\circledast} = Y_\beta =\ytableaushort{{}{\bullet}{}{},{}{}{},{}{\bullet},{},{}}
$$
\caption{
The boxes with $\bullet$ are irrelevant boxes. The irrelevant boxes in $Y_\alpha^{*}$ come from pairs of the marked boxes $(\mathsf{s},\mathsf{s}')$ with
$a'(\mathsf{s}) > a'(\mathsf{s}')$, while those in $Y_\beta^{\circledast}$ are from pairs $(\mathsf{s},\mathsf{s}')$ with $a'(\mathsf{s}) \leq a'(\mathsf{s}')$. 
The total number of the irrelevant boxes is $|S_\alpha| \times |S_\beta| =6$.}
\end{figure}
\begin{figure}[t]
$$
(Y_\alpha, S_\alpha) = (Y_\beta, S_\beta) = \ytableaushort{{}{}{}{}{}{\times},{}{}{},{}{\times},{},{\times}},
\qquad \qquad |S| = |S_\alpha| = |S_\beta| =3
$$

$$
Y_\alpha^{*} = Y_\alpha \setminus S_\alpha=\ytableaushort{{\bullet}{\bullet}{}{}{},{}{}{},{\bullet},{}} \qquad \qquad \qquad
Y_\beta^{\circledast} = Y_\beta =\ytableaushort{{\bullet}{\bullet}{}{}{}{\bullet},{}{}{},{\bullet}{\bullet},{},{\bullet}}
$$
\caption{
The irrelevant boxes in the case $(Y_\alpha, S_\alpha) = (Y_\beta, S_\beta)$;
the numbers of irrelevant boxes in $Y_\alpha^{*}$ and $Y_\beta^{\circledast}$ are $\frac{1}{2}|S|(|S|-1)=3$ 
and $\frac{1}{2}|S|(|S|+1)=6$, respectively.
In total the number of irrelevant boxes is $|S|^2 =9$. }
\end{figure}
%%%%%%%%%%%%%%%%%%%%%%%%%%%%%%%%%%%%%%%%%%%%%%%%%%%%%%%%%%%%%%%%%%%%%%%%%%%%%%%%%%%%%%%%%

In \cite{Nakajima:2008ss}, it is argued that the contribution from the irrelevant boxes is
\begin{equation}
\operatorname{ch} \operatorname{Hom}(S_\alpha, S_\beta) = \sum_{\mathsf{s} \in S_\alpha, \mathsf{t} \in S_\beta} 
t^{\ell'(\mathsf{s}) - \ell'(\mathsf{t})} q^{a'(\mathsf{t}) - a'(\mathsf{s})},
\end{equation}
where $\ell'$ and $a'$ are co-leg length and co-arm length, respectively;
\begin{equation}
\ell'(i,j) = i-1, \qquad a'(i,j) = j-1.
\end{equation}
Hence, we can remove the prime on the sum in \eqref{NY-1} and recast the formula as follows;\footnote{In the second term of the formula \eqref{NY-1}
we can replace $\mathsf{t} \in Y_\beta$ with $\mathsf{t} \in Y_\beta \setminus S_\beta$, since the marked boxes are irrelevant. However, in the formula \eqref{NY-2}
we cannot replace $\mathsf{t} \in  Y_\beta^{\circledast}$ with $\mathsf{t} \in  Y_\beta^{*}$. For example, when $Y_\alpha= Y_\beta, S_\alpha= S_\beta$ 
and $m=|S_\alpha|=|S_\beta|$, the number of the irrelevant boxes in the first term is $\frac{1}{2}m(m-1)$ and for the  second term it is $\frac{1}{2}m(m+1)$.
In total there are $m^2$ terms to be subtracted. In fact it agrees with the number of terms in the last summation of \eqref{NY-2}.}
\begin{align}\label{NY-2}
& % \operatorname{ch} \operatorname{Ext}^1 (E_\alpha, E_\beta(-\ell_\infty))  
\chi_{(\Lambda_\alpha, \Lambda_\beta)} (q,t) \CR
& = {\sum_{\mathsf{s} \in Y_\alpha^{*}}}t^{-\ell_{ Y_\beta^{\circledast}}(\mathsf{s})} q^{-a_{Y_\alpha^{*}}(\mathsf{s})-1} +
 {\sum_{\mathsf{t} \in  Y_\beta^{\circledast}} t^{\ell_{Y_\alpha^{*}}(\mathsf{t})+1} q^{a_{Y_\beta^{\circledast}}(\mathsf{t})}}
 - \sum_{\mathsf{s} \in S_\alpha, \mathsf{t} \in S_\beta} t^{\ell'(\mathsf{s}) - \ell'(\mathsf{t})} q^{a'(\mathsf{t}) - a'(\mathsf{s})}.
\end{align}
It is this formula which allows us to express the Nakrasov factor in an infinite product form;\footnote{Later we will introduce
 the spectral parameter $u$. But here we set $u=1$ for simplicity.} 
\begin{align}\label{infinite}
%& \operatorname{ch} \operatorname{Ext}^1 (E_\alpha, E_\beta(-\ell_\infty)  
& \mathsf{N}_{\Lambda_\alpha, \Lambda_\beta} (q, t) \CR
& = \prod_{i,j =1}^\infty \frac{(q^{(Y_\beta^{\circledast})_j - (Y_\alpha^{*})_i} t^{i-j+1};q)_\infty ( t^{i-j};q)_\infty }
{(q^{(Y_\beta^{\circledast})_j - (Y_\alpha^{*})_i}  t^{i-j} ;q)_\infty (t^{i-j+1};q)_\infty}
 \cdot \prod_{\mathsf{s} \in S_\alpha, \mathsf{t} \in S_\beta} \frac{1}{1-t^{\ell'(\mathsf{s}) - \ell'(\mathsf{t})} q^{a'(\mathsf{t}) - a'(\mathsf{s})}}.
\end{align}
When both $S_\alpha$ and $S_\beta$ are empty and hence $Y^{\circledast} = Y^{*}$,
\eqref{infinite} agrees with Eq.(2.12) in \cite{Awata:2008ed} by the exchange $Y_\alpha \leftrightarrow Y_\beta$.
Hence, it is reasonable to regard \eqref{infinite} as  a generalization of the Nekrasov factor to super partitions. 
Note that the set of marked boxes are labelled by sequences $1 \leq i_1 < i_2 < \cdots < i_{f_\alpha} \leq \ell(Y_\alpha) +1$,
where $f_\alpha = |S_\alpha|$ is the fermion number of the super partition $Y_\alpha$ and $\ell(Y_\alpha)$ is the length
of the super partition. The marked boxes are $(i_1, (Y_\alpha)_{i_1}), (i_2, (Y_\alpha)_{i_2}), \ldots, (i_{f_\alpha}, (Y_\alpha)_{i_{f_\alpha}})$
with $(Y_\alpha)_1 \geq (Y_\alpha)_{i_1} > (Y_\alpha)_{i_2}  > \cdots > (Y_\alpha)_{i_{f_\alpha}} \geq 1$.
With these notations the second finite product in \eqref{infinite} can be written as 
\begin{equation}\label{fermions}
\prod_{a=1}^{f_\alpha} \prod_{b=1}^{f_\beta}  \frac{1}{1- t^{i_a -j_b}  q^{(Y_\beta)_{j_b}- (Y_\alpha)_{i_a}}}.
\end{equation}
Hence, the Nekrasov factor of the super partitions may be further rewritten as 
\begin{align}\label{infinite-new}
& \mathsf{N}_{\Lambda_\alpha, \Lambda_\beta} (q, t) \CR
& = \prod_{i,j =1}^\infty \frac{(q^{(Y_\beta^{\circledast})_j - (Y_\alpha^{*})_i} t^{i-j+1};q)_\infty (t^{i-j};q)_\infty }
{(q^{(Y_\beta^{\circledast})_j - (Y_\alpha^{*})_i}  t^{i-j} ;q)_\infty (t^{i-j+1};q)_\infty}
% \cdot \prod_{s \in S_\alpha, t \in S_\beta} \frac{1}{1-t_1^{\ell'(s) - \ell'(t)} t_2^{a'(s) - a'(t)}}.
\prod_{a=1}^{f_\alpha} \prod_{b=1}^{f_\beta}  \frac{1}{1- t^{i_a -j_b} q^{(Y_\beta^{\circledast})_{j_b} - (Y_\alpha^{\circledast})_{i_a}}} \CR
& = \prod_{(i,j)  \notin F_\alpha \times F_\beta} \frac{(q^{(Y_\beta^{\circledast})_j - (Y_\alpha^{*})_i} t^{i-j+1};q)_\infty ( t^{i-j};q)_\infty }
{(q^{(Y_\beta^{\circledast})_j - (Y_\alpha^{*})_i}  t^{i-j} ;q)_\infty (t^{i-j+1};q)_\infty}
\prod_{{(i,j ) \in F_\alpha \times F_\beta}}\frac{(q^{(Y_\beta^{\circledast})_j - (Y_\alpha^{*})_i} t^{i-j+1};q)_\infty ( t^{i-j};q)_\infty }
{(q^{(Y_\beta^{\circledast})_j - (Y_\alpha^{\circledast})_i}  t^{i-j} ;q)_\infty (t^{i-j+1};q)_\infty} \CR
& = \prod_{i,j =1}^\infty \frac{(q^{(Y_\beta^{\circledast})_j -(Y_\alpha^{\circledast})_i} t^{i-j+1};q)_\infty (t^{i-j};q)_\infty }
{(q^{(Y_\beta^{\circledast})_j - (Y_\alpha^{\circledast})_i}  t^{i-j} ;q)_\infty (t^{i-j+1};q)_\infty}
\prod_{a=1}^{f_\alpha} \prod_{b=1}^{f_\beta}  \frac{1}{1- t^{i_a -j_b+1}  q^{(Y_\beta^{\circledast})_{j_b} - (Y_\alpha^{\circledast})_{i_a}}},
\end{align}
where $F_\alpha$ and $F_\beta$ are the sets of fermionic rows in $Y_\alpha$ and $Y_\beta$, respectively. 
Namely, when both $i$ and $j$ correspond to fermionic rows; $(Y_\alpha^{\circledast})_i - (Y_\alpha^{*})_i =1,
(Y_\beta^{\circledast})_j - (Y_\beta^{*})_j =1$,
the contribution of the irrelevant boxes makes the change $Y_\alpha^{*} \to Y_\alpha^{\circledast}$ in the denominator or the numerator. 
In the last expression, where $Y_\alpha^{*}$ is eliminated, 
the first factor is the same as the standard Nekrasov factor and the second factor is regarded as the correction 
for the super partitions. It seems natural to expect that the correction factor
associated with a pair of fermionic rows comes from the OPE of free fermions. 

%%%%%%%%%%%%%%%%%%%%%%%%%%%%%%%%%%%%%%%%%%%%%%%%%%%%%%%%%%%%%%%%%%%%%%%%%%%%%%%%%%%%%%%%%%%%%%%%%%%%%%%%%%%%%

\section{Pieri formula and matrix elements of $E_{i,0}$}

In this section we provide a proof of Proposition \ref{main-prp}.

\subsection{Integral form of the super Macdonald polynomials}

Recall that the norm of the Macdonald polynomials is given by
\begin{equation}
\langle P_\lambda \vert P_\lambda \rangle = \frac{c'_\lambda}{c_\lambda},
\end{equation}
where
\begin{equation}
c_\lambda = \prod_{\mathsf{s} \in \lambda} (1- q^{a(\mathsf{s})} t^{\ell(\mathsf{s})+1}), 
\qquad c'_\lambda = \prod_{\mathsf{s} \in \lambda} (1- q^{a(\mathsf{s})+1} t^{\ell(\mathsf{s})}). 
\end{equation}

In \cite{Blondeau-Fournier:2011sft} they conjectured the norm of super Macdonald polynomials $\mathcal{M}_\Lambda$;
\begin{equation}
|| \mathcal{M}_\Lambda ||^2 = (-1)^{\binom{m}{2}} \langle\!\langle \mathcal{M}_\Lambda \vert \mathcal{M}_\Lambda \rangle\!\rangle_{q,t}
= q^{|\Lambda^F|} \frac{w_\Lambda(q,t)}{w_{\Lambda^\vee}(t,q)},
\end{equation}
where
\begin{equation}\label{superMac-norm}
w_\Lambda(q,t) := \prod_{\mathsf{s} \in \mathcal{B}(\Lambda)} (1- q^{a_{\Lambda^{*}}(\mathsf{s})+1} t^{\ell_{\Lambda^{\circledast}}(\mathsf{s})}),
\end{equation}
and {\it $\mathcal{B}(\Lambda)$ denotes\footnote{$\mathcal{B}$ stands for the bosonic content.} the set of boxes in the diagram of 
$\Lambda$ that do not appear
at the same time in a row containing a circle and in a column containing a circle}.\footnote{We just quote the definition in \cite{Blondeau-Fournier:2011sft} }
This is nothing but the definition of the relevant boxes by Nakajima-Yoshioka.
In particular we find the diagonal part of the Nekrasov factor is
\begin{align}
\mathsf{N}_{\Lambda \Lambda}(q, t)  &= \prod_{\mathsf{s} \in \mathcal{B}(\Lambda)} 
(1 - t^{-\ell_{\Lambda^{\circledast}}(\mathsf{s})} q^{-a_{\Lambda^{*}}(\mathsf{s}) -1})
(1 - t^{\ell_{\Lambda^{*}}(\mathsf{s})+1} q^{a_{\Lambda^{\circledast}}(\mathsf{s})})  \CR
&= w_{\Lambda}(q^{-1}, t^{-1})  w_{\Lambda^\vee}(t,q). 
\end{align}

It is natural to define the integral form of the Macdonald polynomials by
\begin{equation}\label{integral}
\mathcal{J}_\Lambda := w_{\Lambda^\vee}(t,q) \mathcal{M}_\Lambda .
\end{equation}
In \cite{Blondeau-Fournier:2011sft}  they conjectured that $\mathcal{J}_\Lambda$ are polynomials in $q$ and $t$ with integral
 (not necessarily positive) coefficients. 
We expect that the integral form $\mathcal{J}_\Lambda$ is what corresponds to the fixed point basis.
Then \eqref{integral} is nothing but the base change from the super Macdonald basis to the fixed point basis.

%%%%%%%%%%%%%%%%%%%%%%%%%%%%%%%%%%%%%%%%%%%%%%%%%%%%%%%%%%%%%%%%%%%%%%%%%%%%%%%%%%

\subsection{Matrix elements in the fixed point basis}

From the relation of the Nekrasov factor and the matrix elements of the fixed point basis in the case of Macdonald polynomials,
which is reviewed in Appendix \ref{DIMcase},
We expect the matrix elements of the generators $E_1(z)$ and $E_2(z)$ of the quantum toroidal algebra of type $\mathfrak{gl}_{(1|1)}$ 
are obtained by evaluating the difference of the Nekrasov factor under the change of the marked boxes in the super partition. 
Recall that $E_1(z)$ changes the bosonic rows to fermionic and  $E_2(z)$ changes the fermionic rows to bosonic. 
Namely the change of the row length of super partition by $E_1$ is $\Lambda^{\circledast}_i \to \Lambda^{\circledast}_i +1$, while keeping $\Lambda^{*}_i$. 
Similarly the change of the row length by $E_2$ is $\Lambda^{*}_i \to \Lambda^{*}_i +1$, while keeping $\Lambda^{\circledast}_i$. 
We also have to pay attention to the change of the irrelevant boxes due to the change of marked boxes.

The Nekrasov factor for a pair of the super partitions $\Lambda = (Y_\alpha, S_\alpha)$ and 
$\Pi = (Y_\beta, S_\beta)$ is
\begin{align}
 \mathsf{N}_{\Lambda, \Pi} (u \vert q, t) 
%& = \prod_{i,j =1}^\infty \frac{(t_2^{(Y_\alpha^{*})_i -(Y_\beta^{\circledast})_j} t_1^{i-j+1};t_2^{-1})_\infty ( t_1^{i-j};t_2^{-1})_\infty }
%{(t_2^{(Y_\alpha^{*})_i -(Y_\beta^{\circledast})_j}  t_1^{i-j} ;t_2^{-1})_\infty (t_1^{i-j+1};t_2^{-1})_\infty}
%% \cdot \prod_{s \in S_\alpha, t \in S_\beta} \frac{1}{1-t_1^{\ell'(s) - \ell'(t)} t_2^{a'(s) - a'(t)}}.
%\prod_{a=1}^{f_\alpha} \prod_{b=1}^{f_\beta}  \frac{1}{1- t_1^{i_a -j_b}  t_2^{(Y_\alpha^{\circledast})_{i_a}-(Y_\beta^{\circledast})_{j_b}}} \CR
&= \prod_{(i,j)  \notin F_\alpha \times F_\beta} \frac{(u q^{(Y_\beta^{\circledast})_j - (Y_\alpha^{*})_i} t^{i-j+1};q)_\infty (u t^{i-j};q)_\infty }
{(u q^{(Y_\beta^{\circledast})_j - (Y_\alpha^{*})_i}  t^{i-j} ;q)_\infty (ut^{i-j+1};q)_\infty} \CR
& \qquad \times
\prod_{{(i,j ) \in F_\alpha \times F_\beta}}\frac{(u q^{(Y_\beta^{\circledast})_j - (Y_\alpha^{*})_i} t^{i-j+1};q)_\infty (u t^{i-j};q)_\infty }
{(u q^{(Y_\beta^{\circledast})_j - (Y_\alpha^{\circledast})_i}  t^{i-j} ;q)_\infty (u t^{i-j+1};q)_\infty},
\end{align}
where $F_\alpha$ and $F_\beta$ are the sets of fermionic rows in $\Lambda$ and $\Pi$, respectively. 
We have introduced the spectral parameter $u$.
In terms of the notation $\sigma_i$, where $\sigma_i =0$ for a bosonic row and $\sigma_i=1$ for a fermionic row, we can write down
the infinite product form of the Nekrasov factor;
\begin{equation}\label{Nek-reduced}
 \mathsf{N}_{\Lambda \Pi} (u \vert q, t) = \prod_{i,j=1}^\infty \frac{(u q^{\pi_j - \lambda_i + \sigma_i} t^{i-j+1};q)_\infty (u t^{i-j};q)_\infty}
{(u q^{\pi_j - \lambda_i + \sigma_i(1-\sigma_j)}  t^{i-j} ;q)_\infty (u t^{i-j+1};q)_\infty}.
\end{equation}

Now assume that the $k$-th low of the partition $\Lambda$ is bosonic. Let us add a marked box to this low, which makes the row fermionic. 
To find the change of the Nekrasov factor under this operation, we only have to look at $j=k$ part of \eqref{Nek-reduced}. 
Computing similarly to  the DIM case (See Appendix \ref{DIMcase}), we obtain
\begin{align}\label{matrix-elem-1}
\frac{\mathsf{N}_{\Lambda, \Lambda}(u \vert q,t)}{\mathsf{N}_{\Lambda, \Lambda + \boxtimes_k}(u \vert q,t)} 
&= \prod_{i=1}^\infty \frac{(u q^{\lambda_k - \lambda_i + \sigma_i} t^{i-k+1};q)_\infty}
{(u q^{\lambda_k - \lambda_i + \sigma_i}  t^{i-k} ;q)_\infty}
\frac{(u q^{\lambda_k - \lambda_i + 1}  t^{i-k} ;q)_\infty}
{(u q^{\lambda_k - \lambda_i+ \sigma_i +1} t^{i-k+1};q)_\infty} \CR
&= \frac{1-ut}{1-u}\prod_{\substack{i=1 \\ i \neq k}}^\infty \frac{1- u q^{\lambda_k - \lambda_i + \sigma_i} t^{i-k+1}}
{(1- u q^{\lambda_k - \lambda_i} t^{i-k})^{\delta_{\sigma_i,0}}}.
\end{align}
Note that the denominator becomes trivial for fermionic rows of $\Lambda$. 

Next we assume that the $k$-th row of $\Lambda$ is fermionic. Namely the last box of the row is marked. Let us remove the marking, 
which makes the row bosonic. Similar computations to Appendix \ref{DIMcase} give
\begin{align}\label{matrix-elem-2}
\frac{\mathsf{N}_{\Lambda, \Lambda}(u \vert q,t)}{\mathsf{N}_{\Lambda, \Lambda(\boxtimes_k \to \square_k)}(u \vert q,t)} 
&= \prod_{i=1}^\infty \frac{(u q^{\lambda_k - \lambda_i + \sigma_i} t^{i-k+1};q)_\infty}
{(u q^{\lambda_k - \lambda_i}  t^{i-k} ;q)_\infty}
\frac{(u q^{\lambda_k - \lambda_i + \sigma_i}  t^{i-k} ;q)_\infty}
{(u q^{\lambda_k - \lambda_i+ \sigma_i} t^{i-k+1};q)_\infty} \CR
&= \frac{1}{1-u} \prod_{\substack{i \in F \\ i \neq k}} \frac{1}{1 - u q^{\lambda_k - \lambda_i}  t^{i-k}}.
\end{align}
Hence, only the fermionic rows contribute to the matrix element, which compensates the missing factors in \eqref{matrix-elem-1}.

Similarly looking at $i=k$ part of \eqref{Nek-reduced}, we obtain
\begin{align}
\frac{\mathsf{N}_{\Lambda, \Lambda}(u \vert q,t)}{\mathsf{N}_{\Lambda - \boxtimes_k, \Lambda}(u \vert q,t)} 
&= \frac{1}{1-u}\prod_{\substack{j \in F \\ j \neq k}} \frac{1}{1-u q^{\lambda_j -\lambda_k} t^{k-j}}, \\
\frac{\mathsf{N}_{\Lambda, \Lambda}(u \vert q,t)}{\mathsf{N}_{\Lambda(\square_k \to \boxtimes_k), \Lambda}(u \vert q,t)} 
&=  \frac{1-ut}{1-u} \prod_{\substack{j=1 \\ j \neq k}}^\infty \frac{1- u q^{\lambda_j - \lambda_k} t^{k-j+1}}
{(1- u q^{\lambda_j - \lambda_k} t^{k-j})^{\delta_{\sigma_j,0}}},
\end{align}
which should give the matrix elements of the zero modes of $F_s(z)$.

From the formula of the equivariant character of the tangent space of the correspondence in the Macdonald case, 
which is briefly recalled in Appendix \ref{App:correspondence},  we expect that the matrix elements of the generators 
defined by the correspondence are given by taking the limit $u \to 1$ of the ratios of the Nekrasov factors 
after multiplying the correction factor $\frac{1-u}{(1-ut)(1-uq^{-1})}$. The correction factor cancels the simple pole at $u=1$
and fixes the normalization factor in the limit $u \to 1$.

%%%%%%%%%%%%%%%%%%%%%%%%%%%%%%%%%%%%%%%%%%%%%%%%%%%%%%%%%%%%%%%

\subsection{Base change to the super Macdonald polynomials}

Let us evaluate the change of 
\begin{equation}\label{scaling-superMac}
\widetilde{c}_\Lambda(q,t) := w_{\Lambda^\vee}(t,q) = \prod_{\mathsf{s} \in \mathcal{B}(\Lambda)}
 (1- t^{\ell_{\Lambda^{*}}(\mathsf{s})+1} q^{a_{\Lambda^{\circledast}}(\mathsf{s})}),
\end{equation}
under two operations;
\begin{enumerate}
\item
Add a marked box to the $k$-th row, which is possible only when the row is bosonic.
In this case $\Lambda^{*}$ does not change and only the arm length $a_{\Lambda^{\circledast}}(\mathsf{s})$ 
of  $\mathsf{s}=(k,j)~(1 \leq j \leq \Lambda_k)$ will change.
Let us denote the resulting super partition by $\Lambda + \boxtimes_k$.
\item
Remove the marking in the $k$-th row, which is possible only when the row is fermionic.
In this case $\Lambda^{\circledast}$ does not change and only the leg length $\ell_{\Lambda^{*}}(\mathsf{s})$ of 
$\mathsf{s}=(i, \lambda_k),~(1 \leq i \leq k-1)$  will change.
Let us denote this operation by $\Lambda(\boxtimes_k \to \square_k)$.
\end{enumerate}

%%%%%%%%%%%%%%%%%%%%%%%%%%%%%%%%%%%%%%%%%%%%%%%%%%%%%%%%%%%%%%%%%%%%%%

\begin{lem}
\label{lem:c-recursion}
We have
\begin{align}
\frac{\widetilde{c}_{ \Lambda + \boxtimes_k}(q,t) }{\widetilde{c}_\Lambda(q,t) }
&= \prod_{j=k+1}^{\infty} \frac{1- q^{\lambda_k -\lambda_j + \sigma_j} t ^{j-k}} 
{1- q^{\lambda_k -\lambda_j + \sigma_j}  t^{j-k+1 }} \CR
& \qquad \times \prod_{i<k, i \in F}(1- t^{k-i} q^{\lambda_i - \lambda_k -1})^{-1}
\cdot \prod_{k<j, j \in F} (1- t^{j-k} q^{\lambda_k - \lambda_j+1})^{-1},
\\
\frac{\widetilde{c}_{\Lambda(\boxtimes_k \to \square_k)}(q,t) }{\widetilde{c}_\Lambda(q,t) }
&=  \prod_{i=1}^{k-1} \frac{1-q^{\lambda_i - \lambda_k}t^{k-i+1}}{1- q^{\lambda_i -\lambda_k } t^{k-i}} \CR
& \qquad \times (1-t)\prod_{i<k, i \in F} (1- t^{k-i} q^{\lambda_i - \lambda_k})
\cdot \prod_{k<j, j \in F} (1- t^{j-k} q^{\lambda_k - \lambda_j}).
\end{align}
\end{lem}
Note that when all the rows are bosonic, only the first factors in these formulas survive and we have $\sigma_j=0$. 
Then by taking the product we see
$$
\frac{\widetilde{c}_{ \Lambda + \square_k}(q,t) }{\widetilde{c}_\Lambda(q,t) }
=  t^{k-1} \prod_{i=1}^{k-1} \frac{1-q^{\lambda_k - \lambda_i}t^{i-k-1}}{1- q^{\lambda_k -\lambda_i } t^{i-k}} \
 \prod_{j=k+1}^{\infty} \frac{1- q^{\lambda_k -\lambda_j} t ^{j-k}} {1- q^{\lambda_k -\lambda_j}  t^{j-k+1 }},
$$
which agrees with \eqref{var-c} in Appendix \ref{DIMcase}. 

\begin{proof}
In the definition \eqref{scaling-superMac} the product is taken over the relevant boxes $\mathcal{B}(\Lambda)$ of the super partition $\Lambda$.
The point is that after the above two operation the set of relevant boxes does change.
For simplicity we first evaluate the change of $\widetilde{c}_\Lambda(q,t)$ by neglecting the contributions form the change of irrelevant boxes.
Then the computation is almost similar to Appendix \ref{DIMcase}. 
For the first operation the change of the arm length produces the contributions from the row below the $k$-th row\footnote{
Note that the leg-length refers to $\Lambda^{*}$.};
\begin{equation}
\prod_{j=k+1}^{\infty} \frac{1- q^{\lambda_k -\lambda_j + \sigma_j} t ^{j-k}} 
{1- q^{\lambda_k -\lambda_j + \sigma_j}  t^{j-k+1 }}.
\end{equation}
Similarly in the case of the second operation the change of the leg length produces the contributions from the row above the $k$-th row;
\begin{equation}
\prod_{i=1}^{k-1} \frac{1-q^{\lambda_i - \lambda_k}t^{k-i+1}}{1- q^{\lambda_i -\lambda_k} t^{k-i}}
= t^{k-1} \prod_{i=1}^{k-1} \frac{1-q^{\lambda_k - \lambda_i}t^{i-k-1}}{1- q^{\lambda_k -\lambda_i } t^{i-k}}.
\end{equation}

Now let us take into account the restriction of the product in \eqref{scaling-superMac} to the relevant boxes $\mathcal{B}(\Lambda)$. 
In the case of the first operation, adding a marked box to the $k$-th row creates new irrelevant boxes which are one to one correspondence with fermionic 
row other than the $k$-th row. We do not have to care about the contribution of the new marked box, since it is an irrelevant box. 
Since the new marked box has the coordinate $(k, \lambda_k +1)$, new irrelevant boxes are at
$(i, \lambda_k +1)$ for $i \in F,~i<k$ and $(k,\lambda_j)$ for $j \in F,~k<j$. 
The net contribution is
\begin{equation}
\prod_{i<k, i \in F}(1- t^{k-i} q^{\lambda_i - \lambda_k -1})^{-1}
\prod_{k<j, j \in F} (1- t^{j-k} q^{\lambda_k - \lambda_j+1})^{-1}.
\end{equation}
On the other hand, removing the marking in the $k$-th row annihilates irrelevant boxes which are one to one correspondence with fermionic 
row including the $k$-th row itself. 
Since the marking at $(k, \lambda_k)$ is removed, the boxes which becomes relevant are at $(i, \lambda_k),~i<k$ and
$(k, \lambda_j),~k<j$. The net contribution is
\begin{equation}
(1-t)\prod_{i<k, i \in F} (1- t^{k-i} q^{\lambda_i - \lambda_k})
\prod_{k<j, j \in F} (1- t^{j-k} q^{\lambda_k - \lambda_j}).
\end{equation}
The first factor comes from the box whose marking is removed. 
Taking the product of the above contributions, we obtain the desired formula. 
\end{proof}

Now we claim that the matrix elements in the super Macdonald basis are obtained by multiplying the matrix elements 
in the fixed point basis computed in \eqref{matrix-elem-1} and \eqref{matrix-elem-2}
with the change of the normalization factor \eqref{scaling-superMac} for the integral form,
which is given by Lemma \ref{lem:c-recursion}. Namely, the matrix elements of the zero mode of $E_1$ is
\begin{align}
&(t-1) \prod_{i=1}^\infty \frac{1- q^{\lambda_k - \lambda_i + \sigma_i} t^{i-k+1}}
{(1- q^{\lambda_k - \lambda_i} t^{i-k})^{\delta_{\sigma_i,0}}}
\times \prod_{j=k+1}^{\infty} \frac{1- q^{\lambda_k -\lambda_j + \sigma_j} t ^{j-k}} 
{1- q^{\lambda_k -\lambda_j + \sigma_j}  t^{j-k+1 }} \CR
& \qquad\times \prod_{i<k, i \in F} (1- t^{k-i} q^{\lambda_i - \lambda_k -1})^{-1}
\prod_{k<j, j \in F} (1- t^{j-k} q^{\lambda_k - \lambda_j+1})^{-1} \CR
& = (-q)^{F(k)}(t-1) t^{k-1} \mathsf{f}^{(k)}(q,t)^{-1} \prod_{i=1}^{k-1} \frac{1- q^{\lambda_i - \lambda_k - \sigma_i} t^{k-i-1}}
{1- q^{\lambda_i - \lambda_k-\sigma_i} t^{k-i}}, \label{base-change-e1}
\end{align}
and the matrix elements of the zero mode of $E_2$ is
\begin{align}
&(-1)\cdot\prod_{i \in F, i \neq k} \frac{1}{1 - q^{\lambda_k - \lambda_i}  t^{i-k}}
\times \prod_{i=1}^{k-1} \frac{1-q^{\lambda_i - \lambda_k}t^{k-i+1}}{1- q^{\lambda_i -\lambda_k } t^{k-i}} \CR
& \qquad \times (1-t)\prod_{i<k, i \in F} (1- t^{k-i} q^{\lambda_i - \lambda_k})
\prod_{k<j, j \in F} (1- t^{j-k} q^{\lambda_k - \lambda_j}) \CR
& = (-1)^{F(k)} (t-1) \mathsf{f}^{(k)}(q,t) \prod_{i=1}^{k-1} \frac{1-q^{\lambda_i - \lambda_k}t^{k-i+1}}{1- q^{\lambda_i -\lambda_k} t^{k-i}},
\label{base-change-e2}
\end{align}
where $F(k):= \displaystyle{\sum_{i=1}^{k-1}} \barsigma_i $ is the number of fermionic rows above the $k$-th row and 
\begin{equation}\label{f-framing}
\mathsf{f}^{(k)}(q,t) := \prod_{i \in F, i <k} q^{\lambda_i - \lambda_k} t^{k-i}.
\end{equation}
The sign factor $(-1)^{F(k)}$  arises from the flip of the sign of the powers of $q$ and $t$ for the fermionic lows. 
It is satisfying that the base change produces such sign factor, which implies the fermionic nature of $E_1$ and $E_2$. 
By the change of the convention $(q,t) \to (q^2, t^2)$, the matrix elements \eqref{base-change-e1} and \eqref{base-change-e2}
agree with those in the super Macdonald basis \eqref{matrix-elem-super1} and \eqref{matrix-elem-super2}
up to the monomial factor including \eqref{f-framing}. 
In summary we have arrived at Proposition \ref{main-prp}. 
The additional  monomial factors come from the change of $1- q^{m} t^{n}=-q^mt^n (1- q^{-m}t^{-n})$ 
under the involution $(q,t) \to (q^{-1}, t^{-1})$. 
Hence, one of the ways to remove them is replacing every factor of the form $1- q^{2m} t^{2n}$ by $2\sinh (q^m t^n)$. 

\vspace{5mm}
\newpage
\begin{ack}
We would like to thank Dmitry Galakhov, Alexei Morozov, Hiraku Nakajima, Go Noshita 
and Shintarou Yanagida for useful discussions. 
Our work is supported in part by Grants-in-Aid for Scientific Research (Kakenhi);
23K03087 (H.K.), 21K03180 (R.O. and J.S.) and 24K06753 (J.S.).
The work of R.O. was partly supported by 
%Osaka Central Advanced Mathematical Institute: MEXT Joint Usage/Research Center on Mathematics and
%Theoretical Physics JPMXP0619217849, and 
the Research Institute for Mathematical Sciences,
an International Joint Usage/Research Center located in Kyoto University.
\end{ack}

\vspace{5mm}
%%%%%%%%%%%%%%%%%%%%%%%%%%%%%%%%%%%%%%%%%%%%%%%%%%%%%%%%%%%%%%%%%%%%%%%%%%%%%%
\appendix
\ytableausetup{aligntableaux=top,mathmode,boxsize=0.8em}

\section{Check of the vector representation}
\label{App:vector}

\subsection{Check of \eqref{commu-1}}
 
 \begin{align*}
 K_1^{+}(z) E_2(w) [u]_{k,1} &= \mathcal{E}_2(k) \delta \left( \frac{w}{uq_1(q_1q_2)^k} \right) K_1^{+}(z) [u]_{k,0} \\
 &=  \mathcal{E}_2(k) \delta \left( \frac{w}{uq_1(q_1q_2)^k} \right) \frac{\phi(q_1^{-k-2} q_2^{-k};z,u)} {\phi(q_1^{-k-1} q_2^{-k-1};z,u)}[u]_{k,0}.
  \end{align*}

  \begin{align*}
  E_2(w)  K_1^{+}(z) [u]_{k,1} &=  \frac{\phi(q_1^{-k-1} q_2^{-k+1};z,u)} {\phi(q_1^{-k} q_2^{-k};z,u)}  E_2(w) [u]_{k,1} \\
  &= \mathcal{E}_2(k) \delta \left( \frac{w}{uq_1(q_1q_2)^k} \right) \frac{\phi(q_1^{-k-1} q_2^{-k+1};z,u)} {\phi(q_1^{-k} q_2^{-k};z,u)} [u]_{k,0}.
  \end{align*}
  
 We have 
  \begin{align*}
 \delta \left( \frac{w}{uq_1(q_1q_2)^k} \right) \varphi^{2 \Rightarrow 1}(z,w) 
 &=   \delta \left( \frac{w}{uq_1(q_1q_2)^k} \right)  \frac{\phi(q_1;z, q_1(q_1q_2)^k u)  \phi(q_1^{-1};z,q_1(q_1q_2)^k u)}
 {\phi(q_2;z,q_1(q_1q_2)^k u)  \phi(q_2^{-1};z,q_1(q_1q_2)^k u)} \\
 &=  \delta \left( \frac{w}{uq_1(q_1q_2)^k} \right)  \frac{\phi(q_1^{-k} q_2^{-k} ;z, u)  \phi(q_1^{-k-2}q_2^{-k} ;z,u)}
 {\phi( q_1^{-k-1} q_2^{-k+1} ;z, u)  \phi(q_1^{-k-1} q_2^{-k-1};z, u)},
\end{align*}
where we use the formula \eqref{s-shift}.

Similarly 
 \begin{align*}
 K_2^{+}(z) E_1(w) [u]_{k,0} %&= \mathcal{E}_2(k) \delta \left( \frac{w}{uq_1(q_1q_2)^k} \right) K_1^{+}(z) [u]_{k,0} \\
 &=  \mathcal{E}_1(k) \delta \left( \frac{w}{u(q_1q_2)^{k+1}} \right) \frac{\phi(q_1^{-k-1} q_2^{-k-2};z,u)} {\phi(q_1^{-k-2} q_2^{-k-1};z,u)}[u]_{k+1,1}.
  \end{align*}

  \begin{align*}
  E_1(w)  K_2^{+}(z) [u]_{k,0} %&=  \frac{\phi(q_1^{-k-1} q_2^{-k+1};z,u)} {\phi(q_1^{-k} q_2^{-k};z,u)}  E_2(w) [u]_{k,1} \\
  &= \mathcal{E}_1(k) \delta \left( \frac{w}{u(q_1q_2)^{k+1}} \right) \frac{\phi(q_1^{-k} q_2^{-k-1};z,u)} {\phi(q_1^{-k-1} q_2^{-k};z,u)} [u]_{k+1,1}.
  \end{align*}
  
  \begin{align*}
 \delta \left( \frac{w}{u(q_1q_2)^{k+1}} \right) \varphi^{1 \Rightarrow 2}(z,w) 
 &=   \delta \left( \frac{w}{u(q_1q_2)^{k+1}} \right)  \frac{\phi(q_2;z, (q_1q_2)^{k+1} u)  \phi(q_2^{-1};z,(q_1q_2)^{k+1} u)}
 {\phi(q_1;z, (q_1q_2)^{k+1} u)  \phi(q_1^{-1};z,(q_1q_2)^{k+1} u)} \\
 &=   \delta \left( \frac{w}{u(q_1q_2)^{k+1}} \right)  \frac{\phi(q_1^{-k-1} q_2^{-k} ;z, u)  \phi(q_1^{-k-1}q_2^{-k-2} ;z,u)}
 {\phi( q_1^{-k} q_2^{-k-1} ;z, u)  \phi(q_1^{-k-2} q_2^{-k-1};z, u)}.
\end{align*}

\subsection{Check of \eqref{commu-3}}
 
 \begin{align*}
 \left( K_1^{+}(z) - K_1^{-}(z) \right) [u]_{k,0}
 &= \left( \left[ \frac{z - q_1^{k+2} q_2^{k} u}{z - q_1^{k+1} q_2^{k+1} u} \right]_{+} -  \left[ \frac{z - q_1^{k+2} q_2^{k} u}{z - q_1^{k+1} q_2^{k+1} u} \right]_{-} \right) [u]_{k,0} \\
 &= \delta \left( \frac{z}{q_1^{k+1} q_2^{k+1}u} \right) (1- q_1 q_2^{-1}) [u]_{k,0}, \\
 \left( K_1^{+}(z) - K_1^{-}(z) \right) [u]_{k,1}
 &= \left( \left[ \frac{z - q_1^{k+1} q_2^{k-1} u}{z - q_1^{k} q_2^{k} u} \right]_{+} -  \left[ \frac{z - q_1^{k+1} q_2^{k-1} u}{z - q_1^{k} q_2^{k} u} \right]_{-} \right) [u]_{k,1} \\
 &= \delta \left( \frac{z}{q_1^{k} q_2^{k}u} \right) (1 - q_1 q_2^{-1}) [u]_{k,1}, \\
 \left( K_2^{+}(z) - K_2^{-}(z) \right) [u]_{k,0}
 &= \left( \left[ \frac{z - q_1^{k} q_2^{k+1} u}{z - q_1^{k+1} q_2^{k} u} \right]_{+} -  \left[ \frac{z - q_1^{k} q_2^{k+1} u}{z - q_1^{k+1} q_2^{k} u} \right]_{-} \right) [u]_{k,0} \\
 &= \delta \left( \frac{z}{q_1^{k+1} q_2^{k}u} \right)  (1 - q_1^{-1} q_2)  [u]_{k,0}, \\
\left( K_2^{+}(z) - K_2^{-}(z) \right) [u]_{k,1}
 &= \left( \left[ \frac{z - q_1^{k} q_2^{k+1} u}{z - q_1^{k+1} q_2^{k} u} \right]_{+} -  \left[ \frac{z - q_1^{k} q_2^{k+1} u}{z - q_1^{k+1} q_2^{k} u} \right]_{-} \right) [u]_{k,1} \\
 &= \delta \left( \frac{z}{q_1^{k+1} q_2^{k}u} \right) (1 - q_1^{-1} q_2)  [u]_{k,1}, \\
 \end{align*}
 where we use
\begin{align*}
\left[ \frac{z - \beta u}{z - \alpha u} \right]_{+} - \left[ \frac{z - \beta u}{z - \alpha u} \right]_{-}
&= \left( 1 - \frac{\beta u}{z} \right)  \sum_{n=0}^\infty \left( \frac{\alpha u} {z}\right)^n
- \left(\frac{\beta}{\alpha}- \frac{z}{\alpha u} \right) \sum_{n=0}^\infty \left( \frac{z}{\alpha u} \right)^n 
\\
&= \left( \sum_{n=0}^\infty \left( \frac{\alpha u} {z} \right)^n  +  \sum_{n=1}^\infty \left(  \frac{z}{\alpha u} \right)^n  \right)
- \frac{\beta}{\alpha} \left( \sum_{n=0}^\infty \left( \frac{z}{\alpha u} \right)^n  +  \sum_{n=1}^\infty \left( \frac{\alpha u} {z} \right)^n \right)
\\
&= \left( 1- \frac{\beta}{\alpha} \right) \delta\left( \frac{z}{\alpha u} \right).
\end{align*}

On the other hand
\begin{align*}
\left( E_1(z) F_1(z) + F_1(w) E_1(z) \right)  [u]_{k,0}
&= \mathcal{E}_1(k)  \mathcal{F}_1(k+1) \delta \left( \frac{z}{u(q_1q_2)^{k+1}} \right) \delta \left(  \frac{w}{u(q_1q_2)^{k+1}} \right)   [u]_{k,0} \\
&=  \mathcal{E}_1(k)  \mathcal{F}_1(k+1) \delta \left( \frac{w}{z} \right) \delta \left(  \frac{z}{u(q_1q_2)^{k+1}} \right)   [u]_{k,0} , \\
\left( E_1(z) F_1(z) + F_1(w) E_1(z) \right)  [u]_{k,1}
&= \mathcal{E}_1(k-1)  \mathcal{F}_1(k) \delta \left( \frac{z}{u(q_1q_2)^{k}} \right) \delta \left(  \frac{w}{u(q_1q_2)^{k}} \right)   [u]_{k,1} \\
&=  \mathcal{E}_1(k-1)  \mathcal{F}_1(k) \delta \left( \frac{w}{z} \right) \delta \left(  \frac{z}{u(q_1q_2)^{k}} \right)   [u]_{k,1} , \\
\left( E_2(z) F_2(z) + F_2(w) E_2(z) \right)  [u]_{k,0}
&= \mathcal{E}_2(k-1)  \mathcal{F}_2(k) \delta \left( \frac{z}{uq_1(q_1q_2)^{k}} \right) \delta \left(  \frac{w}{uq_1(q_1q_2)^{k}} \right)   [u]_{k,0} \\
&=  \mathcal{E}_2(k-1)  \mathcal{F}_2(k) \delta \left( \frac{w}{z} \right) \delta \left(  \frac{z}{uq_1(q_1q_2)^{k}} \right)   [u]_{k,0} , \\
\left( E_2(z) F_2(z) + F_2(w) E_2(z) \right)  [u]_{k,1}
&= \mathcal{E}_2(k)  \mathcal{F}_2(k) \delta \left( \frac{z}{uq_1(q_1q_2)^{k}} \right) \delta \left(  \frac{w}{uq_1(q_1q_2)^{k}} \right)   [u]_{k,1} \\
&=  \mathcal{E}_2(k)  \mathcal{F}_2(k) \delta \left( \frac{w}{z} \right) \delta \left(  \frac{z}{uq_1(q_1q_2)^{k}} \right)   [u]_{k,1}.
\end{align*}
Hence we can assume that the normalization constants are independent of $k$ and they satisfy
\beq
\mathcal{E}_1 \mathcal{F}_1 = 1- q_1 q_2^{-1},
\qquad
\mathcal{E}_2 \mathcal{F}_2 = 1- q_1^{-1} q_2.
\eeq

 \subsection{Check of \eqref{commu-4}}

$E_1(z)E_1(w) = E_2(z) E_2(w) =0$ is trivial. Let us look at $E_1(z)E_2(w)$ and $E_2(z)E_1(w)$. 
We have
$$
E_1(z)E_2(w) [u]_{k,0} = E_2(z)E_1(w) [u]_{k,1} =0.
$$
On the other hand
$$
E_2(w)E_1(z) [u]_{k,0}  = \mathcal{E}_1 \mathcal{E}_2 \cdot \delta\left( \frac{z}{u (q_1q_2)^{k+1}} \right)  \delta\left( \frac{w}{uq_1(q_1q_2)^{k+1} } \right) [u]_{k+1,0}.
$$
Since the support on the righthand side is included in  $w=q_1z$, the multiplication of $\varphi^{2 \Rightarrow 1}(z,w)$ makes it vanish. Similarly we have
$$
E_1(w) E_2(z) [u]_{k,1} = \mathcal{E}_1 \mathcal{E}_2 \cdot  \delta\left( \frac{z}{uq_1 (q_1q_2)^k }\right)  \delta\left( \frac{w}{u(q_1q_2)^{k+1}}\right) [u]_{k+1,1}.
$$
The support on the righthand side is included in  $w=q_2z$ and the multiplication of $\varphi^{1 \Rightarrow 2}(z,w)$ makes it vanish. 
Thus, the position of zeros of the structure function fixes the shift of the support of the action of $E_1(z)$ and $E_2(z)$.

%%%%%%%%%%%%%%%%%%%%%%%%%%%%%%%%%%%%%%%%%%%%%%%%%%%%%%%%%%%%%%%%%%%%%%%%%%%%%%%%%%%%%%%%%%%%%%%%%%%%%%%%%%%%%%%%%%%%%%%%%%%%%%%%%%%%%%%%%%%%%%%%%%%%%%%%%

\section{Lower super Macdonald polynomials and Pieri rule}
\label{lower-super-Mac}

To write down the super Macdonald polynomials $\mathcal{M}_{\Lambda}(x, \theta;q,t)$ explicitly,
it is convenient to introduce the bosonic and the fermionic power sum 
polynomials defined by
\begin{equation}
p_k := \sum_{i=1}^N x_i^k, \qquad \pi_k := \sum_{i=1}^N \theta_i x_i^{k-1}.
\end{equation}
In \cite{Galakhov:2024cry} the super Macdonald polynomials of lower levels are 
obtained as follows;\footnote{We have changed the notation $\theta_i$ in \cite{Galakhov:2024cry}
to $\pi_i$, since we use $\theta_i$ as the super partner of $x_i$. We have also adjusted some of 
the coefficients so that they are consistent with the Pieri formula in this paper.}
\begin{align*}
&\mathcal{M}_{(\frac{1}{2})} = \pi_1, \qquad \mathcal{M}_{(1)} =p_1, \\
& \mathcal{M}_{(\frac{3}{2})} = \frac{q^2(1-t^2)}{1-q^2t^2}p_1\pi_1 + \frac{1-q^2}{1-q^2t^2}\pi_2,
\qquad \mathcal{M}_{(1, \frac{1}{2})} = p_1 \pi_1 - \pi_2, \\
& \mathcal{M}_{(2)} = \frac{1}{2}\left( \frac{(1+q^2)(1-t^2)}{1-q^2t^2} p_1^2 + \frac{(1+t^2)(1-q^2)}{1-q^2t^2}p_2 \right), \\
& \mathcal{M}_{(1,1)} = \frac{1}{2}(p_1^2 - p_2), 
\qquad \mathcal{M}_{(\frac{3}{2},\frac{1}{2})} = \pi_2 \pi_1, \\
& \mathcal{M}_{(\frac{5}{2})} = \frac{q^4(1+q^2)(1-t^2)^2}{2(1-q^2t^2)(1-q^4t^2)}p_1^2 \pi_1
+ \frac{q^4(1-q^2)(1-t^4)}{2(1-q^2t^2)(1-q^4t^2)} p_2 \pi_1 \\
& \qquad \qquad + \frac{q^2(1-q^4)(1-t^2)}{(1-q^2t^2)(1-q^4t^2)} p_1 \pi_2 
+ \frac{(1-q^2)^2(1+q^2)}{(1-q^2t^2)(1-q^4t^2)} \pi_3, \\
& \mathcal{M}_{(2, \frac{1}{2})} = \frac{(1+q^2)(1-t^2)}{2(1-q^2t^2)} p_1^2 \pi_1
+ \frac{(1-q^2)(1+t^2)}{2(1-q^2t^2)} p_2 \pi_1 - \frac{q^2(1-t^2)}{1-q^2t^2} p_1 \pi_2 
- \frac{1-q^2}{1-q^2t^2} \pi_3 , \\
& \mathcal{M}_{(\frac{3}{2},1)} = \frac{q^2(1-t^4)}{2(1-q^2t^4)}(p_1^2 \pi_1 - p_2 \pi_1) 
+ \frac{1-q^2}{1- q^2 t^4}( p_1 \pi_2 - \pi_3), \\
& \mathcal{M}_{(1,1, \frac{1}{2})} = \frac{p_1^2 \pi_1}{2} - \frac{p_2 \pi_1}{2} - p_1\pi_2 + \pi_3.
\end{align*}
There are two level $3$ super Macdonald polynomials other than the (bosonic) Macdonald polynomials
$\mathcal{M}_{(3)}, \mathcal{M}_{(2,1)}$ and $\mathcal{M}_{(1,1,1)}$;
\begin{equation}\label{level3}
\mathcal{M}_{(\frac{5}{2},\frac{1}{2})} = \frac{1-q^2}{1-q^2t^2} \pi_3 \pi_1 + \frac{q^2(1-t^2)}{1-q^2t^2} p_1 \pi_2 \pi_1, \qquad
\mathcal{M}_{(\frac{3}{2} ,1, \frac{1}{2})} = p_1 \pi_2 \pi_1 - \pi_3 \pi_ 1.
\end{equation}
\newpage
We note that in contrast to the Macdonald polynomial the super Macdonald polynomials do depend on the parameter even if we set $q=t$.
Another different feature from the Macdonald polynomials is that
the super Macdonald polynomials are not invariant under the involution $(q,t) \to (q^{-1}, t^{-1})$.
\vspace{-5mm}
\begin{figure}[htb]
\begin{center}
\begin{picture}(300, 200)
 \setlength{\unitlength}{0.6mm}
\put(0,45){$\varnothing$}
\put(10,47){\vector(1,0){16}}
\put(15,50){$E_1$}
\put(30,45){\ytableaushort{{\hbox{$\times$}}}}
\put(40,47){\vector(1,0){16}}
\put(45,50){$E_2$}
\put(60,45){\ytableaushort{{}}}
\put(70,47){\vector(1,1){16}}
\put(72,62){$E_1$}
\put(70,47){\vector(1,-1){16}}
\put(72,30){$E_1$}
\put(90,60){\ytableaushort{{}{\hbox{$\times$}}}}
\put(90,30){\ytableaushort{{},{\hbox{$\times$}}}}
\put(105,65){\vector(1,1){12}}
\put(106,77){$E_2$}
\put(105,65){\vector(1,-1){12}}
\put(106,50){$E_1$}
\put(100,30){\vector(1,1){12}}
\put(100,40){$E_1$}
\put(100,30){\vector(1,-1){12}}
\put(100,17){$E_2$}
\put(120,75){\ytableaushort{{}{}}}
\put(120,45){\ytableaushort{{}{\hbox{$\times$}},{\hbox{$\times$}}}}
\put(120,15){\ytableaushort{{},{}}}
\put(140,77){\vector(1,1){12}}
\put(141,90){$E_1$}
\put(140,77){\vector(1,-1){12}}
\put(141,65){$E_1$}
\put(138,45){\vector(1,1){12}}
\put(138,55){$E_2$}
\put(138,45){\vector(1,-1){12}}
\put(138,30){$E_2$}
\put(135,10){\vector(1,1){12}}
\put(136,20){$E_1$}
\put(135,10){\vector(1,-1){12}}
\put(136,-5){$E_1$}
\put(160,95){\ytableaushort{{}{}{\hbox{$\times$}}}}
\put(160,60){\ytableaushort{{}{},{\hbox{$\times$}}}}
\put(160,30){\ytableaushort{{}{\hbox{$\times$}},{}}}
\put(160,-5){\ytableaushort{{},{},{\hbox{$\times$}}}}
\end{picture}
\end{center}
\caption{$E_1$ adds \ytableaushort{{\hbox{$\times$}}}, white $E_2$ deletes $\times$ in the box.
There is at most a single \ytableaushort{{\hbox{$\times$}}}~~in each row and each column.}
\end{figure}
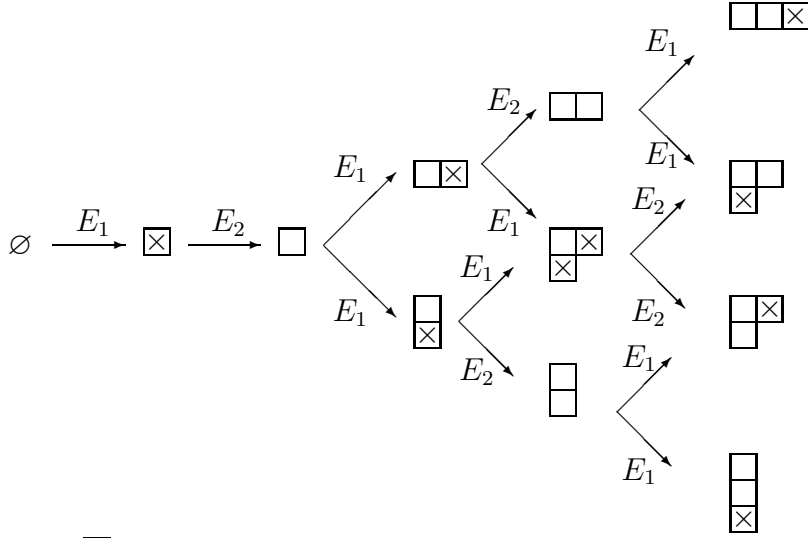

Since $\mathcal{M}_{(\frac{1}{2})}= \pi_1$, let us assume that the zero mode of $E_1(z)$ acts 
on the supersymmetric polynomials as the multiplication with $\pi_1$. The Pieri rule from super DIM algebra tells
\begin{align}
\pi_1 \cdot \mathcal{M}_{(1)} &= \mathcal{M}_{(\frac{3}{2})} + \frac{1-q^2}{1 - q^2t^2} \mathcal{M}_{(1,\frac{1}{2})}, \\
\pi_1 \cdot \mathcal{M}_{(\frac{3}{2})} &= (-1) \cdot \frac{1-q^2}{1 - q^2t^2} \mathcal{M}_{(\frac{3}{2},\frac{1}{2})}, \\
\pi_1 \cdot \mathcal{M}_{(1,\frac{1}{2})} &= \mathcal{M}_{(\frac{3}{2},\frac{1}{2})}. 
\end{align}
It is easy to confirm that the above list of the super Macdonald polynomials satisfies the Pieri rule. 
%Let us temporally neglect the monomial factor $t^{\pm 1}$. 
%We have
%\begin{align}
%\mathcal{M}_{(\frac{3}{2})} + \frac{1-q^2}{1 - q^2t^2} \mathcal{M}_{(1,\frac{1}{2})} 
%&= p_1\pi_1,
%\\
%\pi_1 \cdot \mathcal{M}_{(\frac{3}{2})}
%&= \frac{1-q^2}{1 - q^2t^2} \pi_1 \pi_2,
%\\
%\pi_1 \cdot \mathcal{M}_{(1,\frac{1}{2})}
%&= \pi_1 \pi_2.
%\end{align}
The Pieri rules of the next level are
\begin{align}
\pi_1  \mathcal{M}_{(2)} &=  \mathcal{M}_{(\frac{5}{2})} + \frac{1-q^4}{1-t^2q^4}  \mathcal{M}_{(2,\frac{1}{2})}, \\
\pi_1  \mathcal{M}_{(1,1)} &=  \mathcal{M}_{(\frac{3}{2},1)} + \frac{1-q^2}{1-t^4q^2} 
\mathcal{M}_{(1,1,\frac{1}{2})}. 
\end{align}
We have
\begin{align}
\mathcal{M}_{(\frac{5}{2})} + \frac{1-q^4}{1-t^2q^4}  \mathcal{M}_{(2,\frac{1}{2})} 
&= \frac{q^4(1+q^2)(1-t^2)^2}{2(1-q^2t^2)(1-q^4t^2)}p_1^2 \pi_1 + \frac{q^4(1-q^2)(1-t^4)}{2(1-q^2t^2)(1-q^4t^2)} p_2 \pi_1 \CR
& \qquad + \frac{1-q^4}{1-t^2q^4} \left( \frac{(1+q^2)(1-t^2)}{2(1-q^2t^2)} p_1^2 \pi_1
+ \frac{(1-q^2)(1+t^2)}{2(1-q^2t^2)} p_2 \pi_1 \right) \CR
&= \frac{(1+q^2)(1-t^2)}{2(1-q^2t^2)(1- t^2q^4)} \left[ q^4(1-t^2) + (1-q^4) \right] p_1^2 \pi_1 \CR
& \qquad + \frac{(1-q^2)(1+t^2)}{2(1-q^2t^2)(1- t^2q^4)} \left[ q^4(1-t^2) + (1-q^4)\right] p_2 \pi_1 \CR
&= \pi_1 \cdot \mathcal{M}_{(2)}.
\end{align}
and 
\begin{align}
\mathcal{M}_{(\frac{3}{2},1)} + \frac{1-q^2}{1-t^4q^2} \mathcal{M}_{(1,1,\frac{1}{2})} 
&= \frac{q^2(1-t^4)}{2(1-q^2t^4)}(p_1^2 \pi_1 - p_2 \pi_1) + \frac{1-q^2}{1- q^2 t^4}( p_1 \pi_2 - \pi_3) \CR
& \qquad + \frac{1-q^2}{1-t^4q^2}\left( \frac{p_1^2 \pi_1}{2} - \frac{p_2 \pi_1}{2} - p_1\pi_2 + \pi_3 \right) \CR
&= \frac{q^2(1-t^4)+(1-q^2)}{2(1-t^4q^2)}(p_1^2 \pi_1 - p_2 \pi_1) = \pi_1 \cdot \mathcal{M}_{(1,1)}.
\end{align}
Hence, the above list of the super Macdonald polynomials at level $\frac{5}{2}$ is consistent with the Pieri rule of $E_1$.

The Pieri rule at level $\frac{5}{2}$ gives;
\begin{align}
\pi_1 \cdot \mathcal{M}_{(\frac{5}{2})} &= (-1) \cdot \frac{1-q^4}{1-t^2q^4} \mathcal{M}_{(\frac{5}{2},\frac{1}{2})} \CR
&= \frac{q^2(1-q^4)(1-t^2)}{(1-t^2q^4)(1-q^2t^2)} p_1 \pi_1\pi_2 + \frac{(1+q^2)(1-q^2)^2}{(1-t^2q^4)(1-q^2t^2)} \pi_1\pi_3, \\
\pi_1 \cdot \mathcal{M}_{(2,\frac{1}{2})} &= \mathcal{M}_{(\frac{5}{2},\frac{1}{2})}
= - \frac{q^2(1-t^2)}{1-q^2t^2} p_1 \pi_1\pi_2 - \frac{1-q^2}{1-q^2t^2} \pi_1\pi_3, \\
\pi_1 \cdot \mathcal{M}_{(\frac{3}{2},1)} &= (-1) \cdot \frac{1-q^2}{1-t^4q^2} \mathcal{M}_{(\frac{3}{2},1,\frac{1}{2})}
= \frac{1-q^2}{1-t^4q^2} \pi_1 (p_1\pi_2 - \pi_3), \\
\pi_1 \cdot \mathcal{M}_{(1,1,\frac{1}{2})} &= \mathcal{M}_{(\frac{3}{2},1,\frac{1}{2})}
= - p_1\pi_1 \pi_2 + \pi_1 \pi_3.
\end{align}
Note that $\mathcal{M}_{(\frac{5}{2},\frac{1}{2})}$ and $\mathcal{M}_{(\frac{3}{2},1,\frac{1}{2})}$ are 
consistently determined as \eqref{level3}.

%%%%%%%%%%%%%%%%%%%%%%%%%%%%%%%%%%%%%%%%%%%%%%%%%%%%%%%%%%%%%%%%%%%%%%%%%

\section{Dimensional reduction of the quiver with potential}
\label{App:quiver-reduction}

For any toric Calabi-Yau 3-fold we can associate a quiver with potential.  
The fundamental examples are $\mathbb{C}^3$ and 
the resolved conifold $\mathcal{O}(-1) \oplus \mathcal{O}(-1)  \longrightarrow \mathbb{P}^1$.
The corresponding quivers with potential are given as follows  (see for example \cite{Rapcak:2020ueh});

\begin{center}
\vspace{-10mm}
\includegraphics{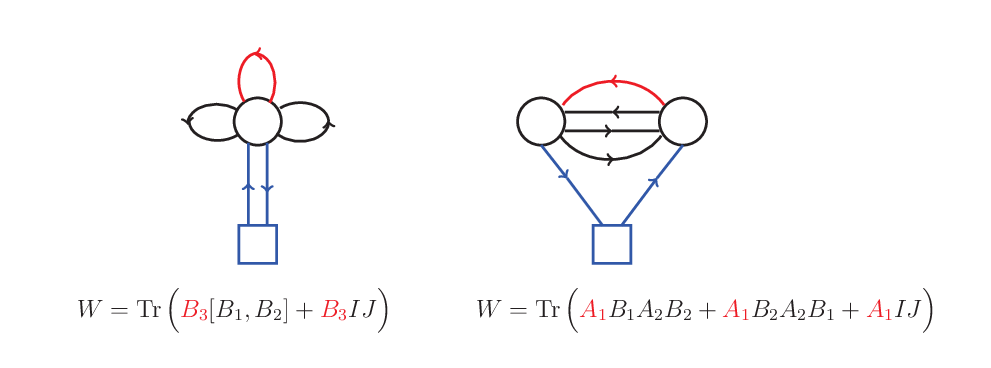}
\end{center}

%%%%%%%%%%%%%%%%%%%%%%%%%%%%%%%%%%%%%%%%%%%%%%%%%%%%%%%%%%%%%%%%%%%%%%%%%%%%%%%%%%%%%%%%%%%%%%%%%%%%%%%
\noindent
Another basic example is $\mathbb{C}^3$ with $\mathbb{Z}_2$ orbifold action describing a surface defect;

\begin{center}
\includegraphics{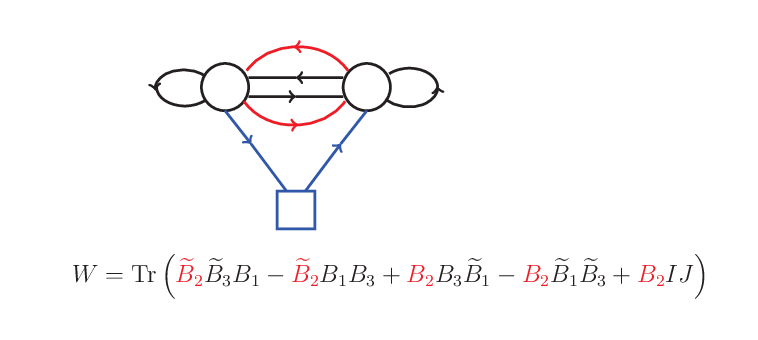}
\vspace{-10mm}
%\begin{tikzpicture}[very thick, scale=0.8]
%
%\draw(1.5,3)circle[radius=0.5];
%\draw(4.5,3)circle[radius=0.5];
%\draw[blue](2.6,0)rectangle(3.4,0.8);
%%
%\draw[blue](2.8,0.8)--(2.05,1.8);
%\draw[->, blue](3.2,0.8)--(3.95,1.8);
%\draw[<-,blue](2.05,1.8)--(1.5,2.5);
%\draw[blue](3.95,1.8)--(4.5,2.5);
%%
%\draw[->](2,2.8)--(3,2.8);
%\draw(3,2.8)--(4,2.8);
%\draw(2,3.2)--(3,3.2);
%\draw[<-](3,3.2)--(4,3.2);
%%
%\draw[->,red](4.1,3.35)arc[start angle = 30, end angle = 95, x radius = 1.2, y radius = 1];
%\draw[red](3,3.85)arc[start angle = 90, end angle = 150, x radius = 1.2, y radius = 1];
%\draw[->,red](1.9,2.7)arc[start angle = 210, end angle = 275, x radius = 1.2, y radius = 1];
%\draw[red](3,2.2)arc[start angle = 270, end angle = 330, x radius = 1.2, y radius = 1];
%
%
%\draw[->](1.05,3.25)arc[start angle=45, end angle =185, x radius=0.6, y radius=0.4];
%\draw(0.05,3)arc[start angle=180, end angle =315, x radius=0.6, y radius=0.4];
%\draw[->](4.9,2.75)arc[start angle=215, end angle =365, x radius=0.6, y radius=0.4];
%\draw(6,3)arc[start angle=0, end angle =135, x radius=0.6, y radius=0.4];
%
%\draw(5,-1) node {$W = \operatorname{Tr} \Big( {\color{red} \widetilde{B}_2} \widetilde{B}_3 B_1 
%- {\color{red}\widetilde{B}_2} B_1 B_3  + {\color{red} B_2} B_3 \widetilde{B}_1 
%- {\color{red} B_2} \widetilde{B}_1 \widetilde{B}_3 + {\color{red} B_2} IJ \Big)$};
%%
%\end{tikzpicture}
\end{center}

In the above examples, deleting the framing sector which is indicated in blue, we obtain the quiver data for
the structure functions of the quiver quantum toroidal algebras of type $\mathfrak{gl}_1, \mathfrak{gl}_{1|1}$.
and $\mathfrak{gl}_2$, respectively.
On the other hand the dimensional reduction is defined by choosing a subset of arrows
which is homogeneous degree one in the potential with some assignment of degree for each arrow \cite{Rapcak:2020ueh}
(See also \cite{Butson:2023eid} for the variety of examples).
For example we can choose red arrow(s) in each quiver. In the case of $\mathbb{C}^3$, we recover the standard ADHM quiver for instanton counting on $\mathbb{C}^2$.
From the resolved conifold we obtain the quiver for instanton counting on the blow-up $\widehat{\mathbb{P}}^2$, which is given in subsection \ref{subsec:quiver}.
The last example gives the chain saw quiver of type $A_1$.
In the terminology of the BPS state counting of $D$-brane system the first two quivers correspond to $D6$-$D0$ system ($\mathbb{C}^3$ case) 
and $D6$-$D2$-$D0$ (the conifold case). The dimensional reduction replaces $D6$-brane by $D4$-brane. Hence, we obtain 
$D4$-$D0$ system and $D4$-$D2$-$D0$, respectively. 

%%%%%%%%%%%%%%%%%%%%%%%%%%%%%%%%%%%%%%%%%%%%%%%%%%%%%%%%%%%%%%%%%%%%%%%%%%%%%%%%%%%%%%

\section{Matching of the equivariant characters}

The fixed points of the torus action on the moduli space of instantons with rank $r$ are indexed by
$r$-tuples of Young diagrams. Each pair $(\lambda, \mu)$ of the Young diagrams in the $r$-tuple 
gives the following contribution 
to the the equivariant character of the tangent space at a fixed point is  \cite{Nakajima:2003pg} (Th.2.11);
%\begin{equation}
%T_{(E, \Phi)} M(r,n) = \sum_{\alpha, \beta=1}^r N_{\alpha, \beta}(t_1, t_2),
%\end{equation}
%where
\begin{equation}\label{ch1}
\chi_{(\lambda, \mu)}(u \vert t_1, t_2)  = u \cdot \left[
\sum_{\mathsf{s} \in \lambda} t_1^{-\ell_{\mu}(\mathsf{s})} t_2^{a_{\lambda}(\mathsf{s})+1} 
+ \sum_{\mathsf{t} \in \mu} t_1^{\ell_{\lambda}(\mathsf{t}) +1} t_2^{-a_{\mu}(\mathsf{t})}  \right],
\end{equation}
where $u$ is the spectral parameter associated with the pair $(\lambda, \mu)$.
The arm length and the leg length at $\mathsf{s}=(i,j)$ are defined by $a_\lambda(i,j) = \lambda_i -j$ and $\ell_\lambda(i,j)= \lambda^\vee_j -i$.
This is the most basic equivariant character which gives the Nekrasov factor $\mathsf{N}_{(\lambda, \mu)}(u \vert t_1, t_2)$.
In this appendix  we discuss several variants of \eqref{ch1}. 
It is helpful to remember that in the convention of \cite{MacD} the parameters $(q,t)$ of the Macdonald polynomial are given by $(q,t) = (t_2^{-1}, t_1)$.

\subsection{$U(1)$ instantons on the blow-up}
\label{App:matching}
The fixed points of the torus action on the moduli space of $U(1)$ instantons on the blow-up $\widehat{\mathbb{P}}^{2}$ 
are labeled by two ways;
\begin{enumerate}
\item
Pairs of partitions $(\lambda, \mu)$,
\item
Super partitions $(Y,S)$.
\end{enumerate}
In the first case the equivariant character of the tangent space at the fixed points is computed in  \cite{Nakajima:2003pg};
\begin{align}\label{eqch-1}
\widehat{\chi}_{(\lambda, \mu)}(t_1, t_2) &= \sum_{\mathsf{s} \in \lambda} \left( t_1^{-\ell_\lambda(\mathsf{s})} (t_2/t_1)^{a_\lambda(\mathsf{s}) +1} 
+ t_1^{\ell_\lambda(\mathsf{s}) +1} (t_2/t_1) ^{-a_\lambda(\mathsf{s})} \right) \CR
& \qquad + \sum_{\mathsf{t} \in \mu} \left( (t_1/t_2)^{-\ell_\mu(\mathsf{t})} t_2^{a_\mu(\mathsf{t}) +1} 
+ (t_1/t_2)^{\ell_\mu(\mathsf{t}) +1} t_2^{-a_\mu(\mathsf{t})} \right).
\end{align}
In the rank one case we may set $u=1$ and $\lambda=\mu$ in \eqref{ch1}. The equivariant character is reduced to
\begin{equation}
\chi_{\lambda}^{U(1)}(t_1, t_2)  = \chi_{(\lambda,\lambda)}(1 \vert t_1, t_2) 
= \sum_{\mathsf{s} \in \lambda}  \left( t_1^{-\ell_{\lambda}(\mathsf{s})} t_2^{a_{\lambda}(\mathsf{s})+1} 
+ t_1^{\ell_{\lambda}(\mathsf{s}) +1} t_2^{-a_{\lambda}(\mathsf{s})} \right).
\end{equation}
Then we have
\begin{equation}\label{twisted}
\widehat{\chi}_{(\lambda, \mu)}(t_1, t_2) = \chi_{\lambda}^{U(1)}(t_1, t_1^{-1}t_2)  + \chi_{\mu}^{U(1)}(t_2^{-1}t_1, t_2).
\end{equation}
The pair $(\lambda, \mu)$ of Young diagrams corresponds to two fixed points of the torus action on the exceptional curve $C \simeq \mathbb{P}^1$.
The twist of the equivariant parameters $(t_1, t_2)$ in \eqref{twisted} comes from the induced torus action on the tangent space at
the fixed points $[1,0]$ and $[0,1]$ on $\mathbb{P}^1 \simeq \mathbb{C}^2 / \sim$.

On the other hand in the case of super partitions we have \cite{Nakajima:2008ss}
\begin{equation}\label{eqch-2}
\widehat{\chi}_{(Y,S)}(t_1, t_2) = \sum_{\mathsf{s} \in \mathcal{B}(Y)} \left( t_1^{-\ell_Y(\mathsf{s})} t_2^{a_{Y\setminus S}(\mathsf{s}) +1}
+  t_1^{\ell_{Y \setminus S}(\mathsf{s}) +1} t_2^{- a_Y(\mathsf{s})} 
\right),
\end{equation}
where $\mathcal{B}(Y)$ denotes the set of relevant boxes in the super partition $(Y,S)$. 
This is the reduction of $\chi_{(\Lambda_\alpha, \Lambda_\beta)}(q, t)$ defined by \eqref{NY-1} 
to the rank one case. We have $\widehat{\chi}_{(Y,S)}(t_1, t_2) = \chi_{(\Lambda, \Lambda)}(t_2^{-1}, t_1)$.
Let $n = |\lambda| + |\mu|$ and $m=|S|$. We have one to one correspondence between
a super partition and the data $(\lambda, \mu, m)$.
Let $Y^{B}$ and $Y^{F}$ be two Young diagrams obtained by decomposing $(Y,S)$ to bosonic rows and fermionic rows. 
As we explained in subsection \ref{sec:super=pair},
$\lambda = Y^{B}$ and $\mu=Y^{F} - \delta_m$, where $\delta_m =(m, m-1, \ldots, 2, 1)$ is the staircase 
partition corresponding to the fermion vacuum with the fermion number $m$. 
Since the number of irrelevant boxes of $(Y,S)$ is $\frac{1}{2}m(m+1) = |\delta_m|$. We have $|\mathcal{B}(Y)| = n$. 
Hence the numbers of terms in the equivariant characters \eqref{eqch-1} and \eqref{eqch-2} are the same.
Recall that we call the region with $m \geq n$ stable sector.

\begin{prp}\label{super=pair}
In the stable sector we have $\widehat{\chi}_{(\lambda, \mu)}(t_1, t_2) = \widehat{\chi}_{(Y,S)}(t_1, t_2)$,
the equivariant characters agree under the correspondence described above.
\end{prp}

To prove the proposition, note that if $\mathsf{s}=(i,j) \in \mathcal{B}(Y)$ is in the fermionic row, 
the last box in the $i$-th row is marked and the last box in the $j$-th column is unmarked. 
In the stable sector, by Lemma \ref{transpose}
the relevant boxes in the bosonic rows and the fermionic rows are exchanged by the transpose of the super partition. 
%Namely $\lambda$ and $\mu$ are exchanged under the transpose of the super partition. 
Hence, we have
\begin{lem}\label{BF-exchange}
In the stable sector, if $\mathsf{s}=(i,j) \in \mathcal{B}(Y)$ is in the bosonic row, namely if the last box in the $i$-th row is unmarked
then the last box in the $j$-th column should be marked. 
\end{lem}
%Clearly this lemma is false if $m=0$. Thus for this lemma to be valid, there should be enough number of fermionic rows.
%This is the reason why we have to assume $m \geq N$.
By Lemma \ref{BF-exchange}, we prove Proposition \ref{super=pair};
\begin{proof}
When $\mathsf{s}\in \mathcal{B}(Y)$ is in the bosonic row, we have the corresponding box $\tilde{\mathsf{s}} \in \lambda$
and $a_Y(\mathsf{s}) = a_{Y \setminus S}(\mathsf{s}) = a_\lambda(\tilde{\mathsf{s}})$. On the other hand,
since there are $a_Y(\mathsf{s})$ relevant boxes on the right to $\mathsf{s}$, 
by the lemma there are $a_Y(\mathsf{s})+1$ fermionic rows below the box $\mathsf{s}$.
Hence we have $\ell_{\lambda}(\tilde{\mathsf{s}})= \ell_{Y}(\mathsf{s}) - a_Y(\mathsf{s}) -1 = \ell_{Y \setminus S}(\mathsf{s}) - a_Y(\mathsf{s})$. 
Substituting these relations to \eqref{eqch-2}, we obtain the first line of \eqref{eqch-1}. 

Similarly, when $\mathsf{s}\in \mathcal{B}(Y)$ is in the fermionic row, we have the corresponding box $\tilde{\mathsf{s}} \in \mu$. 
Let us look at the column of $Y$ to which $\mathsf{s}$ belongs. Since the last box in the column is unmarked, all the boxes below $s$ is relevant
and should be in the fermionic row, otherwise it contradicts to the lemma. 
Hence, we have $\ell_Y(\mathsf{s}) = \ell_{Y \setminus S}(\mathsf{s}) = \ell_\mu(\tilde{\mathsf{s}})$. 
On the other hand, since there are $\ell_Y(\mathsf{s})$ fermionic rows between the row $s$ belongs to and the row
the last box of the column of $\mathsf{s}$ belongs to, there are $\ell(\mathsf{s})+1$ irrelevant boxes 
to the right of $\mathsf{s}$. Since they are deleted in the Young diagram $\mu$, 
$a_\mu(\tilde{\mathsf{s}}) = a_Y(\mathsf{s}) - \ell_Y(\mathsf{s}) -1 = a_{Y \setminus S}(\mathsf{s}) - \ell_Y(\mathsf{s})$. 
Substituting these relations to \eqref{eqch-2}, we obtain the second line of \eqref{eqch-1}. 
\end{proof}

%
%The relations in the case $s\in \mathcal{B}(Y)$ is in the fermionic row are obtained 
%from those in the case $s\in \mathcal{B}(Y)$ is in the fermionic row by exchanging the arm length and the leg length,
%which is consistent with the expectation that the relevant boxes in the bosonic rows and the fermionic rows 
%are exchanged by the transpose of the super partition. 

\subsection{Tangent space of the correspondence}
\label{App:correspondence}

Recall that the moduli space of the rank one instantons on $X$ is given by the Hilbert scheme $X^{[n]}$ of
$n$ points on $X$, where $n$ is identified with the instanton number  \cite{Nakajima-AMS}. 
When $X=\mathbb{C}^2$, the fixed points of the induced torus
action on $X^{[n]}$ are parametrized by the Young diagrams $\lambda$ with $|\lambda|=n$.
Let $I_\lambda$ the monomial ideal corresponding to a fixed point $\lambda$. 
In general the correspondence is defined by the set of triplets;
\begin{equation}
P[i] := \{ (I_1, I_2, x) \mid I_1 \subset I_2,~\operatorname{Supp}(I_2/I_1) = x \},
\end{equation}
which is a subvariety of $X^{[n]} \times X^{[n-i]} \times X$.
There are natural projections $p : P[i] \longrightarrow X^{[n]}$ and $q : P[i] \longrightarrow X^{[n-i]}$
and they define the push-forward $p_{*}$ and the pull-back $q^{*}$ on the equivariant $K$ groups
of the Hilbert scheme. 
The case $i=\pm 1$ is of our interest. Then the fixed points of the induced torus action on $P[1]$ are labeled by 
pairs $(\lambda, \mu)$ of Young diagrams with $\mu \subset \lambda$ and the skew diagram $\lambda \setminus \mu$ 
consists of a single box $\blacksquare$. Combined with the inclusion map of the torus fixed points, the composition $p_{*} \circ q^{*}$
defines the action of the generator $e_0$ of the DIM algebra \cite{Nakajima-AMS},\cite{FT}. 
Note that the composition of the reverse direction $q_{*} \circ p^{*}$ defines the action of $f_0$. 

In \cite{Nakajima-lecture} (Proposition 3.26), the equivariant character of the tangent space 
at a fixed point $(I_\lambda, I_\mu)$ is evaluated as follows;\footnote{Note that $I_\lambda \subset I_\mu$.}
\begin{equation}\label{ch2}
\operatorname{ch}T_{(I_\lambda, I_\mu)} P[1] = t_1 + t_2 + \sum_{\mathsf{s} \in \mu} \left( t_1^{-\ell_\lambda(\mathsf{s})} t_2^{a_\mu(\mathsf{s}) +1} 
+ t_1^{\ell_\mu(\mathsf{s})+1} t_2^{-a_\lambda(\mathsf{s})} \right),
\end{equation}
where the first two terms $t_1 + t_2$ come from $X=\mathbb{C}^2$. 
One may notice a similarity of \eqref{ch1} and \eqref{ch2}.
we have\footnote{Note that we have exchanged $\lambda$ and $\mu$ in  \eqref{ch1}.}
\begin{align}
\chi_{(\mu, \lambda)}(u \vert t_1, t_2) &= u \cdot \left[
\sum_{\mathsf{s} \in \mu} \left( t_1^{-\ell_{\lambda}(\mathsf{s})} t_2^{a_{\mu}(\mathsf{s})+1} 
+ t_1^{\ell_{\mu}(\mathsf{s}) +1} t_2^{-a_{\lambda}(\mathsf{s})}\right) + t_1^{\ell_\mu(\blacksquare)+1}t_2^{-a_\lambda(\blacksquare)} \right]\CR
&=
u \cdot \left[
\sum_{\mathsf{s} \in \mu} \left( t_1^{-\ell_{\lambda}(\mathsf{s})} t_2^{a_{\mu}(\mathsf{s})+1} 
+ t_1^{\ell_{\mu}(\mathsf{s}) +1} t_2^{-a_{\lambda}(\mathsf{s})}\right) + 1 \right].
\end{align}
Note that the corresponding Nekrasov factor has a  simple zero at $u=1$. 
Hence the equivariant character $\operatorname{ch}T_{(I_\lambda, I_\mu)} P[1]$ is obtained from $\chi_{(\mu,\lambda)}$
by the replacement $1 \to t_1 + t_2$. 
The important  point here is that the replacement is universal in the sense that it is independent of $\mu$ and 
the position of $\blacksquare$.

%%%%%%%%%%%%%%%%%%%%%%%

\section{Nekrasov factor and the matrix elements of DIM generators }
\label{DIMcase}

In \cite{FFJMM} the vertical Fock representation of the quantum toroidal algebra is
constructed algebraically by taking semi-infinite tensor product.
Since the basis $|\lambda \rangle$ of the representation is naturally identified with the Macdonald 
polynomials, let us call it Macdonald basis.
Contrary, let us call the basis $[\lambda]$ of the representation by geometric construction fixed point basis.
As remarked in \cite{FFJMM} and \cite{FT} two bases are relation by the change of the normalization.

In the vertical Fock representation the action of $e(z)$ and $f(z)$ on the Macdonald basis are given by
\begin{align}\label{FFJMM-e}
& (1-q^{-1}) e(z) \vert \lambda \rangle \CR
&= \sum_{k=1}^{\ell(\lambda) +1} \prod_{j=1}^{k-1} \frac{(1- q^{\lambda_j - \lambda_k} t^{k-j-1})(1- q^{\lambda_j - \lambda_k -1}t^{k-j+1})}
{(1- q^{\lambda_j - \lambda_k} t^{k-j})(1- q_1^{\lambda_j - \lambda_k -1} t^{k-j})} 
\delta \left( q^{-\lambda_k} t^{k-1}u/z \right)\vert \lambda + 1_k \rangle,
\end{align}
and 
\begin{align}\label{FFJMM-f}
& - (1-q)  f(z) \vert \lambda \rangle \CR
&=  \sum_{k=1}^{\ell(\lambda)}\frac{1 - q^{\lambda_{k} - \lambda_{k+1}}}{1 - q^{\lambda_{k} - \lambda_{k+1}-1}t} \CR
& \qquad \times  \prod_{j=k+1}^{\infty} 
 \frac{(1- q^{\lambda_k - \lambda_j -1}t^{j-k+1})(1- q^{\lambda_{k} - \lambda_{j+1}} t^{j-k})}
{(1- q^{\lambda_{k} - \lambda_{j+1} -1} t^{j-k+1})(1- q^{\lambda_k - \lambda_j} t^{j-k})} 
\delta \left( q^{1-\lambda_k} t^{k-1}u/z \right) \vert \lambda - 1_k \rangle \CR
&=  \sum_{k=1}^{\ell(\lambda)}\prod_{j=k+1}^{\infty} 
\frac{(1- q^{\lambda_k - \lambda_j -1}t^{j-k+1})(1- q^{\lambda_k - \lambda_j} t^{j-k-1})}
{(1- q^{\lambda_k - \lambda_j - 1} t^{j-k})(1- q^{\lambda_k - \lambda_j} t^{j-k})} 
\delta \left( q^{1-\lambda_k} t^{k-1}u/z \right) \vert \lambda - 1_k \rangle.
\end{align}
where for the matching with the Pieri formula \eqref{Pieri1} we have replaced $q_1 \to q^{-1}, q_3 \to t$ in the original formula \cite{FFJMM}. 

For the base change from the Macdonald basis to the fixed point basis, we employ the following functions 
which appear in the norm of the Macdonald polynomials. 
\begin{equation}\label{Mac-norm}
c_\lambda = \prod_{\mathsf{s} \in \lambda} (1- q^{a(\mathsf{s})} t^{\ell(\mathsf{s})+1}), 
\qquad c'_\lambda = \prod_{\mathsf{s} \in \lambda} (1- q^{a(\mathsf{s})+1} t^{\ell(\mathsf{s})}). 
\end{equation}
Namely, up to the monomial factor\footnote{As we will see below, we need
an additional monomial factor $t^{n(\lambda)}$ for a compete matching.}
the fixed point basis is nothing but the integral form of the Macdonald polynomials. 
In the proof of Lemma 6.1 in \cite{Awata:2011ce}, it was used that $c_\lambda$ and $c'_\lambda$ 
satisfy the recursion relations;\footnote{See also Appendix A of \cite{Awata:2017lqa}.}
\begin{align}\label{var-c}
\frac{c_{\lambda + 1_k}}{c_{\lambda}} &= (1- q^{\lambda_k} t^{\ell(\lambda) -k+1}) 
\prod_{i=1}^{k-1} \frac{1-q^{\lambda_i - \lambda_k-1}t^{k-i+1}}{1- q^{\lambda_i -\lambda_k -1} t^{k-i}}
\prod_{j=k+1}^{\ell(\lambda)} \frac{1- q^{\lambda_k -\lambda_j } t ^{j-k}} {1- q^{\lambda_k -\lambda_j}  t^{j-k+1 }} \CR
&= t^{k-1} \prod_{i=1}^{k-1} \frac{1-q^{\lambda_k - \lambda_i+1}t^{i-k-1}}{1- q^{\lambda_k -\lambda_i +1} t^{i-k}}
\prod_{j=k+1}^{\infty} \frac{1- q^{\lambda_k -\lambda_j } t ^{j-k}} {1- q^{\lambda_k -\lambda_j}  t^{j-k+1 }},
\\
\frac{c_{\lambda - 1_k}}{c_{\lambda}} &= t^{1-k} 
\prod_{i=1}^{k-1} \frac{1- q^{\lambda_k -\lambda_i} t^{i-k}}{1-q^{\lambda_k - \lambda_i}t^{i-k-1}}
\prod_{j=k+1}^{\infty} \frac{1- q^{\lambda_k -\lambda_j-1}  t^{j-k+1 }}{1- q^{\lambda_k -\lambda_j -1} t ^{j-k}},
\\
\frac{c'_{\lambda + 1_k}}{c'_{\lambda}} &= (1- q^{\lambda_k +1} t^{\ell(\lambda) -k})
\prod_{i=1}^{k-1} \frac{1- q^{\lambda_i -\lambda_k} t^{k-i}}{1- q^{\lambda_i -\lambda_k} t^{k-i-1}}
\prod_{j=k+1}^{\ell(\lambda)}  \frac{1- q^{\lambda_k -\lambda_j+1} t^{j-k-1}}{1- q^{\lambda_k -\lambda_j +1} t^{j-k}}.
\end{align}

To prove \eqref{var-c} let us evaluate the change of the corresponding character;
\begin{equation}
\chi _\lambda (q,t) := \sum_{\mathsf{s} \in \lambda} q^{a(\mathsf{s})} t^{\ell(\mathsf{s})+1}.
\end{equation}
If we add a new box in the $k$-th low, the change of the leg or the arm length occurs for $\mathsf{s}=(i, \lambda_k +1),~(1 \leq i \leq k-1)$
and $\mathsf{t}=(k, j),~(1 \leq j \leq \lambda_k)$. There is a new contribution from the box $(k, \lambda_k +1)$, which is $1- t$. 
The variation from the boxes of the first kind is
\begin{equation}
\delta_1 \chi _\lambda (q,t) = \sum_{i=1}^{k-1} q^{\lambda_i - \lambda_k-1} ( t^{k-i+1} - t^{k-i}).
\end{equation}
On the other hand the variation from the boxes of the second kind is more involved. First the contribution from
such boxes is
\begin{align}
\sum_{i=k}^{\ell(\lambda)} \left( t^{i-k+1} \sum_{j= \lambda_{i+1}+1}^{\lambda_i} q^{\lambda_k -j} \right) 
= \sum_{i=k}^{\ell(\lambda)} \left( t^{i-k+1}  \frac{q^{\lambda_k -\lambda_{i+1}-1} - q^{\lambda_k - \lambda_{i}-1}}{1 - q^{-1}} \right).
\end{align}
Hence the variation under $\lambda_k \to \lambda_k +1$ is 
\begin{equation}
\delta_2 \chi _\lambda (q,t) =  \sum_{i=k}^{\ell(\lambda)}t^{i-k+1}  \left(  q^{\lambda_k -\lambda_{i+1}} - q^{\lambda_k - \lambda_{i}} \right).
\end{equation}
Hence we obtain
\begin{align}
\frac{c_{\lambda + 1_k}}{c_\lambda} &= (1-t) \prod_{i=1}^{k-1} \frac{1- q^{\lambda_i - \lambda_k-1} t^{k-i+1}}{1- q^{\lambda_i - \lambda_k-1}t^{k-i}} 
 \prod_{i=k}^{\ell(\lambda)} \frac{1- q^{\lambda_k -\lambda_{i+1}} t^{i-k+1}}{1- q^{\lambda_k -\lambda_i} t^{i-k+1}} \CR
 &=  (1- q^{\lambda_k} t^{\ell(\lambda)-k+1})\prod_{i=1}^{k-1} \frac{1- q^{\lambda_i - \lambda_k-1} t^{k-i+1}}{1- q^{\lambda_i - \lambda_k-1}t^{k-i}}
 \prod_{i=k+1}^{\ell(\lambda)} \frac{1- q^{\lambda_k -\lambda_{i}} t^{i-k}}{1- q^{\lambda_k -\lambda_i} t^{i-k+1}},
\end{align}
which completes the proof of \eqref{var-c}.

Now let us make a base change from the Macdonald basis $\vert \lambda \rangle$ to the fixed point basis $[\lambda]$ 
by the scaling of $c_\lambda$. Let $\psi_k(\lambda)$ be the coefficient of $\vert \lambda + 1_k \rangle$ 
for the zero mode $e_0$ of $e(z)$ in \eqref{FFJMM-e}. Then the corresponding coefficient in the fixed point basis is
\begin{equation}\label{geom-matrix-element}
 (1-q^{-1}) \psi_k(\lambda)\frac{c_{\lambda}}{c_{\lambda + 1_k}} 
= t^{1-k} \prod_{\substack{j=1 \\j \neq k}}^\infty 
\frac{1- q^{\lambda_k - \lambda_j} t^{j-k+1}}{1- q^{\lambda_k - \lambda_j} t^{j-k}}.
\end{equation}
We will show that \eqref{geom-matrix-element} is related to the ratio of the Nekrasov factor.
Recall that the Nekarsov factor allows a formal infinite product formula \cite{Awata:2008ed};\footnote{The Nekrasov factor 
is obtained from the equivariant character \eqref{ch1}. We set $(t_1,t_2) = (t, q^{-1})$ for the matching 
with the convention of the Macdonald polynomials. Compared with the convention of \cite{Awata:2008ed}, 
$\lambda$ and $\mu$ are exchanged.}
\begin{align}\label{Nekrasov}
\mathsf{N}_{\lambda \mu}(u \vert q,t) &= \prod_{(i,j) \in \lambda } (1- u q^{-\lambda_i +j -1} t^{-\mu_j^\vee +i})
\prod_{(i,j) \in \mu} (1- u q^{\mu_i -j} t^{\lambda_j^\vee - i+1})
 \CR
&= \prod_{i,j =1}^\infty \frac{(u q^{\mu_i - \lambda_j}t^{j-i+1};q)_\infty (ut^{j-i};q)_\infty}
{(u q^{\mu_i - \lambda_j}t^{j-i};q)_\infty (ut^{j-i+1};q)_\infty}.
\end{align}
For a technical reason, we have introduced the parameter $u$, which is an equivariant parameter of $U(1)$ (flavor) symmetry.
Physically $u$ is regarded as a mass parameter.
It should be mentioned that after the cancellation of the factors in the numerator and the denominator
all the factors in the denominator of \eqref{Nekrasov} disappear and it is actually a finite product;
\begin{align}\label{finite-Nekrasov}
\mathsf{N}_{\lambda \mu}(u \vert q,t) &=
\prod_{j \geq i \geq 1}\frac{(u q^{-\lambda_i+\mu_{j+1}}t^{i-j};q)_\infty}{(u q^{-\lambda_{i}+\mu_{j}}t^{i-j};q)_\infty}
\cdot \prod_{i \geq j \geq 1}\frac{(u q^{\mu_{j}-\lambda_{i}}t^{i-j+1};q)_\infty}{(u q^{\mu_{j}-\lambda_{i+1}}t^{i-j+1};q)_\infty}
\CR
&= \prod_{j \geq i \geq 1}(u q^{-\lambda_{i}+\mu_{j+1}}t^{j-i};q)_{\mu_{j}-\mu_{j+1}}
\cdot \prod_{i\geq j \geq 1} (u q^{\mu_j - \lambda_i}t^{i-j+1};q)_{\lambda_i- \lambda_{i+1}}.
\end{align}
%Note that the inverse of the Nekrasov factor computes the equivariant character of the symmetric product $S_V$ of the relevant module $V$,
From \eqref {Nekrasov}, we obtain\footnote{It is the inverse of the Nekrasov factor that computes
the equivariant character of the symmetric product $S_V$ of the relevant module $V$.}
\begin{align}
\frac{\mathsf{N}_{\lambda, \lambda}(u \vert q,t)}{\mathsf{N}_{\lambda,\lambda+1_k}(u \vert q,t)}
&= \prod_{j=1}^\infty \frac{(u q^{\lambda_k - \lambda_j}t^{j-k+1};q)_\infty}
{(u q^{\lambda_k - \lambda_j}t^{j-k};q)_\infty} 
\frac{(u q^{\lambda_k - \lambda_j+1}t^{j-k};q)_\infty}{(u q^{\lambda_k - \lambda_j+1}t^{j-k+1};q)_\infty}  \CR
&= \frac{1-ut}{1-u} \prod_{\substack{j=1 \\j \neq k}}^\infty  
\frac{1- u q^{\lambda_k - \lambda_j} t^{j-k+1}}{1- u q^{\lambda_k - \lambda_j} t^{j-k}}. 
%&= \frac{1}{1- q^{\lambda_k} t^{\ell(\lambda)-k+1}}
%\prod_{\substack{j=1 \\ j \neq k}}^{\ell(\lambda)}\frac{1- q^{\lambda_k - \lambda_j} t^{j-k+1}}{1- q^{\lambda_k - \lambda_j} t^{j-k}}.
\end{align}
Motivated by the computation of the equivariant character of the tangent space of the correspondence in Appendix \ref{App:correspondence},
we employ a practical prescription of replacing $(1-u)$ to $(1-ut)(1-uq^{-1})$
in order to obtain the matrix elements of the operators from the correspondence. 
Hence taking the limit $u \to 0$, we arrive at
\begin{equation}
\frac{1}{(1-t)(1-q^{-1})} \operatorname{Res}_{u=0}
\left(\frac{\mathsf{N}_{\lambda, \lambda}(u \vert q,t)}{\mathsf{N}_{\lambda,\lambda+1_k}(u \vert q,t)} \right)
= \frac{1}{1-q^{-1}} \prod_{\substack{j=1 \\j \neq k}}^\infty  
\frac{1- q^{\lambda_k - \lambda_j} t^{j-k+1}}{1- q^{\lambda_k - \lambda_j} t^{j-k}}. 
\end{equation}
Up to the monomial factor $t^{1-k}$, this agrees with \eqref{geom-matrix-element}.
Finally by further multiplying the additional factor $t^{n(\lambda)}$ with
\begin{equation}\label{t-framing}
n(\lambda) := \sum_{\mathsf{s} \in \lambda} \ell_\lambda(\mathsf{s}) = \sum_{(i,j) \in \lambda} (i-1) = \sum_{i=1}^\infty (i-1) \lambda_i,
\end{equation}
we can eliminate the monomial factor $t^{1-k}$ completely.

Similarly for the coefficient $\psi'_k(\lambda)$ of $\vert \lambda - 1_k \rangle$ 
for the zero mode $f_0$ in \eqref{FFJMM-f}, we have
\begin{equation}\label{geom-matrix-element2}
- (1-q) \psi'_k(\lambda) \frac{c_{\lambda}}{c_{\lambda - 1_k}}
= t^{k-1} \prod_{\substack{i=1 \\i \neq k}}^\infty 
\frac{1- q^{\lambda_k - \lambda_i} t^{i-k-1}}{1- q^{\lambda_k - \lambda_i} t^{i-k}}.
\end{equation}
On the other hand the corresponding ratio of the Nekrasov factor should be
\begin{align}
\frac{\mathsf{N}_{\lambda, \lambda}(u \vert q,t)}{\mathsf{N}_{\lambda - 1_k, \lambda}(u \vert q,t)}
&= \prod_{i=1}^\infty \frac{(u q^{\lambda_i - \lambda_k}t^{k-i+1};q)_\infty}
{(u q^{\lambda_i - \lambda_k}t^{k-i};q)_\infty} \frac
{(u q^{\lambda_i - \lambda_k+1}t^{k-i};q)_\infty}{(u q^{\lambda_i - \lambda_k+1}t^{k-i+1};q)_\infty} \CR
& = \frac{1-ut}{1-u} \prod_{\substack{i=1 \\i \neq k}}^\infty \frac{1- u q^{\lambda_i -\lambda_k} t^{k-i+1}}{1- u q^{\lambda_i - \lambda_k} t^{k-i}}.
\end{align}
We have
\begin{equation}
\frac{1}{(1-t)(1-q^{-1})} \operatorname{Res}_{u=0}
\left(\frac{\mathsf{N}_{\lambda, \lambda}(u \vert q,t)}{\mathsf{N}_{\lambda - 1_k, \lambda}(u \vert q,t)}\right)
= \frac{1}{1-q^{-1}} \prod_{\substack{i=1 \\i \neq k}}^\infty  
\frac{1- q^{\lambda_i - \lambda_k} t^{k-i+1}}{1- q^{\lambda_i - \lambda_k} t^{k-i}}. 
\end{equation}
After making $(q,t) \to (q^{-1}, t^{-1})$ and multiplying the additional factor $t^{n(\lambda)}$,
we find the matching with \eqref{geom-matrix-element2}.

%%%%%%%%%%%%%%%%%%%%%%%%%%%%%%%%%%%%%%%%%%%%%%%%%%%%%%%%%%%%%%%%%%%%%%%%%%%%%%
%%%%%%%%%%%%%%%%%%%%%%%%%%%%%%%%%%%%%%%%%%%%%%%%%%%%%%%%%%%%%%

\end{document}